\pdfoutput=1
\documentclass[a4paper,11pt]{article}
\usepackage{amsmath,amssymb,amsfonts}
\usepackage[multiple,stable]{footmisc}
\usepackage[amsmath,amsthm,thmmarks]{ntheorem}
\allowdisplaybreaks[4]
\usepackage[noconfig]{refstyle}
\usepackage[dvipsnames]{xcolor}
\usepackage[hyperfootnotes=false, linktocpage=true, colorlinks, citecolor=blue, linkcolor=blue, urlcolor=Maroon]{hyperref}
\usepackage{array,tabularx,booktabs,geometry,subfigure,graphicx,longtable}
\usepackage[bf]{caption}
\usepackage{tikz}
\usetikzlibrary{shapes,arrows}
\tikzstyle{block} = [rectangle, draw, text width=7em, text centered, rounded corners, minimum height=3em]
\usepackage{graphicx,color}
\usepackage{cite}
\usepackage{wasysym}
\usepackage{arydshln}
\usepackage{float}

\let\eqref=\relax
\newref{eq}{name={eq.~},Name={Eq.~},names={eqs.~},Names={Eqs.~},rngtxt={-},refcmd=(\ref{#1})}
\newref{f}{name={footnote~},Name={Footnote~},names={footnotes~},Names={Footnotes~}}
\newref{app}{name={appendix~},Name={Appendix~},names={appendixes~},Names={Appendixes~}}
\newref{tab}{name={table~},Name={Table~},names={tables~},Names={Tables~}}
\newref{ch}{name={chapter~},Name={Chapter~},names={chapters~},Names={Chapters~}}
\newref{sec}{name={section~},Name={Section~},names={sections~},Names={Sections~}}
\newref{fig}{name={figure~},Name={Figure~},names={figures~},Names={Figures~}}
\numberwithin{equation}{section}
\newcommand{\eref}[1]{(\ref{#1})}

\newcommand{\eeq}{\end{equation}}
\newcommand{\beq}{\begin{equation}}
\newcommand{\ba}{\begin{array}}
\newcommand{\ea}{\end{array}}
\newcommand{\cL}{{\cal L}}

\newcommand{\cV}{{\cal V}}
\newcommand{\cT}{{\cal T}}

\newcommand{\cM}{{\cal M}}

\newcommand{\cO}{{\cal O}}
\newcommand{\cA}{{\cal A}}
\newcommand{\IP}{\mathbb P}
\newcommand{\IC}{\mathbb C}

\def\pp {{\mathbb{P}}}

\newcommand{\be}{\begin{equation}}
\newcommand{\ee}{\end{equation}}
\newcommand{\bea}{\begin{equation}\begin{aligned}}	
\newcommand{\eea}{\end{aligned}\end{equation}}		

\newcommand{\iddots}{\mathinner{\mkern2mu\raise1pt\hbox{.}\mkern2mu \raise4pt\hbox{.}\mkern2mu\raise7pt\hbox{.}\mkern1mu}}

\providecommand{\id}{\leavevmode\hbox{\small$\mathrm{1}$\kern-3.8pt\normalsize$\mathrm{1}$}}
\def\fnote#1#2{\begingroup\def\thefootnote{#1}\footnote{#2}
     \addtocounter{footnote}{-1}\endgroup}

\begin{document}

\vspace{1cm}

\title{
       \vskip 40pt
       {\huge \bf A New Construction of Calabi-Yau Manifolds: Generalized CICYs}}

\vspace{2cm}

\author{Lara B. Anderson${}^{1}$, Fabio Apruzzi${}^{2}$, Xin Gao${}^{1}$, James Gray${}^{1}$ and Seung-Joo Lee${}^{1}$}
\date{}
\maketitle
\begin{center} {\small ${}^1${\it Physics Department, Robeson Hall, Virginia Tech, Blacksburg, VA 24061, USA}}\\
{\small ${}^2${\it Institut f\"ur Theoretische Physik, Leibniz Universit\"at Hannover, Appelstrasse 2, D-30167 Hannover, Germany}}\\
\fnote{}{lara.anderson@vt.edu, fabio.apruzzi@itp.uni-hannover.de, xingao@vt.edu, jamesgray@vt.edu, seungsm@vt.edu}
\end{center}

\begin{abstract}
\noindent
We present a generalization of the complete intersection in products of projective space (CICY) construction of Calabi-Yau manifolds. CICY three-folds and four-folds have been studied extensively in the physics literature. Their utility stems from the fact that they can be simply described in terms of a `configuration matrix', a matrix of integers from which many of the details of the geometries can be easily extracted. The generalization we present is to allow negative integers in the configuration matrices which were previously taken to have positive semi-definite entries. This broadening of the complete intersection construction leads to a larger class of Calabi-Yau manifolds than that considered in previous work, which nevertheless enjoys much of the same degree of calculational control. These new Calabi-Yau manifolds are complete intersections
in (not necessarily Fano) ambient spaces with an effective anticanonical class. We find examples with topology distinct from any that has appeared in the literature to date. The new manifolds thus obtained have many interesting features. For example, they can have smaller Hodge numbers than ordinary CICYs and lead to many examples with elliptic and K3-fibration structures relevant to F-theory and string dualities. \end{abstract}

\thispagestyle{empty}
\setcounter{page}{0}
\newpage

\tableofcontents

\section{Introduction} \label{intro}

Calabi-Yau manifolds constructed as complete intersections of polynomial hypersurfaces in products of projective spaces, abbreviated here as `CICY's, have proven to be a very useful class of compactification manifolds for string theory. The CICY three-folds were first discussed and classified in a series of articles in the 1980's \cite{Hubsch:1986ny,Candelas:1987kf,Green:1986ck,Candelas:1987du}. Since that time, these manifolds have been used in a wide variety of different studies, including recent work on moduli stabilization and model building in string phenomenology (see \cite{Anderson:2011ns,Anderson:2011ty,Anderson:2012yf,Anderson:2013qca,Anderson:2013xka,Anderson:2014hia,Buchbinder:2014qca} for some recent examples). Recently the CICY four-folds have been classified and studied \cite{Brunner:1996bu,Gray:2013mja,Gray:2014kda,Gray:2014fla} and work is underway to use this dataset in the context of F-theory compactifications to $4$-dimensions. One of the reasons why this method of constructing Calabi-Yau manifolds has proven to be so useful is that it is exceptionally simple to study and work with. This simplicity stems from the fact that the complicated Calabi-Yau geometry is embedded within an extremely simple ambient space. Many of the geometric quantities of interest on the Calabi-Yau manifold can be worked out via relations to associated ambient space quantities - leading to an exceptional degree of computational control. In this paper, we will generalize the CICY construction in a manner which is applicable to any dimensionality of Calabi-Yau manifold. 

A  CICY can be described by a configuration matrix which encodes the data essential to the definition of the manifold. An example is as follows,\footnote{Strictly speaking, configuration matrices describe an entire family of varieties and a variety is only determined once a choice has been made of its complex structure within the family. In this sense, it can be misleading to use the equality symbol in characterizing a specific variety by its configuration matrix. When no confusions arise, however, we will employ such a conventional abuse of notation.}
\begin{eqnarray} \label{eg1}
X_1 = \left[ \begin{array}{c||ccc} \mathbb{P}^2 & 1 & 1 & 1 \\ \mathbb{P}^4 & 3& 1& 1\end{array}\right] \ ,
\end{eqnarray}
where this describes a manifold, $X_1$, that is embedded in an ambient space $\mathbb{P}^2 \times \mathbb{P}^4$ as the complete intersection of the solution of three polynomial equations, one for each column of integers in (\ref{eg1}). The integers themselves denote the degree of the defining polynomials in the homogeneous coordinates of the ambient projective factors (here one multi-degree $(1,3)$ and two multi-degree $(1,1)$ defining relations). Note that, in this example, $X_1$ is of complex dimension $3$ since the ambient space is of dimension $6$ and there are $3$ polynomials that define this complete intersection. Equivalently, $X_1$ is defined as a complete intersection of global sections of three line bundles. That is, the common solutions of one element\footnote{See Appendix \ref{line_cohom} for a review of notation and tools to compute line bundle cohomology.} of $H^0(\mathbb{P}^2 \times \mathbb{P}^4 , {\cal O}(1,3))$ and two elements of $H^0(\mathbb{P}^2\times \mathbb{P}^4,{\cal O}(1,1))$. Due to their interpretation as the degrees of polynomial equations, the integer entries in a CICY configuration matrix such as (\ref{eg1}) are taken to be positive semi-definite. In the case of generalized CICYs, we will drop this requirement, in general allowing for negative degrees in the defining relations in the ambient space. In order to explain better the correspondence between the actual variety and the configuration matrix, we illustrate now an example of a generalized CICY.

\subsection*{A simple example}

A generalized CICY, or ``gCICY", will be described by a configuration matrix, similar to (\ref{eg1}), but in which negative integers are now permitted to be used. For example,
\begin{eqnarray} \label{eg2}
X_2 = \left[ \begin{array}{c||cc|cc} \mathbb{P}^1 & 1 & 1 & -1 & 1 \\ \mathbb{P}^1 & 1 & 1 & 1 & -1 \\ \mathbb{P}^5 & 3 & 1 & 1& 1\end{array}\right]\;.
\end{eqnarray}
In this example one cannot define $X_2$ as a simple complete intersection of sections of line bundles on the ambient space $\cA \cong \mathbb{P}^1 \times \mathbb{P}^1 \times \mathbb{P}^5$. This is due to the simple fact that $h^0( \mathbb{P}^1 \times \mathbb{P}^1 \times \mathbb{P}^5, {\cal O}(1,-1,1))=0$, for example. To state the problem in another way, one cannot have a polynomial in ambient space coordinates with negative degree! However if we apply the polynomial conditions sequentially from left to right there is no such problem. Define a manifold, ${\cal M}$, to be the complete intersection of the first two polynomials (columns) in the ambient space. Then the remaining two line bundles do have sections on $\cM$ and as such a manifold can be defined associated to the full configuration matrix in \eref{eg2}. Note that the sections of the last two line bundles should correspond to polynomials in appropriately chosen coordinates for ${\cal M}$ although they are not polynomial in the homogeneous coordinates of the ambient space ${\cal A}$. 

Let us look at what the defining equations of $X_2$ are explicitly in terms of the homogenous coordinates of $\cA$. The first two columns correspond to the vanishing of polynomial equations, $p_1=0$ and $p_2=0$ (which we can take to be generic polynomials of appropriate degrees in the homogeneous coordinates). The remaining two defining relations, $q_1=0$ and $q_2=0$, will each describe the vanishing of a specific rational function. Since we know that these rational functions should correspond to some polynomial defining relations in the coordinates of ${\cal M}$, they should not have poles on that surface (as described by $p_1=p_2=0$). We therefore require that the poles in these rational functions do not lie on the solution set of the first two defining relations.  Denoting the ambient homogenous coordinates in $\mathbb{P}^1\times \mathbb{P}^1\times \mathbb{P}^5$ as $\mathbf x_1 = (x_{1}^0:x_{1}^1)$, $ \mathbf x_2 = (x_{2}^0:x_{2}^1)$, and  $\mathbf x_3 = (x_{3}^0 : x_{3}^1: x_{3}^2: x_{3}^3 : x_{3}^4 : x_{3}^5)$, respectively, one simple way to ensure this is the following. We write
\begin{eqnarray} \label{p2}
p_2 = c_0(\mathbf{x}_1, \mathbf{x}_3)\, x_2^0 + c_1 (\mathbf{x}_1, \mathbf{x}_3) x_2^1 = d_0(\mathbf{x}_2, \mathbf{x}_3) x_1^0 + d_1(\mathbf{x}_2, \mathbf{x}_3) x_1^1\;.
\end{eqnarray}
In this expression, $c_0$ and $c_1$ are generic polynomials of degree $(1,0,1)$ in the homogeneous coordinates $(\mathbf{x}_1,\mathbf{x}_2, \mathbf{x}_3)$ of the three ambient space factors. Likewise, $d_0$ and $d_1$ are generic degree $(0,1,1)$ polynomials in the same variables. Then the following rational expressions for $q_1$ and $q_2$,
\begin{eqnarray} \label{p3p4}
q_1 &=& \frac{d_1}{x_1^0} = - \frac{d_0}{x_1^1} \ , \\ \nonumber
q_2 &=& \frac{c_1}{x_2^0} = - \frac{c_0}{x_2^1} \ ,
\end{eqnarray}
which are of the right multi-degree, can be used to algebraically embed $X_2$ in $\cM$. 
The poles of these expressions miss the locus $p_2=0$, in the sense that, when the denominator vanishes, the numerator has to vanish to the same order. For example, if $x_1^0=0$ in (\ref{p3p4}), then $p_2=0$, as in (\ref{p2}), implies that $d_1=0$ as well (the two $\mathbf{x}_1$ homogeneous coordinates cannot simultaneously vanish).  The fact that we can write two equivalent expressions for each of $q_1$ and $q_2$ is a direct consequence of (\ref{p2}), which implies the second equalities in each line of (\ref{p3p4}) if $p_2=0$.

The expressions (\ref{p3p4}) allow us to gain a very concrete picture of the manifold $X_2$ described by (\ref{eg2}). On an open patch of the ambient space where $x_1^0 \neq 0$ and $x_2^0 \neq 0$ we observe that the manifold is defined by the solutions to the equations $p_1=p_2 =d_1=c_1=0$. Thus on such a coordinate patch the manifold is simply the common solution locus of a set of polynomial equations. Such a description does not hold globally, however. If we look instead at the patch where $x_1^1 \neq 0$ and $x_2^0 \neq 0$, 
the expression $\frac{d_1}{x_1^0}$ for the $q_1$ equation in (\ref{p3p4}) is not the most useful one since there is a locus in this patch where the denominator $x_1^0$ vanishes. Using the second expression, $-\frac{d_0}{x_1^1}$, however, leads again to a polynomial description of our manifold by $p_1=p_2=d_0=c_1=0$. Such a trick works for any given patch and we thus learn that the manifold is defined locally by polynomial equations, although the exact polynomials involved vary as one changes the coordinate chart one is considering. Note in particular that, since no single polynomial $d$ can globally replace $q_1$ for instance, this manifold is {\it not} equivalent to a manifold of the form
\begin{eqnarray} \label{thischap1}
\left[ \begin{array}{c||cccc} \mathbb{P}^1 &1 &1 & 1 & 0 \\ \mathbb{P}^1 &1 &1 &0&1 \\ \mathbb{P}^5 & 3&1&1&1\end{array} \right]\;,
\end{eqnarray}
which one would naively obtain simply by setting a numerator of each rational defining relation to zero globally. This is just as well, as the configuration matrix (\ref{thischap1}) does not describe a Calabi-Yau manifold.

\vspace{0.1cm}

Constructions of the form we are discussing may seem a little strange at first to the reader who is used to dealing with ordinary CICYs. However it is simply an extreme example of a very well studied phenomenon. Many CICYs have so-called  ``non-polynomial" deformations \cite{Green:1987rw,Candelas:1993dm,Candelas:1994hw,Mavlyutov, 2004InMat.157..621M}. That is, they have complex structure deformations which are not visible as changes to the polynomial defining relations in the ambient space coordinates. In the case of generalized CICYs we have taken a logical limit of this situation in which {\it all} of the complex structure associated to certain defining relations are non-polynomial. These new manifolds are also very similar to previous constructions in that they are defined simply as algebraic hypersurfaces (or complete intersections) in simple ambient spaces. However, unlike in previous constructions, here the ambient spaces themselves are not necessarily toric nor Fano. Instead, these ambient spaces themselves are defined as complete intersections in products of projective spaces and are characterized by the fact that their anticanonical divisor is {\it effective but not, in general, ample} (as in the Fano case). We refer the readers to Refs.~\cite{hacon,Mori88, clemens89} for a discussion of the importance of the form of the anticanonical divisor in recent mathematical classifications in the context of the Minimal Model program \cite{Fujino09}.
	
This class of Calabi-Yau manifolds shares many of the advantages of the ordinary CICYs, including much of the same ease in computation, a preponderance of elliptically fibered examples and so forth. It is, however, clearly a more general class of manifolds. In addition gCICYs have many interesting features; for example a tendency to lead to Calabi-Yau manifolds with smaller Hodge numbers than the former construction. Finally, the fact that, unlike the ordinary CICYs, not every gCICY is smooth, leads to the possibility of the dataset giving rise to simple, computable examples with, in the language of the F-theory literature, manifest non-Higgsable clusters \cite{Morrison:2012np,Grassi:2014zxa,Morrison:2014lca}. Such global constructions of this phenomenon could greatly aid current explorations of the F-theory landscape of vacua. We hope to obtain a complete classification of the gCICY three- and four-folds in future work \cite{US_FUTURE}. In the present work we generate an initial dataset of $2,761$ three-fold geometries many of which are manifestly new CY manifolds with distinct topology not appearing before in existing datasets of Calabi-Yau three-folds. For comparison, we refer the reader to
\cite{Candelas:1987kf,Kreuzer:2000xy,batyrev_grassman,batyrev_flag,Batyrev:2008rp,Klemm:1992bx,2008arXiv0802.3669K,Hori:2006dk,Donagi:2007hi,Hori:2011pd,Jockers:2012zr,Jockers:2012dk,rodland2,Lynker:1998pb} for other classes of constructions of CY manifolds.

The rest of this paper is structured as follows. In Section \ref{constrandtop} we describe the construction of gCICYs in detail. We begin with our conventions for defining a gCICY configuration matrix in Subsection \ref{constr}. In Subsection \ref{subsec:topology} we describe how to compute the topology associated to such a gCICY configuration matrix, assuming that the variety itself is smooth. In Section \ref{secsmooth} we describe how to construct the gCICY geometry explicitly via its defining relations, and in the process detail how a smoothness check can be performed on the variety. In Section \ref{redun} we detail some of the redundancies that can occur relating different configuration matrices to the same Calabi-Yau manifold. Some of these redundancies mirror the standard CICY case and some do not. While all the techniques described in Sections~\ref{constrandtop}-~\ref{redun} apply to gCICYs of an arbitrary dimension, in Section \ref{classcases} we focus on Calabi-Yau three-folds and take the first steps towards a classification of gCICYs, by studying and classifying certain low co-dimensional sub-classes and detail their geometry. Finally in Section \ref{phys_outlook} we summarize the physics applications of this new dataset, including obvious elliptic fibration structures~\cite{Gray:2014fla} of gCICYs, as well as fibration structures involving K3 and higher-dimensional Calabi-Yau varieties and provide an outlook of interesting future directions of investigation.

\section{The Construction and Topology} \label{constrandtop}

\subsection{The construction} \label{constr}

Our construction of gCICYs is based on a multi-step process. We first impose all of the semi-positive defining relations to obtain an ordinary complete intersection, $\cM$, of a set of polynomial relations in the ambient product of projective spaces, while the rest of the defining relations (including negative entries) are to be sequentially imposed on $\cM$. 

Let us be more specific and begin by constructing ${\cal M}$ as the complete intersection of $K$ polynomials, $p_{\alpha}$ where $\alpha=1, \cdots, K$, in a product of projective spaces, $\cA= \IP^{n_1} \times \cdots \times \IP^{n_m}$. We choose the dimension of $\cM$,
\beq
{\rm dim}_\IC \; \cM = \sum\limits_{r=1}^m n_r - K  \ ,
\eeq  
such that it is strictly greater than the desired dimension, $N$, of the Calabi-Yau manifold to be constructed. Note that, as in a standard CICY, the polynomials $p_\alpha$ are sections of an appropriate line bundle $\cO_\cA(a^{1}_\alpha, \cdots, a^{m}_\alpha)$, with $a^{r}_{\alpha} \geq 0$ specifying the non-negative homogeneous degree of $p_\alpha$ in the $r$-th projective piece. Here the indices $r, s, \cdots  = 1, \cdots, m$ are used to label the projective ambient space factors $\IP^{n_r}$, and the indices $\alpha, \beta, \cdots = 1, \cdots, K$, to label the polynomials, $p_\alpha$. A family of such geometries can be characterized by a configuration matrix of the form, 
\bea\label{confForM}
\left[ \,\mathbf n\, ||\, \{\mathbf a_\alpha\} \,\right] = \def\arraystretch{1.2}\left[\ba{c||ccc} 
\IP^{n_1}&  a^{1}_{1} & \cdots &a^{1}_{K} \\
\IP^{n_2} &  a^{2}_{1} & \cdots & a^{2}_{K}\\
\vdots &  \vdots &\ddots&\vdots\\
\IP^{n_m} & a^{r}_{1} & \cdots& a^{r}_{K}\\ 
\ea\right]  \ , 
\eea
with non-negative integer entries $a^r_\alpha$. 

\vspace{0.1cm}

Inside ${\cal M}$ we define the generalized CICY, $X$, in terms of an ordered list of line bundles ${\cal L}_{\mu}$ where $\mu=1,\ldots,L$. These are used to iteratively describe hypersurfaces inside a nested sequence of spaces as follows. First, we demand that $h^0({\cal M},{\cal L}_1)\neq0$ and define a hypersurface ${\cal M}_1$ inside ${\cal M}$ as the vanishing locus of a global section in $H^0(\cM, \cL_1)$. Second, we demand that $h^0({\cal M}_1,{\cal L}_2)\neq 0$ and define a hypersurface ${\cal M}_2$ inside ${\cal M}_1$ as the vanishing locus of a global section in $H^0(\cM_1, \cL_2)$. We sequentially repeat this procedure for all the line bundles $\cL_\mu$ until the gCICY is obtained as $X=\cM_L$. Note that, in general, one needs not have a section of ${\cal L}_{\mu}$ on the ambient space ${\cal A}$, or even on ${\cal M}$ for $\mu \geq 2$. We need only have sections of ${\cal L}_{\mu}$ on ${\cal M}_{\mu-1}$, i.e. we simply require that $h^0({\cal M}_{\mu-1},{\cal L}_{\mu})\neq0$. As such, as we have already discussed, {\it gCICYs are not described by the solution set of a system of simple polynomial equations in the ambient space ${\cal A}$}.

By writing ${\cal L}_{\mu}=\cO_{\cM_{\mu-1}}(b^1_{\mu}, \cdots, b^m_{\mu})$, $\mu=1,\ldots L$, with $\cM_0=\cM$, we present the defining data of a gCICY by the following matrix, 
\bea\label{conf}
\left[ \,\mathbf n\, ||\, \{\mathbf a_\alpha\}\,|\, \{\mathbf{b}_\mu\}\,\right] = \def\arraystretch{1.2}\left[\ba{c||ccc|ccc} 
\IP^{n_1}&  a^{1}_{1} & \cdots &a^{1}_{K} & b^1_1 & \cdots & b^1_{L} \\
\IP^{n_2} &  a^{2}_{1} & \cdots &a^{2}_{K} & b^2_1 & \cdots & b^2_{L}\\ 
\vdots &  \vdots &\ddots&\vdots & \vdots & \ddots & \vdots \\
\IP^{n_m} & a^{m}_{1} & \cdots&a^{m}_{K}  & b^m_1 & \cdots & b^m_{L} \\ 
\ea\right]  \ . 
\eea
This looks very much like an ordinary CICY configuration matrix with just two important differences. Firstly, the entries $b^r_\mu$ can be negative, although the $a^r_\alpha$ are still positive semi-definite. Secondly, the order of the columns involving $b$'s is important and we establish the convention that one restricts to sections of the line bundles in an order taken from left to right in constructing the Calabi-Yau $N$-fold. Clearly, not every such matrix describes a gCICY, as the ${\cal L}_{\mu}$ must be chosen so as to have global sections over ${\cal M}_{\mu-1}$. However, by definition, every gCICY can be described by such a configuration matrix and we will see that much of the familiar utility of describing manifolds with such objects is preserved in this case. 

Clearly, the number, $L$, of the line bundles on $\cM$ has to be, 
\beq
L={\rm dim}_\IC\; \cM -  N = \sum\limits_{r=1}^m n_r - K - N\ , 
\eeq
if we want the resulting subvariety, $X \subset \cM$, to be an $N$-fold. The $N$-dimensional variety $X$ has codimension $K+L$ in $\cA$. However, since the $K$ sections $p_\alpha$ and the $L$ sections $q_\mu$ are of a different type, we say that $X$ is of codimension $(K,L)$ in this case. The condition that the resulting variety is Calabi-Yau remains unaltered: namely that the sum of the degrees of the entries in each row must equal $n_r+1$ for its respective $\mathbb{P}^{n_r}$ factor. This will  be discussed in detail in the next subsection. In the case of a single hypersurface with $L=1$, this is the familiar condition that $X$ must be the anticanonical hypersurface of ${\cal M}$.

\vspace{0.1cm}

It should be noted that there is a sub-type of gCICYs which are particularly easy to describe and which display many of the interesting properties of the dataset. These are the cases where $h^0({\cal M},{\cal L}_{\mu})\neq 0$ for all the $\mu$'s. In this case it frequently does not matter in which order one restricts to sections of the ${\cal L}_{\mu}$ as they all have global sections already on ${\cal M}$ \footnote{Because the number of sections of a given line bundle may vary on the different ${\cal M}_{\mu}$, it should be noted that the order in which the line bundles are placed may still be important in questions such as deciding smoothness of the resulting variety.}. We refer to such a generalized CICY as a minimally generalized CICY or an ``mgCICY." Many of the concrete examples that we discuss in what follows will be of this type.
	
\subsection{Determining the topology of generalized CICY manifolds} \label{subsec:topology}

In this subsection tools are provided to compute topological properties of a gCICY from its configuration matrix. For the moment, we will make the assumption that the associated variety is smooth\footnote{In fact we can somewhat relax this assumption, as we will describe in later sections.}. Unlike in the CICY case, gCICYs are not always generically smooth and this property must be explicitly tested as will be explored in Section~\ref{secsmooth}. Nevertheless we find it useful to describe the computation of topological properties of the dataset first and then proceed on to verifying smoothness in a second step. This allows us to preserve the separation between topology, as derived by integer manipulation of the configuration matrix, and geometry, as determined by the detail of the defining relations, that is seen in the discussion of standard CICYs. In any case, it is useful to determine which configuration matrices could have interesting topology, such as vanishing first Chern class, before proceeding on to the more detailed question of whether or not they are smooth.

For the rest of this section, we explore such defining topological quantities as the Hodge numbers, Chern classes, and triple intersection numbers of our new class of Calabi-Yau manifolds. We will also discuss the computation of other geometrical data such as the K\"ahler and Mori cones and line bundle cohomology which is crucial in order to extract the effective physics in a string compactification. One important omission that we do not consider here is a study of the first fundamental group, $\pi_1(X)$, of these geometries. Unlike in the case of ordinary CICY manifolds, it is not guaranteed that the gCICY manifolds are simply connected. Due to the absence of such useful tools as the Lefschetz Hyperplane theorem (employed in the case of CICYs \cite{Hubsch:1992nu}) we leave a complete study of this to a future classification.

Much of the formalism that we describe below applies to any Calabi-Yau three-fold described as an algebraic hypersurface inside some K\"ahler ambient space and is not unique to the construction at hand. For a more complete overview of these tools, we refer the reader to such useful references as \cite{Hubsch:1992nu,huybrechts}. Below we will review some of these general results and focus on how they can be implemented for generalized complete intersection manifolds.

\subsubsection{The Koszul complex}\label{kosz_sec}

To begin, consider a generalized complete intersection manifold, $X$, defined as a single hypersurface in the compact, K\"ahler ambient space ${\cal M}$. This construction will be iterated below to describe more general complete intersection manifolds. As described previously, these ambient spaces themselves are each constructed as complete intersection manifolds in products of projective spaces \cite{Hubsch:1986ny,Candelas:1987kf,Green:1986ck,Candelas:1987du}. By Bertini's theorem (see \cite{hartshorne, GH}), ${\cal M}$ constructed in this way are generically smooth K\"ahler manifolds. As in many other constructions of CY manifolds, we begin by constructing $X$ as an {\it anticanonical hypersurface} in ${\cal M}$, and thus we will demand that ${\cal K}^{\vee}_{\cal M}$ has global sections ($h^0({\cal M}, {\cal K^{\vee}}_{\cal M}) >0$). However, unlike in many constructions of CY manifolds, ${\cal M}$ will is not required to be Fano (that is, ${\cal K}^{\vee}_{\cal M}$ is not necessarily ample).

To construct the hypersurface, $X$, we will impose on the coordinates of ${\cal M}$, collectively denoted by $x$, the algebraic equation $p(x)=0$ in the class $D$ (i.e. $p$ is a global, holomorphic section of the line bundle ${\cal O}_{\cal M}(D)$ and $D=[p(x)=0]$). Then we have the familiar Adjunction Formula \cite{hartshorne, GH}:
\beq
0 \rightarrow TX \rightarrow  T{{\cal M}}|_{X} \rightarrow  \cO_\cM(D)|_{X} \rightarrow 0 \ . \label{adjunction}
\eeq
As will be argued below, $X$ will satisfy $c_1(X)=0$ and thus, is a CY manifold when ${\cal O}(D)={\cal K^{\vee}}_{\cal M}$ \cite{yau,Yau:1986gu}. Using \eref{adjunction}, the tangent bundle valued cohomology, $H^{\ast}(X, TX)$, can be computed by restriction from ${\cal M}$. The most useful tool to this end is the familiar  Koszul short exact sequence \cite{GH, hartshorne}:
\beq
0 \rightarrow {\mathcal I}_{X|{\cal M}} \rightarrow  {\cal O}_{{\cal M}} \rightarrow  {\cal O_\cM}|_{{X}} \rightarrow 0 \ , \label{koszul_pre}
\eeq
where ${\mathcal I}_{X|{\cal M}}$ is the ideal sheaf of $X \subset {\cal M}$. In the case the divisor $D$ is reduced and irreducible\footnote{More specifically, ${\cal I}$ is invertible if and only if $D$ is cut out by a single equation, not vanishing at any associated point of ${\cal M}$. In this case, $D$ is called an effective Cartier divisor \cite{hartshorne, GH}.}, the ideal sheaf is invertible and a line bundle, leading to
\beq
0 \rightarrow {\cal O}_{{\cal M}}(-D) \rightarrow  {\cal O}_{{\cal M}} \rightarrow  {\cal O}_\cM|_{{X}} \rightarrow 0 \ . \label{koszul}
\eeq
Moreover, for any vector bundle, $\pi: {\cal V} \to {\cal M}$, we can take the tensor product with the above exact sequence to produce another short exact sequence (i.e. the above sequence of sheaves/modules is {\it flat} \cite{hartshorne}):
\beq
0 \rightarrow {\cal O}_{{\cal M}}(-D) \otimes {\cal V} \rightarrow  {\cal V} \rightarrow  {\cal V}|_{{X}} \rightarrow 0 \ . \label{koszul_tensor}
\eeq
From this sequence we can use the cohomology of any bundle $\cV$ on ${\cal M}$ to determine the cohomology of the restricted bundle $\cV|_{X}$. It should be noted that in most known algebraic constructions of CY manifolds \cite{Hubsch:1986ny,Candelas:1987kf,Green:1986ck,Candelas:1987du}, this sequence is defined for {\it ample} divisors $D$. Such a choice guarantees that the variety $X$ swept out by the condition $p(x)=0$ is generically a smooth, irreducible variety. For the case at hand, however, we will simply require that $D$ is effective (and frequently nef) and that the line bundle ${\cal O}_\cM(D)$ has global sections. In general the variety $X$ defined by the global sections of a non-Ample line bundle may be singular, reducible, or non-reduced. To proceed, we must first consider the global sections of $\cO_\cM(D)$ and whether or not they can be used to define smooth manifolds. The details of this analysis will be described in Section \ref{secsmooth}. For now, we will simply assume that such good sections exist and that $X$ is smooth, but it is important to note that the following analysis is only valid for smooth $X$.

To determine the topology of $X$ then, consider the long exact sequence in cohomology associated to \eref{adjunction}:
\beq
0 \to H^{0}(X,TX) \rightarrow H^0(X, T{{\cal M}}|_{X} ) \rightarrow H^0(X,O(D)|_{X} ) \rightarrow H^{1}(X,TX) \rightarrow \cdots \label{les_adj}
\eeq
where the cohomology of bundles restricted to $X$ can be computed using \eref{koszul_tensor} and standard techniques (see Section \ref{line_cohom}).

At this point, determining the tangent bundle valued cohomology $h^\ast(X,TX)$ is as straightforward as any other construction of a Calabi-Yau manifold as a complete intersection manifold and the techniques employed are entirely analogous. This can be concisely summarized by the two-step procedure: \\
\begin{enumerate}
\item [1)] Compute bundle-valued cohomology on the smooth K\"ahler manifold ${\cal M}$.
\item [2)] Use the Koszul \eref{koszul} and Adjunction \eref{adjunction} sequences as well as the long exact sequence \eref{les_adj} to determine bundle valued cohomology on $X$. 
\end{enumerate}
The first of these steps is now well-understood in the physics literature and the reader is referred to \cite{Hubsch:1992nu} for a review. The way that these standard results can be implemented for gCICYs in step 2) is most simply understood in the context of an example.

\subsubsection{An illustration of the determination of Hodge numbers}\label{example_hodgy}
The calculation of Hodge numbers is best understood by illustration and we consider a simple example here. Let $\tiny{{\cal M}= \left[\begin{array}{c||c}
\pp^{1} & 3 \\
\pp^{4} & 2 
\end{array}\right]}$ and $X$ be defined by a generic global section of the line bundle ${\cal O}_{\cal M}(-1,3)$ (where ${\cal O}_\cM(a,b)$ denotes the line bundle associated to the divisor $aH_1+bH_2$ where $H_r$ are the ambient hyperplanes restricted to ${\cal M}$). Note that on the ambient product of projective spaces ${\cal A}=\pp^1 \times \pp^4$ the line bundle ${\cal O}_{\cal A}(-1,3)$ is not ample and has no global sections. However, on the hypersurface ${\cal M}$, $h^0({\cal M}, {\cal O}_\cM(-1,3))=15$ and it is possible to consider an algebraic variety swept out by one such general global holomorphic section. As described above, in analogy to the configuration matrices used to describe ordinary CICY manifolds \cite{Hubsch:1986ny,Candelas:1987kf,Green:1986ck,Candelas:1987du}, we will schematically denote the final CY manifold $X$ here by the following configuration matrix
\begin{equation}\label{firsteg}
X=\left[\begin{array}{c||c|c}
\pp^{1} & 3 & -1\\
\pp^{4} & 2 & 3
\end{array}\right] \ . 
\end{equation}
Using the techniques described in \cite{hartshorne, GH, Hubsch:1992nu} and the Euler sequence for the tangent bundle to projective space,
\beq
 0 \to {\cal O}_\cA^{\oplus 2} \to {\cal O}_\cA(1,0)^{\oplus 2} \oplus {\cal O}_\cA(0,1)^{\oplus 5} \to T{\cal A} \to 0 \ ,
\eeq
and the key sequences \eref{koszul} and \eref{adjunction}:
\begin{align}
& 0 \to {\cal O}_{\cal A}(-3,-2) \to {\cal O}_{\cal A} \to {\cal O}_\cA|_{\cal M} \to 0 \ , \\
& 0 \to T{\cal M} \to T{\cal A}|_{\cal M} \to {\cal O}_\cA(3,2)|_{\cal M} \to 0 \ .
\end{align}

It is straightforward to determine the tangent bundle valued cohomology on ${\cal M}$ to be
\beq
h^\ast({\cal M}, T{\cal M})=(0,32,0,0,0) \ .
\eeq
To move on to the second step described in the previous Subsection, consider next the adjunction formula \eref{adjunction} for the CY manifold itself:
\beq
0 \to TX \to T{\cal M}|_{X} \to {\cal O}_\cM(-1,3)|_X \to 0 \ . \label{adj_here}
\eeq
If the cohomologies, $h^{\ast}(X, T{\cal M}|_{X})$ and $h^\ast(X,  {\cal O}_\cM(-1,3)|_X)$ are each determined, then \eref{les_adj} can be used to find the Hodge numbers of $X$. First, to determine the cohomology of ${\cal O}_\cM(-1,3)|_{X}$ the Koszul sequence yields
\beq
0 \to {\cal O}_{\cal M} \to {\cal O}_{\cal M}(-1,3) \to {\cal O}_\cM(-1,3)|_X \to 0 \ .
\eeq
From the long exact sequence in cohomology associated to this we have 
\beq
h^*(X,{\cal O}(-1,3)|_X)=(14,0,0,0) \ .
\eeq
Likewise, for $T{\cal M}$:
\beq
0 \to T{\cal M} \otimes {\cal O}_\cM(1,-3) \to T{\cal M} \to T{\cal M}|_{X}  \to 0  \ . \label{MtoX}
\eeq
The cohomology of the twisted bundle $T{\cal M} \otimes {\cal O}_\cM(1,-3)$ can be computed in the same manner as that of $T{\cal M}$ described above. Using
\begin{align}
& 0 \to {\cal O}_\cA(1,-3)^{\oplus 2} \to {\cal O}_\cA(2,-3)^{\oplus 2}\oplus {\cal O}_\cA(1,-2)^{\oplus 5} \to T{\cal A}\otimes {\cal O}_\cA(1,-3) \to 0 \ ,  \\
& 0 \to {\cal O}_{\cal A}(-3,-2) \to {\cal O}_{\cal A} \to {\cal O}_\cA|_{\cal M} \to 0 \ , \\
& 0 \to T{\cal M} \otimes {\cal O}_\cM(1,-3) \to (T{\cal A}\otimes {\cal O}_\cA(1,-3) )|_{\cal M} \to {\cal O}_\cM(4,-1)\to 0 \ .
\end{align}
It can be verified that
\begin{align}
& h^\ast({\cal M},{\cal O}_\cM(4,-1))=(0,0,0,0,0) \ ,\\
&h^\ast({\cal M}, (T{\cal A}\otimes {\cal O}_\cA(1,-3) )|_{\cal M})=(0,0,0,2,0) \ ,
\end{align}
which can now be combined in the long exact sequence in cohomology associated to \eref{MtoX}:
\beq
h^\ast(X, T{\cal M}|_{X})=(0,32,2,0) \ .
\eeq
At last then, the pieces can be put together. The long exact sequence in cohomology associated to \eref{adj_here}
\beq
0 \to H^{0}(X,TX) \rightarrow H^0(X, T{{\cal M}}|_{X} ) \rightarrow H^0(X,{\cal O}_\cM(-1,3)|_{X} ) \rightarrow H^{1}(X,TX) \rightarrow \cdots \label{les_adj2} 
\eeq
simplifies to 
\beq
0 \to H^0(X,{\cal O}_\cM(-1,3)|_{X}) \to H^1(X,TX) \to H^1(X, T{\cal M}|_{X}) \to 0 \to H^2(X,TX) \to H^2(X, T{\cal M}|_{X}) \to 0 \ .
\eeq
Thus, the Hodge numbers of the CY are
\begin{align}\label{hodgy}
&h^{2,1}(X)=h^1(TX)=46 \ , \\
&h^{1,1}(X)=h^1(TX^{\vee})=h^2(TX)=2 \ .
\end{align}
An interesting observation at this stage is that once again, $h^1(TX) > h^0({\cal M}, {\cal O}_\cM(-1,3))$. That is, this generalized complete intersection manifold is ``non-polynomial" twice over! Not only can it not be realized as a complete intersection hypersurface in $\pp^1 \times \pp^4$ alone, but moreover, there are more complex structure moduli than are realized as coefficients of the defining relation $p(x)=0$ on ${\cal M}$. This example illustrates how much general structure can potentially be overlooked by considering standard CICY manifolds. 

Having obtained a gCICY manifold with $\chi = -88$ we are led now to consider the remaining topology of $X$. In the next section we will consider the Chern classes and triple intersection numbers of $X$ and perform an independent check of the Euler number given above.

\subsubsection{Chern classes and triple intersection numbers}\label{cherny}

Once again we begin by considering the co-dimension $1$ case, in which $X$ is defined as a single hypersurface in ${\cal M}$. Since $X \subset {\cal M}$ is a K\"ahler submanifold of a K\"ahler space ${\cal M}$, it is easiest to discuss the Chern classes of $X$ via pullback. Let $i: X \hookrightarrow {\cal M}$ denote the embedding of $X$ in ${\cal M}$. In this notation, the Adjunction formula \eref{adjunction} takes the form
\beq
0 \rightarrow TX \rightarrow i^\ast (T{\cal M}) \rightarrow i^\ast ( {\cal K^{\vee}}_{\cal M}) \rightarrow 0 \ .\label{adj_again}
\eeq
By the commutivity of pullback maps and Chern classes, for any bundle $\pi:{\cal V} \to {\cal M}$, we have $c(i^\ast {\cal V})=i^\ast(c({\cal V}))$. Thus, by the short exactness of \eref{adj_again},
\beq
i^\ast (c( T{\cal M}) )=c(TX) \wedge i^\ast (c( ( {\cal K^{\vee}}_{\cal M})) \ .
\eeq
For each of the Chern classes then, we have that
\begin{align}
& c_1(TX)=i^{\ast}(c_1(T{\cal M})- c_1( {\cal K^{\vee}}_{\cal M}))=0 \label{c1eq} \ , \\
&c_2(TX)=i^{\ast}(c_2(T{\cal M})) \label{c2eq} \ ,\\
&c_3(TX)=i^\ast(c_3({\cal M})- c_2(T{\cal M}) \wedge c_1({\cal K}^{\vee}_{\cal M})) \ .
\end{align}

The Chern classes on ${\cal M}$ are given as simple functions of the entries in the 
configuration matrix \cite{Hubsch:1992nu}. For the total Chern class
\beq
c({\cal M})= c_1^r J_r + c_2^{rs} J_r J_s + c_3^{rst} J_r J_s J_t+ \ldots \ ,
\eeq
where $J_r$ is a basis the K\"ahler (i.e. $(1,1)$) forms in ${\mathbb P}^{n_r}$, the configuration matrix \eref{confForM} implies that
\begin{eqnarray}\label{chernX}
c_1^r &=&( n_r+ 1)-\sum_{j=1}^K a^r_j  \ , \nonumber \\
c_2^{rs} &=& \frac12 \left[ -\delta^{rs}(n_r + 1) + 
  \sum_{j=1}^K a^r_j a^s_j + c_1^r c_1^s\right], \\
c_3^{rst} &=& \frac16 \left[\delta^{rst}(n_r + 1) - 
  \sum_{j=1}^K a^r_j a^s_j a^t_j -c_{2}^{(rs}c_1^{t)}+c_1^r c_1^s c_1^t \right] \nonumber \ .
\end{eqnarray}
Using the formulas above and the pullback of the K\"ahler form, these formulas allow us to efficiently calculate the total Chern class of the Calabi-Yau three-fold, $X$. From \eref{c1eq} we see that the Calabi-Yau condition remains unchanged from the case of ordinary CICYs. {\it In order to define a CY manifold, the hypersurface defining the CY must be in the class of the anticanonical divisor of preceeding rows of the configuration matrix}. As mentioned above, this amounts to the familiar condition that the sum of the entries of each row of \eref{conf} must add up to its respective $\mathbb{P}^{n_r}$ dimension plus one.

The third and top Chern class determines the Euler number of the three-fold. This last quantity can be found with a notion of integration over $X$. As described in \cite{Hubsch:1992nu} (see Thm. 1.3), since $X$ is a complex submanifold of ${\cal M}$ (which is in turn a submanifold of a product of projective spaces), there is a closed $(1, 1)$ form, $\mu$, whose restriction to $X$ represents the top Chern class of the normal bundle, $c_1({\cal N})$. Then if $\omega$ is any closed $(3,3)$-form on $X$, then
\beq
\int_{X} \omega = \int_{{\cal M}} \mu_{X} \wedge \omega \ .
\eeq
Likewise, viewing ${\cal M}$ as a complex subspace of the ambient multi-projective space, integration over ${\cal M}$ can be defined with respect to a measure $\mu_{\cal M}$ and pulled back to a simpler integration over the ambient space ${\cal A}=\pp^{n_1} \times \pp^{n_2}\ldots$:
\beq\label{integration}
\int_{\cal M} \cdot = \int_{\cal A} \mu_{\cal M} \wedge \cdot \ , \qquad
\mu_{\cal M} := \wedge_{j=1}^K \left( \sum_{r=1}^m a^j_r J_r \right) \ .
\eeq
Putting these pieces together then, we see that we can define integration over $X$ via integration on the simple ambient space ${\cal A}$ with a suitable choice of normal form:
\beq
\int_X \cdot  = \int_{\cal A} \mu_{X}\wedge \mu_{{\cal M}} \wedge \cdot \ .
\eeq
We will explore the exact form of $\mu_{X}$ in more detail in Section \ref{secsmooth}. As an application of the above formula, the triple intersection numbers can be 
computed as
\beq\label{tripleintersec}
d_{rst}=\int_X J_{r} \wedge J_{s} \wedge J_{t}  = \int_{\cal A}J_{r} \wedge J_{s} \wedge J_{t}\wedge  \mu_{X}\wedge \mu_{{\cal M}} \ .
\eeq

\subsubsection*{An Example}
As an example of the utility of these techniques, we can once again consider the example manifold given in Section \ref{example_hodgy}. As observed in \eref{hodgy}, the Hodge numbers of this manifold are $(h^{11},h^{21})=(2,46)$. It is natural to ask, from this topology alone can we conclude whether or not \eref{firsteg} defines a previously unknown CY three-fold? A comparison with the literature shows that this Hodge number pair appears twice\footnote{Note that this Hodge number pair also appears in \cite{Batyrev:2008rp} but the remaining topology (i.e. Chern classes and triple intersection numbers) has not been calculated for that construction of manifolds.} in the original CICY list \cite{Candelas:1987kf} and not at all in the Kreuzer-Skarke list  \cite{Kreuzer:2000xy}. From the Hodge data alone then, we cannot decide if this is a new manifold. However, the further topology described above in this case can help us to distinguish this geometry. Using \eref{c2eq}, the second Chern class can be written as
\beq
{c_2(TX)}^{rs}d_{rst}=(24,46) \ ,
\eeq
and from \eref{tripleintersec} the only non-vanishing intersection numbers are
\beq
d_{122}=6,~~~d_{222}=7 \ .
\eeq
With this in hand it is easy compare this second Chern class and triple intersection numbers to the two manifolds in the CICY list with the same Hodge numbers. We can ask whether there exists any basis change for which they can be made equivalent? A straightforward calculation verifies that no such basis change exists and in fact, this manifold is inequivalent to the other known CYs with this Hodge data. It consists of a previously unknown Calabi-Yau three-fold.

\subsubsection{The K\"ahler and Mori cones}\label{kahlersec}
A key component of the CY geometry that is crucial to understand is the structure of the K\"ahler and Mori cones (i.e. the cone of ample divisors and that of effective curves). Practically, we define the K\"ahler cone of a Calabi-Yau three-fold to be defined by the requirements
\beq\label{kahlerity}
\int_{X} J \wedge J \wedge J >0~~,~~\int_{S} J\wedge J >0~~,~~ \int_{C} J >0 \ ,
\eeq
for all $S, C \subset X$ homologically non-trivial, irreducible, reduced proper surfaces and curves in $X$, respectively. The K\"ahler cone is associated to ample divisors on $X$ (and the interior of the nef cone \cite{GH}). If the K\"ahler forms are well understood they can be used to define the Mori cone (or cone of effective curves) as the cone of curves intersecting positively with the divisors dual to $(1,1)$-forms in the K\"ahler cone.

An important case distinction must be made between those forms/divisors which descend from the ambient space -- referred to as ``favorable" in the context of \cite{Anderson:2007nc,Anderson:2008uw,Anderson:2011ns,Anderson:2012yf,Anderson:2013xka} -- and those that only exist on $X$. In general, for any construction of CY manifolds and $X \subset {\cal A}$, when $h^{1,1}(X) > h^{1,1}({\cal A})$ very few general tools exist for explicitly determining the K\"ahler and Mori cone and each example must be handled on a case-by-case basis. Even in the case when $X$ itself is fully favorable and $h^{1,1}(X)=h^{1,1}({\cal A})$, there remain important questions that must be addressed in determining the effective and ample cones. We restrict ourselves to the case of favorable manifolds in the following discussion.

In the case of ordinary CICYs, a key tool in determining the K\"ahler and Mori cones is the Lefschetz Hyperplane Theorem (see the statement given in \cite{Bott}) which guarantees that given an ample divisor $D \subset {\cal A}$ it is possible to construct an isomorphism between the spaces of K\"ahler forms, $H^{1,1}(D)$ to $H^{1,1}({\cal A})$. In the case at hand, gCICYs are defined via effective but not ample divisors in the ambient spaces ${\cal M}$ and thus, we cannot always apply this standard theorem. To make progress then, we must directly explore the conditions in \eref{kahlerity} directly.

To determine the K\"ahler and Mori cones of $X$, a necessary first step is the determination of these cones on the ambient space ${\cal M}$. To begin, it must be noted that in general, the cone of {\it effective divisors} of ${\cal M}$ is {\it larger} than that of the ambient product of projective spaces. It is precisely this enhancement that provides the necessary freedom to build gCICYs. As an example, consider again the manifold given in \eref{firsteg} in Section \ref{example_hodgy}:
\begin{equation}\label{firsteg_oncemore}
X=\left[\begin{array}{c||c|c}
\pp^{1} & 3 & -1\\
\pp^{4} & 2 & 3
\end{array}\right] \ ;
\end{equation}
On the ambient space ${\cal A} =\mathbb{P}^1 \times \mathbb{P}^4$ all effective divisors are in the class $aH_1 +b H_2$ with $a,b \geq 0$ where $H_i$, $i=1,2$ are the hyperplanes of the ambient projective space factors. Since the defining equation of ${\cal M} \subset {\cal A}$ is associated to the ample line bundle ${\cal O}_{\cal A}(3,2)$, the Lefschetz hyperplane theorem \cite{Hubsch:1992nu} can be applied and $h^{1,1}({\cal M})=h^{1,1}({\cal A})=2$. However, it is straightforward to verify that although the dimension of the effective cone stays the same in going from ${\cal A}$ to ${\cal M}$, the mapping between these spaces increases the ``width" of the effective cone. 

An analysis of the line bundle cohomology on ${\cal M}$ using the techniques described in Appendix \ref{line_cohom} reveals that $h^0({\cal M}, {\cal O}_{\cal M}(aH_1 + bH_2))>0$ for
\beq\label{eff_cone}
\{a,b \geq 0 \}~~|~|~~\{ a=-1,~ b \geq 2 \}~~ |~|~~\{ a \leq -2, b \geq \frac{1}{6} (-5 + 8 |a|) + \frac{1}{6} \sqrt{-119 + 64 |a| + 64 a^2} \} \ .
\eeq
Thus, the effective cone has enhanced to include divisors with $a<0$. Each of these may be used to define $3$ (complex) dimensional subvarieties of ${\cal M}$. Those subvarities of ${\cal M}$ that satisfy $c_1(X)=0$ and give rise to smooth Calabi-Yau three-folds are listed in Section \ref{classcases}. If the effective cone can change, it is natural to ask next, what about the ample cone?

Before testing \eref{kahlerity} directly, it should be recalled that there is a useful standard cohomological criteria for ampleness of a divsior $D \subset {\cal M}$ \cite{hartshorne}: \\

\emph{$D$ is ample if and only if, for each coherent sheaf ${\cal F}$ on ${\cal M}$, there is an integer, $n_0$, depending on ${\cal F}$, such that, for each $i>0 $ and each $n \geq n_0$, $H^i({\cal M}, {\cal F} \otimes {\cal O}_{\cal M}(nD))=0$.} \\

While this is not of practical use to determine the ample cone (since we do not a priori know the full set of coherent sheaves on ${\cal M}$), it can be used to rapidly constrain the ample cone. For example, it clearly follows from this result that if $D$ is ample, it must be the case that
\beq
h^i({\cal M},  {\cal O}_{\cal M}(nD))=0~~~\text{for}~~i>0 \ , 
\eeq
for some $n> n_0$ (taking ${\cal F}={\cal O}_{\cal M}$). It is easy to check that for the example given above, all the divisors determined by the conditions \eref{eff_cone} also satisfy this necessary (but not sufficient) condition for ampleness.

To proceed further then, we must use our understanding of the effective cone to produce a basis of subvarieties of ${\cal M}$ to test the equivalent of \eref{kahlerity} directly. For ${\cal M}$ a fourfold these conditions take the form
\beq\label{mconds}
\int_{\cal M} J \wedge J \wedge J \wedge J >0~~,~~\int_{D} J\wedge J \wedge J >0~~,~~\int_{S} J\wedge J>0~~,~~ \int_{C} J >0 \ .
\eeq

For the example above, we will see that in fact, the ample cone of ${\cal M}$ is identical to that of ${\cal A}$. Each divisor satisfying \eref{eff_cone} can be tested directly. As a first step, we can consider a putative $J$ associated to a line bundle in \eref{eff_cone} and compare it to the subvarieties obtained by intersecting strictly positive divisors (i.e. $aH_1+ bH_2$ with $a,b>0$, those descending from ample divisors on ${\cal A}$). In each case, we find that all of the divisors in \eref{eff_cone} satisfy the positivity conditions of the ample cone. However, since the effective cone has enlarged there are \emph{more} algebraic sub-varieties to be considered and hence in this example,  we find that the K\"ahler cone can restrict back to the strictly positive range for $a,b$.

To illustrate this, the anticanonical class of ${\cal M}$ itself, $K_{\cal M}^{-1}= {\cal O}(-1,3)$, is a useful example. $D=-H_1 + 3H_2$ is an effective divisor and satisfies the the conditions in \eref{mconds} when compared to sub-varieties obtained via restriction from ${\cal A}$. However, there exist subvarieties of ${\cal M}$ which do not descend from ${\cal A}$, over which the volume can fail to be positive. For example, $-H_1 + 2H_2$ is an effective divisor on ${\cal M}$ (with $h^0({\cal M},{\cal O}(D)=3$
). Consider the smooth surface defined by the intersection of two divisors in this class
\beq
S=\{p_1=0\} \cap \{p_2=0\} \ ,
\eeq
where $[p_i=0] \in [-H_1+2H_2]$. Then denoting $J_{K^{-1}}$ to be the form dual to ${K_{\cal M}}^{-1}$ it can be verified that
\beq
\int_{S} J_{K^{-1}} \wedge J_{K^{-1}}=\int_{\cal A} J_{K^{-1}} \wedge J_{K^{-1}} \wedge \mu_{S} \wedge \mu_{\cal M}<0 \ ,
\eeq
with $\mu_S=(-J_1 +2 J_2) \wedge (-J_1 +2 J_2)$. Thus, in this case the anticanonical class of ${\cal M}$ itself is effective but not ample.

Finally, to compute the K\"ahler and Mori cones of the Calabi-Yau three-fold itself, we proceed in a step-wise manner. Using the tools described above we first understand the change in the effective and ample cones in moving from ${\cal A}$ to ${\cal M}$ before finally exploring the same conditions on $X \subset {\cal M}$ itself. In each step, the restriction of K\"ahler forms are K\"ahler, but linear independence must be verified\footnote{And indeed, in many cases independent K\"ahler forms on the ambient space can become dependent upon restriction to the hypersurface.} and the constraints on coefficients generically change as illustrated above.


\section{Construction of Sections and Smoothness} \label{secsmooth}

In this section we return to the question of smoothness of gCICYs. As a precursor to discussing such issues, we must describe the defining relations of these manifolds in more detail than was required for the investigation of topology in the preceding section.

\vspace{0.1cm}

Given a configuration matrix of the form (\ref{conf}), one can construct the gCICY, $X$, in the manner described in Section \ref{constr}. For this section we will, for simplicity, restrict ourselves to the case of mgCICYs  which can be constructed in two steps, 
\beq
X \hookrightarrow \cM \hookrightarrow \cA \ , 
\eeq
where each of the two embeddings is a complete intersection. Much of what we will describe in this section generalizes to more complicated examples, however. 

The embedding, $\cM \hookrightarrow \cA$, is manifestly algebraic in the coordinates of $\cA$ in that all the sections over $\cA$ involved in forming the complete intersection are a polynomial of the ambient homogenous coordinates, $\mathbf x_r = (x_{r}^0:x_{r}^1:\cdots:x_{r}^{n_r})$, $r=1, \cdots, m$. The other embedding, $X \hookrightarrow \cM$, is also an algebraic complete intersection and as such must have a polynomial description when an appropriate coordinate system is used for $\cM$. However, much of the computational power comes from relating various quantities to the simple ambient space geometry for $\cA$. In this paper, therefore, the homogeneous coordinates, $\mathbf x_r$, $r=1, \cdots, m$, will be employed whenever possible, and all the global sections for this latter (gCICY) embedding, $X \hookrightarrow \cM$, will be constructed in terms of them, using the idea of {\it ``tuning"} in the following sense. 

Let us start with a general set up and consider a section $q \in \Gamma(\cM, \;\cO_\cM(\mathbf b))$, with $\mathbf b = (b^1, \cdots, b^m)$. In view of its zero and pole structure, $q$ may take the rational form,\footnote{To be precise, the line bundle degree rules the behaviour of the entire object $q$ and therefore, $N$ and $D$ may have a common degree shift. Thus, we are not guaranteed to find all the sections by demanding the rational form as in eq.~(\ref{q}). To make sure all the sections have been constructed, one should compare the number of linearly independent $q$'s and the dimension of the section space, $H^0(\cM, \cO(\mathbf b))$.}
\beq\label{q}
q = \frac{N(\mathbf x_1, \cdots, \mathbf x_m)}{D(\mathbf x_1, \cdots, \mathbf x_m)} \ , 
\eeq
where $N$ and $D$ are a polynomial in $\mathbf x_1, \cdots, \mathbf x_m$, of multi-degree $\left[ \mathbf b\right]_+$ and $\left[ \mathbf b\right]_-$, respectively. Here, $\left[ \mathbf b\right]_+$ is a vector of length $m$, extracting only the positive entries from $\mathbf b$. Similarly, $\left[ \mathbf b \right]_-$ only extracts the negative entries from $\mathbf b$, but with the signs inverted so that $\mathbf b = \left[\mathbf b \right]_+ - \left[ \mathbf b\right]_-$ ({\it e.g.}, for $\mathbf b = (1,2,0,-3)$ of length $m=4$, we have $\left[\mathbf b \right]_+=(1,2,0,0)$ and $\left[\mathbf b \right]_-=(0,0,0,3)$). 
Note that the expression, (\ref{q}), for a general numerator $N$ of the right multi-degree, is not regular on the divisor $D=0$ of the manifold $\cM$. However, $q$ can remain regular on this divisor if $N$ takes a specific form such that it also vanishes there. 
Thus, the key idea in making the rational expression a holomorphic object is to {\it ``tune"} the coefficients of the numerator polynomial, $N$, so that vanishing of the denominator, $D$, is completely cancelled by that of $N$. To be precise, we demand that
\beq\label{nconstraint}
N \in \left<D\right> \cap \IC\left[\mathbf x_1, \cdots, \mathbf x_m\right]   \ , 
\eeq
should hold, where $\left<D\right>$ is the ideal generated by $D$ in the homogeneous coordinate ring of $\cM$,\beq
R(\cM):=\IC\left[\mathbf x_1, \cdots, \mathbf x_m\right] / \left<p_1, \cdots, p_K\right> \ .
\eeq 
The regularity condition,~(\ref{nconstraint}), says that $q = N/D$ is in fact a polynomial section over $\cM$.

Once we have all the sections, $q_\mu \in \Gamma(\cM, \;\cO_\cM(\mathbf b_\mu))$ written down for $\mu = 1, \cdots, L$, the gCICY, $X$, is then constructed as the common vanishing locus,
\beq
X=\{x \in \cM\;|\; q_1 (x) = \cdots = q_L(x) = 0 \} \ . 
\eeq 
To be more explicit, let us illustrate the numerator tuning with the two earlier examples. For the first gCICY,~(\ref{eg2}), the configuration matrices for $X$ and $\cM$ are, respectively,
\begin{eqnarray}\label{e.g.1}
X = \left[ \begin{array}{c||cc|cc} \mathbb{P}^1 & 1 & 1 & 1 & -1 \\ \mathbb{P}^1 & 1 & 1 & -1 & 1 \\ \mathbb{P}^5 & 3 & 1 & 1& 1\end{array}\right]\; ; \quad  \cM = \left[ \begin{array}{c||cc} \mathbb{P}^1 & 1 & 1 \\ \mathbb{P}^1 & 1 & 1  \\ \mathbb{P}^5 & 3 & 1 \end{array}\right]\;  \ .
\end{eqnarray}
Using the methods described in Section \ref{kosz_sec}, we can determine the Hodge number of this manifold to be $h^{1,1}(X_2)=3, h^{1,2}(X_2)=81$. 
The rational expressions for $q_1 \in H^0(\cM, \cO_\cM(1,-1,1))$ and $q_2 \in H^0(\cM, \cO_\cM(-1,1,1))$ have already been written in eq.~(\ref{p3p4}), where they have their numerators completely fixed by one of the defining equations, $p_2 \in H^0(\cA, \cO_\cA(1,1,1))$, for the embedding $\cM \hookrightarrow \cA$. Note that for the right multi-degree of the numerator polynomial of $q_1$, for instance, $12$ independent monomials that are bi-linear in $\mathbf x_1=(x_1^0:x_1^1)$ and $\mathbf x_3 =(x_3^0: \cdots : x_3^5)$ can be used. It is the holomorphicity of $q_1$ in $\cM$ that rules out all but one particular linear combination of them (up to scaling). Such a restriction of the numerator polynomial is the result of the aforementioned tuning. Indeed, one can compute the dimension of the line bundle cohomology:
\beq
h^0(\cM, \cO_\cM(1,-1,1)) = 1 \ ; \quad h^0(\cM, \cO_\cM(-1,1,1))=1 \ , 
\eeq
which shows that eq.~(\ref{p3p4}) is the complete answer. 

For the second gCICY,~(\ref{q}), whose topological properties have been computed in Section~{2}, the configuration matrices for $X$ and $\cM$ are, respectively, 
\begin{equation}\label{e.g.2}
X=\left[\begin{array}{c||c|c}
\pp^{1} & 3 & -1\\
\pp^{4} & 2 & 3
\end{array}\right] \ ; \quad \cM=
\left[\begin{array}{c||c}
\pp^{1} & 3 \\
\pp^{4} & 2 
\end{array}\right] \ .
\end{equation} 
Again, the two sections, $p_1 \in H^0(\cA, \cO_\cA(3,2))$ and $q_1 \in H^0(\cM, \cO_\cM(-1,3))$ can be sequentially constructed so that the rational expression for $q_1$ is determined by the choice of $p_1$. For concreteness, let us write $p_1$ as,
\beq\label{p1}
p_1 (\mathbf x_1, \mathbf x_2) = (x_1^0)^3 \, P_{11}(\mathbf x_2) + (x_1^0)^2 x_1^1\,P_{12}(\mathbf x_2) + x_1^0 (x_1^1)^2 \, P_{13}(\mathbf x_2) + (x_1^1)^3 \, P_{14}(\mathbf x_2) \ , 
\eeq
where $P_{1a}$, for $a=1,2,3,4$, are a quadric in $\mathbf x_2$.
According to the degree-splitting rule,~(\ref{q}), $q_1$ has the rational form,
\bea
q_1 = \frac{N(\mathbf x_1, \mathbf x_2)}{D(\mathbf x_1, \mathbf x_2)} \ , 
\eea
where $N$ and $D$ are a polynomial of multi-degree $(0,3)$ and $(1,0)$, respectively. For the numerator tuning, we first need to choose a denominator. With the choice, $D(\mathbf x_1, \mathbf x_2) = x_1^0$, for instance, the corresponding numerator polynomial should vanish when $x_1^0=0$ in $\cM$. On the other hand, since $p_1$ vanishes on $\cM$, so it does on the divisor $x_1^0=0$ of $\cM$. Therefore, substituting $x_1^0=0$ in eq.~(\ref{p1}), we obtain an identically-vanishing section $P_{13}(\mathbf x_2) \equiv 0$ over that divisor. Now, $P_{13}$ is a quadric in $\mathbf x_2$ while we need a cubic expression for the numerator, $N$. Filling in the missing degree with a linear polynomial in $\mathbf x_2$, we obtain the following $5$ global sections 
\beq\label{1-5}
s_i := \frac{P_{14} (\mathbf x_2)\, x_2^i }{x_1^0}\ , ~\quad i=0, \cdots, 4 \ , 
\eeq
of the line bundle $\cO_\cM (-1,3)$. On the other hand, by computing the dimension of the section space one sees that $h^0(\cM, \cO_\cM(-1,3))=15$, which shows that for a complete basis one needs to obtain $10$ more independent sections. This can be achieved by starting with different choices for the denominator polynomial. For instance, following the same procedure with $D(\mathbf x_1, \mathbf x_2) = x_0 -x_1$, one obtains $5$ more sections
\beq\label{6-10}
t_i :=\frac{\sum\limits_{a=1}^4 P_{1a} (\mathbf x_2)\, x_2^i}{x_1^0 - x_1^1} \ , ~\quad i=0, \cdots, 4 \ , 
\eeq
and similarly, with $D(\mathbf x_1, \mathbf x_2) = x_0 +x_1$, one obtains
\beq\label{11-15}
u_i:=\frac{\sum\limits_{a=1}^4 (-1)^a P_{1a} (\mathbf x_2)\, x_2^i}{x_1^0 + x_1^1} \ , ~\quad i=0, \cdots, 4 \ , 
\eeq
as additional $5$ global sections. 
A total of $15$ sections have thus been obtained, as in eqs.~(\ref{1-5}),~(\ref{6-10}), and~(\ref{11-15}), each of which lies in the $15$-dimensional section space of $\cO_\cM(-1,3)$. Interestingly, given that $p_1$ is a generic polynomial, one can show that these $15$ sections are linearly independent. Furthermore, any other rational expression for $q_1$, obtained by starting with yet another choice of $D$, turns out to be expressible as their linear combination. Therefore, we come to the conclusion that these $15$ sections do form a complete basis. 

For certain gCICY configuration matrices, especially for those involving entries strictly smaller than $-1$, the analytic construction of sections may become trickier than for the two example cases above. We have thus developed a method to obtain the sections numerically in Mathematica, details of which can be found in Appendix~\ref{numsec}. 

\vspace{0.5cm}

\subsection{Smoothness}
With such an explicit description for the defining relations of a gCICY we can now return to the question of its smoothness. Let us recall that for an ordinary CICY with $b_\mu^r$ non-negative, $X$ is embedded in $\cA$ as a smooth complete intersection if the $(K+L)$-form,
\beq\label{nfXA}
\Theta_{X / \cA} = {\rm d}p_1\wedge \cdots \wedge {\rm d}p_K \wedge {\rm d} q_1 \wedge \cdots \wedge {\rm d} q_L \ , 
\eeq
is nowhere vanishing on $X$. 
Let us first illustrate how vanishing of this $(K+L)$-form, $\Theta_{X/\cA}$, remains the correct singularity criterion for a gCICY case, even when there is no algebraic embedding of $X$ in $\cA$ and $\Theta_{X/\cA}$ cannot be thought of as a normal form any longer.

The starting point is that the manifold $\cM$, with its generic complex structure, is a smooth complete intersection of $K$ polynomials in $\cA$. In particular, this requires that the corresponding normal form,
\beq\label{nfMA}
\Theta_{\cM/\cA}={\rm d} p_1 \wedge \cdots \wedge {\rm d} p_K \ ,
\eeq
is nowhere vanishing on $\cM$. The smoothness criterion for $X$ then follows in terms of its own normal form,
\beq\label{nfXM}
\Theta_{X/\cM}={\rm d} q_1 \wedge \cdots \wedge {\rm d} q_L \ .
\eeq
Now, the claim we need to establish is that vanishing of $\Theta_{X/\cM}$ at a point in $X$ is equivalent to that of $\Theta_{X/\cA}$.  
Note that at a practical level, we write these expressions in the coordinates $\mathbf x_r$, and in taking exterior derivatives of $q_\mu$, some misleading components in ${\rm d}q_\mu$ that do not lie in $\cT_x^* \cM$ can appear. However, when we eventually form the $(K+L)$-form $\Theta_{X/\cA}$, all these components normal to $\cT_x^* \cM$ disappear due to the presence of the prefactor, $\Theta_{\cM/\cA}$. Furthermore, because that prefactor is nowhere vanishing on $\cM$ and in particular on $X$, we are led to the desired conclusion that at each point in $X$ vanishing of $\Theta_{X/\cM}$ implies that of $\Theta_{X/\cA}$, and vice versa.

Let us illustrate this smoothness check based on the $(K+L)$-form, $\Theta_{X/\cA}$, with the two gCICYs,~(\ref{e.g.1}) and~(\ref{e.g.2}), for which the relevant global sections have already been constructed. For the former gCICY,~(\ref{e.g.1}), $p_1$ and $p_2$ are each a generic homogeneous polynomial of the right multi-degree and with respect to that choice, $q_1$ and $q_2$ are given as eq.~(\ref{p3p4}). Given these 4 sections, $p_1$, $p_2$, $q_1$ and $q_2$, one can form the system of equations,
\beq\label{syst1}
p_1=0=p_2=q_1=q_2\ ; \quad {\rm d} p_1 \wedge {\rm d} p_2 \wedge {\rm d} q_1 \wedge {\rm d} q_2 =0 \ , 
\eeq
whose solutions are to be found for the homogeneous coordinates, 
\beq
\mathbf x_1=(x_1^0:x_1^1),\,\quad \mathbf x_2=(x_2^0:x_2^1), \,\quad \mathbf x_3=(x_3^0:\cdots:x_3^5) \ . 
\eeq
Similarly, for the latter gCICY,~(\ref{e.g.2}), $p_1$ is a generic homogenous polynomial of the bi-degree $(3,2)$ and a generic $q_1$ can be obtained by linearly combining the $15$ independent sections, $s_i$, $t_i$ and $u_i$, for $i=0, \cdots, 4$, explicitly given in eqs.~(\ref{1-5}), (\ref{6-10}) and (\ref{11-15}), respectively. In other words, we take 
\beq\label{lc15}
q_1= \sum\limits_{i=0}^4 \alpha_i \, s_i + \sum\limits_{i=0}^4\beta_i \, t_i + \sum\limits_{i=0}^4\gamma_i \, u_i \ , 
\eeq
for a generic choice of the $15$ coefficients, $\alpha_i$, $\beta_i$ and $\gamma_i$, $i=0, \cdots, 4$. Given these 2 sections, $p_1$ and $q_1$, one can form the system of equations,
\beq\label{syst2}
p_1 = 0 = q_1 \ ; \quad {\rm d} p_1 \wedge {\rm d} q_1 = 0 \ , 
\eeq
which have to be solved for the homogeneous coordinates, 
\beq
\mathbf x_1=(x_1^0:x_1^1),\,\quad\mathbf x_2=(x_2^0: \cdots : x_2^4) \ . 
\eeq

Given the system of constraints for a singularity, such as eqs.~(\ref{syst1}) and (\ref{syst2}), one needs to test whether the system admits a solution or not. However, it can be extremely time-consuming on a computer if tackled in a brute-force manner. 
For the case of an ordinary CICY, this can be efficiently achieved by the Gr\"obner basis method, as implemented in the Mathematica package ``Stringvacua"~\cite{stringvacua}, for instance. For the case of a gCICY, however, due to the presence of negative entries in the configuration matrix, not every constraining equation is algebraic when written in terms of the $\mathbf x_r$ and the Gr\"obner basis method does not apply as straight-forwardly.
One way to avoid this obstacle is to separately analyze different regions of $\cM$, by rewriting the system of constraints as an equivalent polynomial system in each region. More specifically, we subdivide $\cM$ into various regions according to which denominators that appear in the constraining equations vanish. For instance, the first region to consider is the one where none of the denominators vanish; in this region, we may clear the denominators to obtain an equivalent algebraic system of constraints. The next simplest regions are those where only a single denominator vanishes; any rational expressions involving this vanishing denominator can then be rewritten, in the coordinate ring of $\cM$, as an alternative algebraic expression. One thereby obtains an equivalent algebraic system also in these regions, upon clearing the remaining denominators. Such a strategy works for the common zero loci of an arbitrary collection of denominator expressions, and that is how we subdivide $\cM$ into various regions, in each of which the usual Gr\"obner basis method may apply to efficiently test the existence of a singular point. 

Again, let us illustrate this with our two examples. For the first gCICY example,~(\ref{e.g.1}), the system in question,~(\ref{syst1}), has rational expressions originating from $q_1=d_1/x_1^0$ and $q_2=c_1/x_2^0$, as in eq.~(\ref{p3p4}). This case is particularly simple that one does not even have to clear denominators. However, the subdivision idea is still useful so let us proceed. We consider the subdivision of $\cM$ into the following four regions,
\bea
R_1&:&\quad  x_1^0 \neq 0, \quad x_2^0 \neq 0 \ ,  \\  
R_2 &:&\quad x_1^0 =0, \quad x_2^0 \neq 0 \ , \\ 
R_3 &:&\quad x_1^0 \neq 0,  \quad x_2^0 = 0 \ , \\ 
R_4 &:&\quad x_1^0 = 0, \quad x_2^0 = 0 \ , 
\eea
depending on which of the two denominators, $x_1^0$ and $x_2^0$, vanish. 
Firstly, in the region $R_1$, both $x_1^0$ and $x_2^0$ can be set to 1 using the scaling of the two $\IP^1$ coordinates. Therefore, $q_1$ and $q_2$ both reduces to a polynomial, $d_1$ and $c_1$, respectively, and the system becomes completely algebraic. In $R_2$, one may still set $x_2^0$ to 1 so that $q_2 = c_1$ is a polynomial but $q_1$ involves a non-trivial denominator. However, as shown in eq.~(\ref{p3p4}), $q_1=d_1/x_1^0$ can also be written as $-d_0/x_1^1$ and $x_1^1$ can be set to 1 since  $x_1^0$ and $x_1^1$ can not vanish simultaneously. Therefore, in this region, $q_1$ also reduces to the polynomial $-d_0$ and the system becomes algebraic again. Similarly, one can use $q_1 = d_1$, $q_2 = -c_0$ in the region $R_3$, and $q_1=-d_0$, $q_2=-c_0$ in the region $R_4$. 

Now, for the second gCICY example,~(\ref{e.g.2}), let us first recall that the generic section $q_1$ in eq.~(\ref{lc15}) had 3 denominator expressions, $x_0$, $x_0 - x_1$ and $x_0 + x_1$. Thus, we consider the subdivision of $\cM$ into the following 4 regions,
\bea
R_1&:&\quad  x_1^0 \neq 0, \quad x_1^0-x_1^1 \neq 0, \quad x_1^0+x_1^1 \neq 0 \ ,  \\  
R_2&:&\quad  x_1^0 = 0, \quad x_1^0-x_1^1 \neq 0, \quad x_1^0+x_1^1 \neq 0 \ ,  \\  
R_3&:&\quad  x_1^0 \neq 0, \quad x_1^0-x_1^1 = 0, \quad x_1^0+x_1^1 \neq 0 \ ,  \\  
R_4&:&\quad  x_1^0 \neq 0, \quad x_1^0-x_1^1 \neq 0, \quad x_1^0+x_1^1 = 0 \ ,  
\eea
depending again on which denominators vanish. Note that at most one denominator can be set to zero at a time on $\IP^1_{\mathbf x_1}$. 
In region $R_1$, none of the denominators vanish and hence, one can clear all the denominators from the system~(\ref{syst2}) and test if it admits a solution. With the denominators cleared, however, naively testing the existence of a solution might lead to a point outside of $R_1$ and the test result would become ambiguous. To avoid such an ambiguity, we add to the system~(\ref{syst2}) three auxiliary variables, let us call them $y_1$, $y_2$ and $y_3$, together with the three constraints, 
\beq\label{aux}
y_1 x_1^0 -1 = 0; \quad y_2 (x_1^0-x_1^1) - 1 =0; \quad y_3 (x_1^0 + x_1^1) -1 =0 \ ,  
\eeq
so that none of the denominators can vanish subject to the new system of algebraic equations. In region $R_2$, on the other hand, $s_i$ in eq.~(\ref{1-5}) have their denominator vanishing in that region, and therefore the part in $q_1$ that involves $s_i$ need to be carefully taken care of. A proper way to deal with this is to rewrite them as,
\bea
s_i &= \frac{P_{14}(\mathbf x_2) \, x_2^i }{x_1^0} \\
 &= - \frac{ (x_1^0)^2 \, P_{11}(\mathbf x_2) + x_1^0 x_1^1\,P_{12}(\mathbf x_2) + (x_1^1)^2 \, P_{13}(\mathbf x_2) }{(x_1^1)^3} \, x_2^i \ , 
\eea 
where we have made use of $p_1=0$, \eref{p1}. Since $x_1^0=0$ in this region, the new denominator $x_1^1$ cannot vanish and hence, we may set $x_1^1 = 1$ using the scaling. Then the expressions for $s_i$ become a polynomial and if we use them in eq.~(\ref{lc15}), the denominator clearing does not lead to any ambiguity once we consider two auxiliary variables $y_1$, $y_2$, together with the two constraints,
\beq
y_1 (x_1^0-x_1^1) - 1 =0; \quad y_2 (x_1^0 + x_1^1) -1 =0 \ ,  
\eeq
To complete the story, let us also work out the details for regions $R_3$ and $R_4$. In region $R_3$, we rewrite $t_i$ in eq.~(\ref{6-10}) as
\bea
t_i &=\frac{\sum\limits_{a=1}^4 P_{1a} (\mathbf x_2)\, x_2^i}{x_1^0 - x_1^1}  \\
 &= - \frac{P_{11}(\mathbf x_2) ((x_1^0)^2 + x_1^0 x_1^1 + (x_1^1)^2) + P_{12}(\mathbf x_2) (x_1^0 + x_1^1) x_1^1 + P_{13}(\mathbf x_2) (x_1^1)^2}{(x_1^1)^3} \, x_2^i \ , 
\eea
where vanishing of $p_1$ has been used. Since $x_1^1 \neq 0$ in this region, we may set $x_1^1=1$ by scaling and use the resulting polynomial expressions for $t_i$ to rewrite $q_1$ in eq.~(\ref{lc15}). 
Again, we clear all the denominators and, to make sure only the points in region $R_3$ are analyzed, add two auxiliary variables $y_1$ and $y_2$, together with the constraints,
\beq
y_1 x_1^0 -1 = 0; \quad y_2 (x_1^0 + x_1^1) -1 =0 \ .  
\eeq
Finally in region $R_4$, we rewrite $u_i$ in eq.~(\ref{11-15}) as
\bea
u_i &=\frac{\sum\limits_{a=1}^4 (-1)^a P_{1a} (\mathbf x_2)\, x_2^i}{x_1^0 + x_1^1}  \\
 &= - \frac{P_{11}(\mathbf x_2) ((x_1^0)^2 - x_1^0 x_1^1 + (x_1^1)^2) + P_{12}(\mathbf x_2) (x_1^0 - x_1^1) x_1^1 + P_{13}(\mathbf x_2) (x_1^1)^2}{(x_1^1)^3} \, x_2^i \ , 
\eea
where vanishing of $p_1$ has been used. Since $x_1^1 \neq 0$ in this region, we may set $x_1^1=1$ by scaling and use the resulting polynomial expressions for $u_i$ to rewrite $q_1$ in eq.~(\ref{lc15}). 
Again, we clear all the denominators and, to make sure only the points in region $R_4$ are analyzed, add two auxiliary variables $y_1$ and $y_2$, together with the constraints,
\beq
y_1 x_1^0 -1 = 0; \quad y_2 (x_1^0 - x_1^1) -1 =0 \ .  
\eeq

Therefore, the singularity criteria, in each of the subdivided regions of $\cM$, have become a purely algebraic system and the techniques from numerical algebraic geometry applies straight-forwardly to test the existence of a singular point in each region. This way, it has been shown that both of the above gCICY examples,~(\ref{e.g.1}) and~(\ref{e.g.2}), lead to a smooth Calabi-Yau three-fold. 

For gCICY cases, it is by now clear that when constructing the sections for the line bundles involving a negative degree, one needs to go through the ``tuning'' process based on~(\ref{nconstraint}) and as a result, not every candidate expression satisfying the degree constraint can be a global holomorphic section. It is such a tuning that leads to distinctive features of gCICYs not observed for the ordinary CICY cases. 

One interesting feature is that there may arise singular gCICYs even for a generic complex structure while the ordinary CICYs are all smooth~\cite{Green:1986ck,Hubsch:1992nu}. Note that factorisation of a defining equation is a potential source for singularities. For CICY cases, a generic defining equation with a given multi-degree does not factorize. However, for gCICYs, even a generic defining equation may factorize (i.e. the associate divisor may not be base point free). Let us illustrate such a section-space factorization with the following example,
\begin{equation}\label{e.g.singular}
X=\left[\begin{array}{c||cc|c}
\pp^{3} & 2 & 0 &2\\
\pp^{1} & 0 & 1&1\\
\pp^{1} & 0 & 2 & 0 \\
\pp^{1} & 1 & 2 & -1 
\end{array}\right] \ ; \quad \cM=
\left[\begin{array}{c||cc}
\pp^{3} & 2 & 0  \\
\pp^{1} & 0 & 1 \\
\pp^1 & 0 & 2 \\
\pp^{1} & 1 & 2 
\end{array}\right] \ ,
\end{equation} 
for which we may write 
\beq
p_1(\mathbf x_1, \mathbf x_2, \mathbf x_3, \mathbf x_4) = x_4^0 \, P_{11}(\mathbf x_1) + x_4^1 \, P_{12}(\mathbf x_1) \ , 
\eeq
where $P_{11}$ and $P_{12}$ are a generic quadric in $\mathbf x_1 = (x_1^0:\cdots:x_1^3)$. Thus, the numerator tuning due to eq.~(\ref{nconstraint}) leads to the following holomorphic sections, 
\beq
s_i = \frac{P_{12}(\mathbf x_1)\,x_2^i}{x^0_4} \ , \quad i=0, 1 \ ,   
\eeq 
of the line bundle $\cO_\cM (2,1,0,-1)$ over $\cM$. Furthermore, the computation of line bundle cohomology results in $h^0(\cM,  \cO_\cM (2,1,0,-1)) = 2$, which means that the two sections $s_1$ and $s_2$ should form a complete basis for the section space. Therefore, a generic section $q_1$ should have the following form,
\beq
q_1 = \alpha_0 \, s_0+ \alpha_1 \, s_1 = \frac{P_{12}(\mathbf x_1)  }{x^0_4} \cdot(\alpha_0 \, x_2^0 + \alpha_1 \, x_2^1) \ , 
\eeq
where $\alpha_0$ and $\alpha_1$ are complex constants. Because both factors, $P_{12}/x_4^0$ and $\alpha_0 x_2^0 + \alpha_1 x_2^1$, are an algebraic section, this corresponds to the defining section $q_1$ being legitimately factorized. Furthermore, one can see that there indeed exists a point in $X$ where both factors vanish simultaneously, thereby leading to a singularity.  

\vspace{0.5cm}

\subsection{Reducedness}
Another consequence of the numerator tuning is that the coordinate ring does not have to be reduced. Again, let us illustrate this with a simple example how a non-reduced coordinate ring may arise. We consider,
\beq\label{e.g.nonreduced}
X=\left[\begin{array}{c||cc|c}
\pp^{1} & 0 & 2 &0\\
\pp^{4} & 1 & 1&3\\
\pp^{1} & 1 & 4 &-3 
\end{array}\right] \ ; \quad \cM=
\left[\begin{array}{c||cc}
\pp^{1} & 0 &2  \\
\pp^{4} & 1 & 1 \\
\pp^1 & 1 & 4 
\end{array}\right] \ ,
\eeq 
for which the first defining polynomial can be written as
\beq
p_1 (\mathbf x_1, \mathbf x_2, \mathbf x_3) = x_3^0 \, P_{11}(\mathbf x_2) + x_3^1 \, P_{12}(\mathbf x_2) \ ,  
\eeq
where $P_{11}$ and $P_{12}$ are a generic linear polynomial in $\mathbf x_2 = (x_2^0: \cdots: x_2^4)$. The numerator tuning for the section of the line bundle $\cO_\cM(0,3,-3)$ over $\cM$ results in the following expression,
\beq
\left(\frac{P_{12}(\mathbf x_2)}{x_3^0}\right)^3 \ . 
\eeq 
Furthermore, computation of the dimension of the associated line bundle cohomology results in $h^0(\cM, \cO_\cM(0,3,-3))=1$ and thus, the most general section takes the form, 
\beq
q_1 = \alpha \,\left(\frac{P_{12}(\mathbf x_2)}{x_3^0}\right)^3 \ , 
\eeq
for a complex constant $\alpha$. Note that this can be though of as the cube of another holomorphic section, 
\beq
s=\frac{P_{12}(\mathbf x_2)}{x_3^0} \ , 
\eeq
of the line bundle $\cO_\cM(0,1,-1)$. 
Now, the coordinate ring of $X$ defined naively as
\beq
R(X)=R(\cM)/ \left<s^3\right> \ , 
\eeq 
is not reduced. However, since the tools developed in Subsection~\ref{subsec:topology} for working out the topological properties assume that the coordinate ring is reduced, it is better to consider
\beq
R(X_{\rm red}) = R(\cM)/\left<s\right> \ , 
\eeq
which describes $X_{\rm red}$ defined as the hypersurface $s=0$ in $\cM$. 
Set-theoretically, the gCICY configuration,~(\ref{e.g.nonreduced}), for $X$ can be equivalently thought of as
\beq\label{e.g.reduced}
X_{\rm red}=\left[\begin{array}{c||cc|c}
\pp^{1} & 0 & 2 &0\\
\pp^{4} & 1 & 1&1\\
\pp^{1} & 1 & 4 &-1 
\end{array}\right] \ ; \quad \cM=
\left[\begin{array}{c||cc}
\pp^{1} & 0 &2  \\
\pp^{4} & 1 & 1 \\
\pp^1 & 1 & 4 
\end{array}\right] \ ,
\eeq 
and only with this reduced configuration can we directly apply the tools in Subsection~\ref{subsec:topology} to compute the topological quantities of the resulting classical variety. One immediate observation from this reduction is that the gCICY, $X \sim X_{\rm red}$, is not a Calabi-Yau manifold. For ordinary CICY cases, the Calabi-Yau criterion was that each row in the configuration matrix should have its entries sum up to match the degree of the anticanonical bundle of the corresponding ambient projective space. However, from the illustration above, we learn that this could be a misleading criterion for gCICYs when the configuration leads to a non-reduced coordinate ring. 

Furthermore, in such a non-reduced case, the smoothness check has to be carefully performed, too. In the example,~(\ref{e.g.nonreduced}), if we naively form the system of singularity constraints,
\beq\label{syst-nonreduced}
p_1=0=p_2=q_1\ ; \quad {\rm d} p_1 \wedge {\rm d} p_2 \wedge {\rm d} q_1 = 0 \ , 
\eeq 
with $q_1 = \alpha \, s^3$, then we come to the incorrect conclusion that $X$ is singular everywhere as $s=0$ in $\cM$ is already sufficient to satisfy eq.~(\ref{syst-nonreduced}).

\vspace{0.5cm}

\section{Redundancies} \label{redun}

Just as in the case of ordinary CICYs, configuration matrices of gCICYs also exhibit {\it redundancies}. A redundancy is simply a situation where two different configuration matrices describe the same manifold. In fact, the structure of known redundancies is even richer in the gCICY case than in that of the CICYs. Many of the same relations between configuration matrices seen in the simpler case still hold. In addition, new redundancies do arise, such as those relating seemingly good configuration matrices to a description of the empty set. In this section, we present possible sources of such redundancies, one in each subsection, emphasizing some distinctive features that are only seen for gCICYs. 

\vspace{0.5cm}
\subsection{Splitting transitions}
Between ordinary CICY manifolds, there is an important process known as a {\it splitting transition}~\cite{Candelas:1987kf}, which can essentially be thought of as a combination of deformation and blowup. The splitting, in its most general version, relates two configuration matrices of the following form,
\beq\label{split-rel}
\left[\begin{array}{c||cccc}
\pp^{n} & 1 & \cdots & 1 & 0 \\
A & \mathbf u_1 & \cdots &\mathbf u_{n+1} & C\\
\end{array}\right] \quad \longleftrightarrow \quad 
\left[\begin{array}{c||cc}
A & \sum\limits_i^{n+1} \mathbf u_i & C  \\
\end{array}\right] \ ,
\eeq 
where $A$ is a product of projective spaces, $\mathbf u_i$'s are a column vector of degree entries each, $C$, a submatrix of an appropriate size, and $0$, a row of zeros. 
A simple example, as studied in Ref.~\cite{Candelas:1987kf}, is the following pair of CICY three-folds,
\beq\label{e.g.split}
\left[\begin{array}{c||cc}
\pp^{1} & 1 & 1 \\
\pp^{2} & 3 & 0 \\
\pp^{2} & 0 & 3 \\
\end{array}\right] \quad \longleftrightarrow \quad 
\left[\begin{array}{c||c}
\pp^2 & 3 \\
\pp^2 & 3 \\
\end{array}\right] \ ,
\eeq
where the bi-cubic column of the right hand side is splitted via a $\IP^1$ row of the left hand side. We may write the two defining equations for the splitted configuration as,
\bea\label{e.g.split.de}
p_1 (\mathbf x_1, \mathbf x_2, \mathbf x_3) &= x_1^0 \,P_{11}(\mathbf x_2) + x_1^1\, P_{12} (\mathbf x_2) \ , \\  
p_2 (\mathbf x_1, \mathbf x_2, \mathbf x_3) &= x_1^0 \,P_{21}(\mathbf x_3) + x_1^1 \,P_{22}(\mathbf x_3) \ , 
\eea
where $P_{11}$ and $P_{12}$ are a cubic polynomial in $\mathbf x_2$, and $P_{21}$ and $P_{22}$ are a cubic in $\mathbf x_3$. In order for the system~(\ref{e.g.split.de}) to have a non-trivial solution in $\mathbf x_1$, the determinental expression,
\beq
\Delta = P_{11}(\mathbf x_2) \, P_{22}(\mathbf x_3) - P_{12}(\mathbf x_2) \, P_{21}(\mathbf x_3) \ , 
\eeq
should vanish, leading to a bi-cubic relation in $\mathbf x_2$ and $\mathbf x_3$, as indicated by the RHS of eq.~(\ref{e.g.split}). In general, however, there exist point-like singularities in the vanishing locus of $\Delta$ and one arrives, via a deformation, at a smooth bi-cubic three-fold.  

In Ref.~\cite{Candelas:1987kf}, splitting was called {\it effective} if the Euler numbers of the two manifolds differ and {\it ineffective} if they are the same. It was also argued there that one should consider two CICY three-fold configurations equivalent if they are related by an ineffective splitting because that guarantees the process is completely free of singularities. In the example~(\ref{e.g.split}), the Euler numbers of the LHS and the RHS are $0$ and $-162$, respectively, and thus the splitting is effective. 

The idea of splitting and the consequence of its effectiveness straight-forwardly generalizes to gCICY manifolds. The rule for relating two gCICY configuration matrices remains the same as eq.~(\ref{split-rel}), except that now the $1$'s of the $\IP^n$ row can either be in the non-negative sector or in the sector where negative entries are allowed. However, one needs to be careful about certain new aspects of the gCICY construction. Firstly, although the splitting process is most clear in terms of the configuration matrices, what we have in mind is how the corresponding gCICY geometries are related and it does not make sense, for the purpose of this paper, to naively relate two configuration matrices by the splitting of the form, eq.~(\ref{split-rel}), when one or both of them lead to an empty set due, for instance, to absence of sections. Secondly, the Euler-number comparison only makes sense if both configurations correspond to a smooth gCICY three-fold and if we were to make use of the topology tools of Section~\ref{constrandtop} in computing the Euler number, both configuration matrices would have to result in a reduced geometry. Therefore, to be able to say that two gCICY configurations related by an ineffective splitting lead to the same manifold, one first needs to verify that both are a nice configuration matrix, in that the corresponding geometry is a non-empty smooth three-fold with a reduced coordinate ring.\footnote{To be precise, there is another subtlety one needs to be careful about. As it is clear from the example,~(\ref{e.g.split.de}), the derivation of splitting relies on the fact that each of the sections from the $n$ columns involved in the process has a linear dependence in the coordinates of the splitted $\IP^n$. For ordinary CICYs, this is surely the case and even for gCICYs, any sections constructed as a rational form, eq.~(\ref{q}), will meet this property. However, because our gCICY construction bases on a sequential section construction, eq.~(\ref{q}) might not give rise to a most general section and sometimes one might need to investigate a common degree shift in the numerator and the denominator, in which case the derivation of splitting might not work. Although this does not seem to happen due to the controlled cohomology structure corresponding to the splitted columns of the configuration matrix, we do not have a complete proof of that and instead, for given configurations we explicitly ensure the linearity of sections. }

\vspace{0.5cm}
\subsection{Identities and reduction rules}
In addition to the ineffective splitting, there are many other sources of redundancies. Most notably, as already described in Ref.~\cite{Candelas:1987kf}, one can make use of the identities between various manifolds for part of the full configuration matrices; there, an exhaustive list is given of identities between one-folds and between two-folds, and also a partial list of those between three-folds and between manifolds of a higher dimension. 
For instance, the second identity in the list for one-folds~\cite{Candelas:1987kf} is
\beq\label{e.g.identity}
\left[\begin{array}{c||c}
\pp^{1} & 1  \\
\pp^{1} & 1 \\
\end{array}\right] =
\IP^1 \ ,
\eeq
which can be applied to show that
\beq\label{e.g.identity.app}
\left[\begin{array}{c||cc}
\pp^{1} & 1  & a\\
\pp^{1} & 1 & b\\
A & \mathbf 0 & C \\
\end{array}\right] =
\left[\begin{array}{c||c}
\pp^{1} &  a+b \\
A & C \\
\end{array}\right] \ , 
\eeq
where $a$ and $b$ are a row vector of degree entries, $A$ is a product of projective spaces, $\mathbf 0$, a column of zeros, and $C$, a submatrix of an appropriate size. 

All such reduction rules are based on the identities between part of the two configuration matrices as well as the resulting relationship between the holomorphic line bundles on the two sides. Therefore, they straight-forwardly generalize to the gCICY cases and we may retrieve all the rules from Section 4 and Appendix of Ref.~\cite{Candelas:1987kf}. 

\vspace{0.5cm}
\subsection{Trivial line bundles}
The trivial line bundle over a connected base manifold has a unique global section, the constant function. For the ordinary CICY case, even when the configuration matrix is viewed as that of a gCICY, the line bundle corresponding to each column of the configuration is non-trivial on the sequentially constructed ambient space over which the sections are constructed. For the gCICY case, however, due to the presence of negative entries, one might end up describing a trivial line bundle by a non-trivial column. Let us begin with an illustrative toy example for which this happens:
\beq\label{e.g.trivial}
X=\left[\begin{array}{c||c|c}
\pp^{1} & 1 & -1\\
\pp^{1} & 1 & 1
\end{array}\right] \ ; \quad \cM=
\left[\begin{array}{c||c}
\pp^{1} & 1  \\
\pp^{1} & 1 
\end{array}\right] \ .
\eeq 
The ambient space $\cM$, where the line bundle $\cL=\cO_\cM(-1,1)$ is defined, is a $\IP^1$, which has a rank-one Picard lattice. Therefore, its configuration matrix in eq.~(\ref{e.g.trivial}) is non-favourable since the Picard lattice of its ambient space, $\IP^1 \times \IP^1$, is of rank two. There are several ways to see that $\cL$ is a trivial line-bundle over $\cM=\IP^1$; one can directly compute its cohomology dimensions, resulting in
\beq
h^{0}(\cM, \cL) = 1 \ ; \quad h^1(\cM, \cL) = 0 \ , \eeq
which implies that $\cL= \cO_{\cM}$, or alternatively, by constructing the unique holomorphic section of $\cL$ over $\cM$ in its ration form, eq.~(\ref{q}), one can also see that the vanishing locus is empty.  
An immediate consequence of this is that any gCICY configuration matrices with the two-by-two sub-block for $X=\tiny{\left[\begin{array}{c||c|c}
\IP^1 &1 & -1 \\
\IP^1 & 1 & 1 \ 
\end{array}\right]}$ appearing as
\beq\label{trivial}
\def\arraystretch{1.2}\left[\ba{c||ccc|ccc} 
\vdots & \ddots & \mathbf 0& \ddots & \ddots & \mathbf 0 & \ddots  \\
\IP^1 &  \cdots & 1 &\cdots & \cdots & -1 & \cdots \\ 
\vdots &  \ddots &\mathbf 0  & \ddots & \ddots & \mathbf 0 & \ddots  \\
\IP^1 & \cdots & 1&\cdots  & \cdots & 1 & \cdots \\ 
\vdots & \ddots &\mathbf 0 & \ddots & \ddots &\mathbf 0 & \ddots  
\ea\right]  \ , 
\eeq
corresponds to an empty set, given that only a single section (up to scaling) arises from the mixed-sign column with $-1$ and $1$ along the two $\IP^1$ directions.
Note that such reductions to an empty set can only happen for a non-favourable situation and hence, the minimal number of rows involved is $2$. 
One can then see that the most general way in which an empty set arises via a two-by-two description of $\cO_{\IP^1}$, as in eq.~(\ref{trivial}), is either of the following two types, 
\beq\label{trivialgeneral}
\def\arraystretch{1.2}\left[\ba{c||ccc|ccc} 
\vdots & \ddots & \mathbf 0& \ddots & \ddots & \mathbf 0 & \ddots  \\
\IP^1 &  \cdots & 1 &\cdots & \cdots & \pm 1 & \cdots \\ 
\vdots &  \ddots &\mathbf 0  & \ddots & \ddots & \mathbf 0 & \ddots  \\
\IP^1 & \cdots & n&\cdots  & \cdots & \mp n & \cdots \\ 
\vdots & \ddots &\mathbf 0 & \ddots & \ddots &\mathbf 0 & \ddots  
\ea\right] \ . 
\eeq
There are various other ways that lead to an empty-set configuration via the appearance of $\cO_{\IP^1}$ involving a bigger sub-block and even via that of $\cO_\cM$ for a more general base $\cM$. Although classification of all such possibilities can be an interesting direction to pursue further, the idea should already be clear from these toy instances,~(\ref{trivial}) and~(\ref{trivialgeneral}), and we do not attempt in this paper to complete the job.  

\vspace{0.5cm}

\subsection{Multiple components}
Configuration matrices of a gCICY may describe multiple Calabi-Yau components and, in particular, multiple copies of a Calabi-Yau manifold. One way this can occur is through the gCICY configuration matrix of the following form,
\beq\label{mcgeneral}
\def\arraystretch{1.2}\left[\ba{c||ccc|c} 
\vdots & \ddots & \mathbf 0& \ddots & \ddots\\
\IP^1 &  \cdots & n &\cdots & \cdots\\
\vdots &  \ddots &\mathbf 0  & \ddots & \ddots   \\
\ea\right] \ , 
\eeq 
where the column involving $n$ gives rise to $n$ points in the $\IP^1$ and the rest of the configuration matrix, obtained by deleting that column as well as the $\IP^1$ row, leads to a gCICY three-fold, for each of these $n$ points on the $\IP^1$. 
An example can be found in the third entry in Table \ref{t:(1,1)14}, which describes copies of a quintic three-fold. 

Note that such a multiple-copy structure via configurations of the form,~(\ref{mcgeneral}), could also have arisen for ordinary CICYs. Since negative entries are not allowed for them, however, the entry, $n$, in eq.~(\ref{mcgeneral}) is either $0$, $1$, or $2$, all of which can then be neglected for the following reason. The column involving $n$ is redundant for $n=0$ and $n=1$ cases, as the former gives no defining relation at all and the latter only fixes a point in the $\IP^1$, which should then be substituted to the defining relation dependent on this $\IP^1$ direction, resulting in the CICY three-fold whose configuration matrix is a reduced one from eq.~(\ref{mcgeneral}) by deleting the column involving the $n=1$ and the $\IP^1$ row. Similarly, $n=2$ case will lead to two copies of a Calabi-Yau three-fold, as the column involving the $n=2$ gives a quadratic equation in the $\IP^1$ coordinates, leading to two zero points. However, such case was neglected in Ref.~\cite{Candelas:1987kf}(see eq.~(1.40) there) as it corresponds to a product manifold. 

For a similar reason, gCICY configuration matrices of the form,~(\ref{mcgeneral}), give $n$ copies of a Calabi-Yau topology, each of which has a different shape in general. Nevertheless, if we only search for a connected geometry, such a multiple-copy case, or a multiple-component case in general, is another source of redundancy and can thus be excluded. 


\vspace{0.5cm}


\section{Beginning a Classification of  Generalized CICYs} \label{classcases}
Although a full classification of gCICY manifolds is beyond the scope of the current work, in this Section we take the first steps towards such a categorization and build an initial dataset of $2,761$ configuration matrices. We proceed to systematically classify and/or scan several classes of low co-dimension gCICY three-folds.

\subsection{Codimension (1,1) generalized CICYs}
\label{subsec:codim11}
The simplest type of generalized CICYs is a class given by the configuration matrix:
\begin{equation} \label{eq:X(1,1)}
X_{(1,1)}=
\left[\begin{array}{c||c|c}
\IP^{n_1} & a^{1} & b^{1}\\
\IP^{n_2} & a^{2} & b^{2}\\
\vdots & \vdots &\vdots\\
\IP^{n_m} & a^{m} & b^{m}\\
\end{array}\right] \ ,
\end{equation}
where we recall that the $a^i$'s are all semi-positive, whereas some of the $b^i$'s can take negative values. We call this class codimension (1,1) gCICYs. To begin, it must be understood first what kind of new gCICY configuration matrices of this type are well-defined varieties -- that is, under what conditions the second column of the configuration matrix defines a line bundle with global sections on a manifold defined by the first positive column in \eref{eq:X(1,1)}. We will impose the CY condition on the first Chern class in a second step. More concretely, we will study how many negative entries we have in the second column, $b^i$'s, and how much negative they can go, by requiring the existence of a global section of the line bundle 
\begin{equation}
\mathcal{L}_{\mathcal{M}}\equiv\mathcal{O}_{\mathcal{M}}(b^{1},\ldots,b^m) \ ,
\end{equation}
where $\mathcal{L}_{\mathcal{M}}$ defines a hypersurface in a new ambient manifold, given by the complete intersection manifold $\mathcal{M}$,
\begin{equation}\label{mdf}
\mathcal{M}=
\left[\begin{array}{c||c}
\IP^{n_1} & a^{1} \\
\IP^{n_2} & a^{1} \\
\vdots & \vdots \\
n_m & a^{m} \\
\end{array}\right] \ .
\end{equation}
We will reduce the problem from $\mathcal{M}$ to the original ambient space by using the twisted Koszul short-exact sequence
\begin{equation} \label{kszlcod11}
0\longrightarrow \mathcal{N}^{\vee}\otimes \mathcal{L}\longrightarrow\mathcal{L \longrightarrow} \mathcal{L}|_{\mathcal{M}}\longrightarrow 0 \ ,
\end{equation}
where the normal bundle of $\mathcal{M}$ is the line bundle $\mathcal{N}^{\vee} =  \mathcal{O}(-a^{1},\ldots,-a^m)$ in the original ambient space. The values of $b^i$'s will be practically all constrained by the cohomology of $\mathcal{L}_{\mathcal{M}}$, which we evaluate using the long-exact sequence in cohomology associated to \eref{kszlcod11}. In fact, all we need to do now is to check when the condition $h^0(\mathcal{M},\mathcal{L})\neq 0$ is satisfied.

\subsubsection{A negative bound } \label{sec:negbound}
The number of global sections, $h^0(\mathcal{M},\mathcal{L})$, depends on the ambient space cohomology groups of  $\mathcal{N}^{\vee} \otimes \mathcal{L} =  \mathcal{O}(-a^{1}+b^1,\ldots,-a^m+b^m)$ and $\mathcal{L}=  \mathcal{O}(b^{1},\ldots,b^m)$. The dimensions $h^*({\cal A},\mathcal{L})$ and $h^*({\cal A},\mathcal{N}^{\vee} \otimes \mathcal{L})$ can be evaluated with the Bott-Borel-Weyl formula for line bundle cohomology for products of projective spaces (see Appendix \ref{line_cohom} and \cite{Hubsch:1992nu,Anderson:2008ex} for reviews) and by looking then at the long exact sequence in cohomology associated to \eref{kszlcod11}. 
We will begin our analysis at this last sequence. Using the results of Appendix \ref{line_cohom} it can be verified that non-ample line bundles on this class of ${\cal M}$ will arise only from negative entries in $\pp^1$ directions and as result, the only non-vanishing cohomology groups associated to the first two terms in \eref{kszlcod11} are $H^1({\cal A}, \cdot)$. The map between them is given by the defining relation of $\mathcal{M}$:
\begin{equation}
q: h^{1}({\cal A},\mathcal{N}^{\vee} \otimes \mathcal{L}) \to h^1({\cal A},\mathcal{L}) \ ,
\end{equation}
In the long exact sequence associated to \eref{kszlcod11} this map is {\it generically} injective when $h^{1}({\cal A},\mathcal{N}^{\vee} \otimes \mathcal{L})< h^1({\cal A},\mathcal{L})$, and hence we can generically\footnote{We have strong evidences that the map is injective when the defining relation of $\mathcal{M}$ are general enough. We gain these evidences by computing explicitly the line bundle cohomologies and by analysing their behaviours when the entries of $\mathcal{L}$ and $\mathcal{N}^{\vee} \otimes \mathcal{L}$ change.} exclude this possibility since we demand that $h^0(\mathcal{M},\mathcal{L})\neq 0$ . For this reason and for the fact that there is only one non-vanishing line bundle cohomology in an ambient space, which is a product of projective space, we have just two different possibilities for $h^0(\mathcal{M},\mathcal{L})\neq 0$,
\begin{enumerate}
\item $h^1({\cal A},\mathcal{N}^{\vee}\otimes \mathcal{L})=0 \; \Rightarrow \; h^0({\cal A},\mathcal{L})>h^0({\cal A},\mathcal{N}^{\vee}\otimes \mathcal{L})$, with $h^0({\cal A},\mathcal{L})\neq 0$, however, if the zero-cohomology of $\mathcal{L}$ is non-vanishing in the ambient space, it reduces to the case where all the integer entries of $\mathcal{L}$ are positive: $b^i\geq 0, \; \forall i=1,\ldots, m$.
\item $h^0({\cal A},\mathcal{L})= 0$ and $h^1({\cal A},\mathcal{N}^{\vee}\otimes \mathcal{L})\neq 0$, this last condition, together with the Bott-Borel-Weyl formula in \eref{bottformula} and \eref{kunneth} \cite{Hubsch:1992nu,Anderson:2008ex} yields the following sufficient\footnote{These conditions are always sufficient and generically necessary for the existence of non-trivial global sections of a line bundle on ${\cal M}$ for suitably ``generic" defining relations in \eref{mdf}. However, for all configuration matrices (including ordinary CICYs) these conditions can be weakened and line bundle cohomology can ``jump" \cite{Anderson:2009ge,Anderson:2009mh,Donagi:2004qk,Donagi:2004ub,Braun:2005xp} when special defining polynomials are chosen for ${\cal M}$. In these special cases new effective divisors exist. We will not consider such cases in the following discussion. }  (and generically necessary) conditions for $h^0({\cal M}, {\cal L}) >0$:
\begin{equation} \label{eq:primcond}
\exists ! \, i=1 \; | \; n_1=1, \qquad (b^{1}-a^{1})\leq -2, \qquad \forall i \neq 1\;\; b^i>0, \;\; (b^i-a^i)\geq 0 \ .
\end{equation}
where we have set the first $n_1=1$ for convenience, and, moreover, 
\begin{equation} \label{eq:boundcond}
h^1({\cal A},\mathcal{N}^{\vee}\otimes \mathcal{L})>h^1({\cal A},\mathcal{L}) \ .
\end{equation}
\end{enumerate}
We are of course interested in extensions of the standard class of CICYs, hence we will study further only the second case. Equation \eref{eq:primcond} tells us that in codimension (1,1) only a negative entry, $b^1$, in a $\pp^1$ factor is possible. Finally, we expect that \eref{eq:boundcond} gives a bound on the negative value for $b^1$. We then apply the Bott-Borel-Weyl theorem \cite{Hubsch:1992nu,Anderson:2008ex}, together with all the conditions in \eref{eq:primcond}, and we obtain the following expressions for the line bundle first-cohomologies, $h^1$, in the ambient space,
\begin{eqnarray}
h^1({\cal A},\mathcal{L})&=&
(-b^{1}-1)\prod^{m}_{i=2} \binom{b^i+n_i}{n_i} \ , \label{eq:h1L}\\
h^1({\cal A},\mathcal{N}^{\vee}\otimes \mathcal{L})&=&
(a^1-b^1-1)\prod^{m}_{i=2} \binom{b^{i}-a^{i}+n_i}{n_i} \ .\label{eq:h1NL}
\end{eqnarray}
Plugging in (\ref{eq:h1L}-\ref{eq:h1NL}) into \eref{eq:boundcond}, we get the following bounding inequality:
\begin{equation} \label{eq:boundin}
\frac{(a^1-b^1-1)}{(-b^{1}-1) }>R \ ,
\end{equation}
where we rearranged the expression and $R$ is a ratio that depends on the line bundle cohomologies in $\pp^{n_i}$, with $i>1$,
\begin{equation} \label{eq:ratio}
R \equiv \frac{\prod^{m}_{i=2} \binom{b^i+n_i}{n_i} }{ \prod^{m}_{i=2} \binom{b^i-a^i+n_i}{n_i}} \ .
\end{equation}
Reducing \eref{eq:boundin} in terms of $b^1$, we get 
\begin{eqnarray}
1<R<1+a^1\quad &\& &\quad -\frac{a^1}{R-1}-1<b^1< -2 \label{eq:bound1}\\
R=1\quad &\& &\quad b^1<- 2 \label{eq:unbound1} \ .
\end{eqnarray}
In the second case \eref{eq:unbound1}, we do not have any negative bound, however this apparently infinite class is actually finite. As will be shown in Section \ref{infinite_sec} these are just copies of a known CICY, when we impose the right condition on the Chern-classes of the manifold $X_{(1,1)}$. From \eref{eq:bound1}, we get an expression for the negative bound  of $b^1$ in terms of $R$ and $a^1$.
\begin{equation}
b^1=-[1+\frac{a^1}{R-1}] \ .
\end{equation}

Finally we want to impose the Calabi-Yau condition on the first Chern class $c_1(TX)=0$, which restricts $b^{1}=2-a^1$, and $b^i+a^i=n_i+1$ with $i=2\ldots m$. In this way the bound depends only on $R$. We can summarize all the possible cases in the following four cases:
\begin{itemize}
\item $R=4$, the bound is $b^1\geq -2$.
\item $R>4$, the bound is $b^1=-1$.
\item $R=2$, there is no bound, and $b^i\leq -1$. Apparently this is another infinite class, we will show in Section \ref{infinite_sec} it is actually finite.
\item $R=1$, for this case as well, we have no bound, and the subclass appears infinite. However it turns out that the Euler characteristic is a multiple of a Euler characteristic of a known CY, $\tilde{X}$, as well as all the topological data. Moreover, looking in detail at this case, we can see that for $R=1$ the configuration matrix takes the form
\begin{equation}
X_{(1,1)}=
\left[\begin{array}{c||c|c}
\IP^1 & 2+i & i\\
\IP^{n_2} & 0 & b^{2}\\
\vdots & \vdots &\vdots\\
\IP^{n_m} & 0 & b^{m}\\
\end{array}\right],\qquad i \in \mathbb{Z}_{>0} \ .
\end{equation}
We will show in section \ref{TypeI} that $X_{(1,1)}$ corresponds to $(2+i)$ copies of the known Calabi-Yau
\begin{equation}
\tilde{X}=
\left[\begin{array}{c||c}
\IP^{n_2} &  b^{2}\\
\vdots & \vdots\\
\IP^{n_m} &  b^{m}\\
\end{array}\right] \ ,
\end{equation}
with all positive entries for definition. 
These infinite manifolds are all non strictly Calabi-Yau, more specifically each of these is a manifold formed by multiple copies of an already known CICY.

\end{itemize}

\subsubsection{The classification result}
CICYs with codimension (1,1) are determined by two equations in a product of projective space, hence, we can encounter only 5 ambient space which are possible for this class of CICY three-folds. The complete list of generalized codimension (1,1) CICY is given according to the classification in \ref{sec:negbound}. The Euler characteristic together with the Hodge numbers $(h^{(1,1)}, h^{(2,1)})$ are summarized in Tables \ref{t:(1,1)14}-\ref{t:(1,1)11111}.  In the last column,  we present the results of a smoothness check of the manifolds. In the tables below, we will use the symbol $\CIRCLE$ to denote a smooth manifold, $\Circle$ a singular manifold and $\LEFTcircle$ will indicate a manifold for which smoothness is not yet determined.
We remark that the last infinite class of every table cannot be strictly considered as CY manifolds:
\begin{itemize}
\item For each $i$ in the last subclass of manifolds of table \ref{t:(1,1)14}, we have a manifold formed by $(2+i)$ copies of $[ \,\IP^4\, ||\, 5\, ]$.

\item For each $i$ in the last subclass of manifolds of table \ref{t:(1,1)113}, we have a manifold formed by $(2+i)$ copies of $\left[\begin{array}{c||c}
\IP^3 & 4\\
\IP^1 & 2\\
\end{array}\right] $.

\item For each $i$ in the last subclass of manifolds of table \ref{t:(1,1)122}, we have a manifold formed by $(2+i)$ copies of $\left[\begin{array}{c||c}
\IP^2 & 3\\
\IP^2 & 3
\end{array}\right] $.

\item For each $i$ in the last subclass of manifolds of table \ref{t:(1,1)1112}, we have a manifold formed by $(2+i)$ copies of $\left[\begin{array}{c||c}
\IP^2 & 3\\
\IP^1 & 2\\
\IP^1& 2\\
\end{array}\right] $.

\item For each $i$ in the last subclass of manifolds of table \ref{t:(1,1)11111}, we have a formed by $(2+i)$ copies of $\left[\begin{array}{c||c}
\IP^1 & 2\\
\IP^1& 2\\
\IP^1 & 2 \\
\IP^1 & 2 
\end{array}\right] $.

\end{itemize} 

\vspace{0.25in}

The list of $(1,1)$ manifolds is presented in Tables \ref{t:(1,1)14}-\ref{t:(1,1)11111} given below.

\begin{table}[!h]
\begin{center}
\begin{tabular}{|c|c|c|c|c|c|c|}\hline
$X$& R & i & $\chi$&\footnotesize{($h^{1,1}(X), h^{1,2}(X)$)} & Infinite Class & Smoothness \\\hline\hline &&&&&&\\[-4.2mm]

\tiny${\left[
\begin{array}{c||c|c}
 \IP^4 &2 & 3\\
 \IP^1 & 3 & -1 \\
\end{array}
\right]}$& 7 & N/A &$-88$& $(2,46)$ &  N/A & \CIRCLE  \\\hline &&&&&&\\[-4.2mm]
\tiny${\left[
\begin{array}{c||c|c}
\IP^4 &1 & 4\\
\IP^1 &2+i & -i \\
\end{array}
\right]}$& 2 & $i \in \mathbb{Z}_{>0}$&$-168$& $(2,86)$& Type III & \CIRCLE \\\hline&&&&&&\\[-4.2mm]
\tiny${\left[
\begin{array}{c||c|c}
\IP^4 &0 & 5\\
\IP^1 &2+i & -i \\
\end{array}
\right]}$& 1 & $i \in \mathbb{Z}_{>0}$ & $-200 (i+2)$& $(i+2,101(i+2))$ &  Type I &  \CIRCLE \\\hline\end{tabular}
\caption{3 cases in
  $\mathbb{P}^4\times \pp^1$. $\CIRCLE$ indicates they are smooth.}\label{t:(1,1)14}
\end{center}
\end{table}

\begin{table}[!hbtp] 
\begin{center}
\begin{tabular}{|c|c|c|c|c|c|c|}\hline
$X$& R & i & $\chi$&\footnotesize{($h^{1,1}(X), h^{1,2}(X)$)}& Infinite Class & Smoothness \\\hline\hline &&&&&&\\[-4.2mm]

\tiny${\left[
\begin{array}{c||c|c}
  \IP^3 &2 & 2\\
   \IP^1 & 0 & 2 \\
 \IP^1 &3 & -1 \\
\end{array}
\right]}$& 10 & N/A &$-56$& $(3,31)$ & N/A & \CIRCLE  \\\hline &&&&&&\\[-4.2mm]
\tiny${\left[
\begin{array}{c||c|c}
 \IP^3 &2 & 2\\
  \IP^1 &1 & 1 \\
 \IP^1 &3 & -1 \\
\end{array}
\right]}$& 20 &  N/A & $-104$& $(3,55)$ & N/A & \CIRCLE \\\hline &&&&&&\\[-4.2mm]
\tiny${\left[
\begin{array}{c||c|c}
 \IP^3 &1 & 3\\
  \IP^1 &1 & 1 \\
 \IP^1 &2+i & -i \\
\end{array}
\right]}$& 4 & $i =1, 2$ & $ -72, \,\,-48$& $(3,39),\,\,(3,27)$ & N/A & \CIRCLE \\\hline &&&&&&\\[-4.2mm]
\tiny${\left[
\begin{array}{c||c|c}
 \IP^3 &1 & 3\\
  \IP^1 &0 & 2 \\
 \IP^1 &2+i & -i \\
\end{array}
\right]}$& 2 & $i \in \mathbb{Z}_{>0}$ & $ -144$& $(3,75)$ & Type III & \CIRCLE \\\hline&&&&&&\\[-4.2mm]
\tiny${\left[
\begin{array}{c||c|c}
 \IP^3 &0 & 4\\
  \IP^1 &1 & 1 \\
 \IP^1 &2+i & -i \\
\end{array}
\right]}$& 2 & $i \in \mathbb{Z}_{>0}$ & $ -168$& $(2,86)$ & Type II &\CIRCLE \\\hline&&&&&&\\[-4.2mm]
\tiny${\left[
\begin{array}{c||c|c}
 \IP^3 &0 & 4\\
  \IP^1 &0 & 2 \\
 \IP^1 &2+i & -i \\
\end{array}
\right]}$& 1 & $i \in \mathbb{Z}_{>0}$ & $ -168(i+2)$& $(i+2,86(i+2))$ & Type I &\CIRCLE \\\hline

\end{tabular}
\caption{6 cases in
  $\mathbb{P}^3\times \mathbb{P}^1\times\mathbb{P}^1 $. $\CIRCLE$ indicates they are smooth.} \label{t:(1,1)113}
\end{center}
\end{table}

\begin{table}[!hbtp]
\begin{center}
\begin{tabular}{|c|c|c|c|c|c|c|}\hline
$X$& R & i & $\chi$&\footnotesize{($h^{1,1}(X), h^{1,2}(X)$)}&  Infinite Class &Smoothness \\\hline\hline &&&&&&\\[-4.2mm]

\tiny${\left[
\begin{array}{c||c|c}
 \IP^2 & 1 & 2\\
  \IP^2 &1 & 2\\
    \IP^1 &2+i & -i\\
\end{array}
\right]}$& 4 & $i=1,2$ &$ -78, \,\,-60$& $(3,42),\,\,(3,33)$ & N/A & \CIRCLE  \\\hline&&&&&&\\[-4.2mm]
\tiny${\left[
\begin{array}{c||c|c}
 \IP^2 &0 & 3 \\
 \IP^2 &1 & 2\\
  \IP^1 &2+i & -i \\
\end{array}
\right]}$& 2 & $i \in \mathbb{Z}_{>0}$& $-144$& $(3,75)$ & Type III & \CIRCLE \\\hline&&&&&&\\[-4.2mm]
\tiny${\left[
\begin{array}{c||c|c}
 \IP^2 &0 & 3 \\
 \IP^2 &0 & 3 \\
  \IP^1 &2+i & -i \\
\end{array}
\right]}$& 1 & $i \in \mathbb{Z}_{>0}$ & $ -162(i+2)$ & $(2(i+2),83(i+2))$ & N/A & \CIRCLE \\\hline

\end{tabular}
\caption{3 cases in
  $ \pp^2 \times \pp^2 \times \pp^1$. $\CIRCLE$ indicates they are smooth.} \label{t:(1,1)122}
\end{center}
\end{table}

\begin{table}[!hbtp]
\begin{center}
\begin{tabular}{|c|c|c|c|c|c|c|}\hline
$X$& R & i & $\chi$&\footnotesize{($h^{1,1}(X), h^{1,2}(X)$)}& Infinite Class & Smoothness \\\hline\hline&&&&&&\\[-4.2mm]

\tiny${\left[
\begin{array}{c||c|c}
  \IP^2 &1 & 2\\
 \IP^1 & 1& 1 \\
 \IP^1 & 1 & 1\\
 \IP^1 & 3 & -1 \\
\end{array}
\right]}$& 8 & N/A &$-68$& $(4,38)$ & N/A &\CIRCLE  \\\hline&&&&&&\\[-4.2mm]
\tiny${\left[
\begin{array}{c||c|c}
 \IP^2 &1& 2\\
 \IP^1 &0 & 2 \\
 \IP^1 &1 & 1\\
  \IP^1 &2+i & -i \\
\end{array}
\right]}$& 4 &  $i=1,2$& $ -56, \,\,-32$& $(4,32),\,\,(4,20)$ & N/A & \CIRCLE \\\hline&&&&&&\\[-4.2mm]
\tiny${\left[
\begin{array}{c||c|c}
 \IP^2 &0 & 3\\
 \IP^1 &1 & 1 \\
 \IP^1 &1 & 1 \\
 \IP^1 &2+i & -i \\
\end{array}
\right]}$& 4 & $i =1, 2$ & $ -36, \,\,0$& $(9,27),\,\,(11,11)$ &  N/A &\CIRCLE \\\hline&&&&&&\\[-4.2mm]
\tiny${\left[
\begin{array}{c||c|c}
 \IP^2 &0 & 3\\
 \IP^1 &0 & 2 \\
 \IP^1 &1 & 1 \\
  \IP^1 &2+i & -i \\
\end{array}
\right]}$& 2 & $i \in \mathbb{Z}_{>0}$ & $ -144$& $(3,75)$ & Type II & \CIRCLE \\\hline&&&&&&\\[-4.2mm]
\tiny${\left[
\begin{array}{c||c|c}
 \IP^2 &1 & 2\\
 \IP^1 &0 & 2 \\
 \IP^1 &0 & 2 \\
  \IP^1 &2+i & -i \\
\end{array}
\right]}$& 2 & $i \in \mathbb{Z}_{>0}$ & $ -128$& $(4,68)$ & Type III & \CIRCLE \\\hline&&&&&&\\[-4.2mm]
\tiny${\left[
\begin{array}{c||c|c}
 \IP^2 &0 & 3 \\
 \IP^1 &0 & 2 \\
 \IP^1 &0 & 2 \\
  \IP^1 &2+i & -i \\
\end{array}
\right]}$& 1 & $i \in \mathbb{Z}_{>0}$ & $ -144(i+2)$& $(3(i+2),75(i+2))$ & Type I & \CIRCLE \\\hline

\end{tabular}
\caption{6 cases in
  $ \pp^2 \times \pp^1 \times \pp^1 \times \pp^1 $. $\CIRCLE$ indicates they are smooth.} \label{t:(1,1)1112}
\end{center}
\end{table}

\begin{table}[!hbtp]
\begin{center}
\begin{tabular}{|c|c|c|c|c|c|c|}\hline
$X$& R & i & $\chi$&\footnotesize{($h^{1,1}(X), h^{1,2}(X)$)}& Infinite Class &Smoothness \\\hline\hline&&&&&&\\[-4.2mm]

\tiny${\left[
\begin{array}{c||c|c}
 \IP^1 & 1& 1 \\
 \IP^1 & 1 & 1\\
 \IP^1 &  1 & 1\\
 \IP^1 & 1 & 1\\
  \IP^1 & 3 & -1 \\
\end{array}
\right]}$& 16 & N/A &$-80$& $(5,45)$ & N/A & \CIRCLE  \\\hline&&&&&&\\[-4.2mm]
\tiny${\left[
\begin{array}{c||c|c}
 \IP^1 &1 & 1\\
 \IP^1 &1 & 1\\
 \IP^1 &1 & 1\\
 \IP^1 &0& 2\\
  \IP^1 &3& -1 \\
\end{array}
\right]}$& 8 &  N/A & $ -48$& $(5,29)$ & N/A & \CIRCLE \\\hline&&&&&&\\[-4.2mm]
\tiny${\left[
\begin{array}{c||c|c}
 \IP^1 &1 & 1\\
 \IP^1 &1 & 1\\
 \IP^1 &0 & 2\\
 \IP^1 &0& 2 \\
  \IP^1 &2+i & -i \\
\end{array}
\right]}$& 4 & $i =1, 2$ & $ -32, \,\,0$& $(10,26),\,\,(12,12)$ &  N/A & \CIRCLE \\\hline&&&&&&\\[-4.2mm]
\tiny${\left[
\begin{array}{c||c|c}
 \IP^1 &1 & 1\\
 \IP^1 &0 & 2\\
 \IP^1 &0 & 2\\
 \IP^1 &0& 2 \\
  \IP^1 &2+i & -i \\
\end{array}
\right]}$& 2 & $i \in \mathbb{Z}_{>0}$ & $ -128$& $(4,68)$ &  Type II & \CIRCLE \\\hline&&&&&&\\[-4.2mm]
\tiny${\left[
\begin{array}{c||c|c}
 \IP^1 &0 & 2\\
 \IP^1 &0 & 2\\
 \IP^1 &0 & 2\\
 \IP^1 &0& 2\\
  \IP^1 &2+i & -i \\
\end{array}
\right]}$& 1 & $i \in \mathbb{Z}_{>0}$ & $ -128(i+2)$& $(4(i+2),68(i+2))$ &  Type I & \CIRCLE \\\hline

\end{tabular}
\caption{5 cases in
  $\pp^1 \times \pp^1 \times \pp^1 \times \pp^1 \times \pp^1$. $\CIRCLE$ indicates they are smooth.} \label{t:(1,1)11111}
\end{center}
\end{table}

\subsubsection{Apparently infinite classes}\label{infinite_sec}

There are twelve apparently infinite families of Calabi-Yau manifolds in the list given in the previous section. Here we will analyze these cases further and demonstrate that these are, in each case, a set of redundant descriptions of the same, standard CICY. 
These infinite classes of configurations matrices can be grouped in to three distinct types which is  illustrated in Tables \ref{t:(1,1)14}-\ref{t:(1,1)11111} . 

\subsubsection*{Type I} \label{TypeI}

The configurations matrices of Type I all describe multiple, disconnected, copies of a regular CICY. Let us exemplify the structure with the third example in Table \ref{t:(1,1)14},
 \begin{equation}
\left[\begin{array}{c||c|c}
\mathbb{P}^{1} & 2 + i & -i\\
\mathbb{P}^{4} & 0 & 5
\end{array}\right], \qquad i \in \mathbb{Z}_{>0} \ .
\end{equation}
We can solve the first equation, a degree $2+i$ polynomial in $\mathbb{P}^1$ to find $2+i$ points in that first ambient space projective factor. Substituting these into the rational functions described by the second column of the configuration matrix, we obtain $2+i$ quintic polynomials in $\mathbb{P}^4$. At a generic locus in complex structure modulus space, therefore, this configuration describes $2+i$ disconnected quintics in $\mathbb{P}^4$ separated from one another by their location in $\mathbb{P}^1$.

As a confirmation of this analysis, we find that $h^0({\cal O})=2+i$ in these cases and that the euler number is $-(2+i)(200)$, i.e. $(2+i)$ times that of the quintic.

\subsubsection*{Type II}

The configuration matrices of Type II also all describe regular CICYs. We again exemplify the structure with a single case, in this instance the fifth example in Table \ref{t:(1,1)113},
\begin{equation} \label{typeIIeg}
\left[\begin{array}{c||c|c}
\mathbb{P}^{1} & 2+i& -i\\
\mathbb{P}^{1} & 1 & 1\\
\mathbb{P}^{3} & 0 & 4
\end{array}\right], \qquad i \in \mathbb{Z}_{>0} \ .
\end{equation}
These conifiguration matrices all describe the same manifold as the regular CICY,
\begin{eqnarray} \label{thetwofour}
\left[\begin{array}{c||c} \mathbb{P}^1 & 2 \\ \mathbb{P}^3 & 4 \end{array} \right] \ ,
\end{eqnarray}
via a relation analogous to what is called ``ineffective splitting" in the CICY literature.

Let us write the two equations described by the configuration matrices (\ref{typeIIeg}) as follows
\begin{eqnarray}  \label{amb1}
p_1 &=& P_{10} y^0 + P_{11} y^1 =0 \ , \\ \label{amb2}
p_2 &=& \frac{P_{20}}{D_{20}}y^0 +\frac{P_{21}}{D_{21}}y^1=0 \ .
\end{eqnarray}
It is important for us to describe exactly what is meant by the various $P$ polynomials which appear in these expressions. $P_{10}$ and $P_{11}$ are general degree $2+i$ polynomials. The polynomials $P_{20}, P_{21}, D_{20}$ and $D_{21}$ are not generic examples of their degree, however. $P_{20}$ and $P_{21}$ are degree $\left\{0,0,4\right\}$ while $D_{20}$ and $D_{21}$ are degree $\left\{i,0,0 \right\}$. However, the choice of the denominator and numerator polynomials are correlated (and also correlated with $P_{10}$ and $P_{11}$) such that the poles in the associated rational functions miss the hypersurface $p_1$. As described earlier, this structure is associated to the fact that the line bundle ${\cal O}(-i,1,4)$ only has global sections when taken to be a line bundle on the hypersurface defined by $p_1$ and not when interpreted as a line bundle on the ambient projective product. Nevertheless, the final Calabi-Yau three-fold can be described in terms of the intersection of a rational function and a polynomial on the ambient space, once a tuning of the freedom in the $P$'s and $D$'s of this form has been performed.

We may rewrite the ambient space equations (\ref{amb1}) and (\ref{amb2}) in a matrix form as follows
\begin{eqnarray}
\left( \begin{array}{cc}P_{10} &  P_{11} \\   \frac{P_{20}}{D_{20}} & \frac{P_{21}}{D_{21}} \end{array} \right) \left( \begin{array}{c} y^0 \\ y^1 \end{array} \right) =0 \ .
\end{eqnarray}
These equations have a solution, remembering that the $y^i$ are homogeneous coordinates on $\mathbb{P}^1$, if and only if the determinant of the matrix vanishes,
\begin{eqnarray} \label{thischap}
\frac{P_{10} P_{21}}{D_{21}} - \frac{P_{11} P_{20}}{D_{20}} =0 \ .
\end{eqnarray}
These equations have ``net degree" $\left\{2,4\right\}$ in $\mathbb{P}^1\times \mathbb{P}^3$. However, naively, (\ref{thischap}) seems to be still a rational defining relation. In fact, due to the manner in which the singularities in $p_2$ miss the zero locus of $p_1$, the numerators in the two terms of (\ref{thischap}) contain factors of $D_{20}$ and $D_{21}$ respectively, leading to a polynomial, $\left\{2,4\right\}$ defining relation in $\mathbb{P}^1\times \mathbb{P}^3$.

To demonstrate this let us derive (\ref{thischap}) in a slightly different manner (we performed the analysis above to emphasize the similarity with ineffective splits in the case of an ordinary CICY). First, we solve (\ref{amb1}) for either $y^0$ or $y^1$,
\begin{eqnarray}
y^0 &=& -\frac{P_{11}}{P_{10}} y^1 \ , \\  \label{chappy}
y^1 &=& -\frac{P_{10}}{P_{11}} y^0 \ .
\end{eqnarray}
For a generic choice of complex structure, $P_{11}$ and $P_{10}$ are two uncorrelated degree $2+i$ polynomials in $\mathbb{P}^1$. They can thus not both be vanishing simultaneously and thus at least one of the above solutions is always perfectly well defined.

Substituting these results into (\ref{amb2}) we obtain the following two expressions
\begin{eqnarray} \label{chappy2}
\left(\frac{P_{20}}{D_{20}} \frac{P_{11}}{P_{10}} + \frac{P_{21}}{D_{21}} \right) y^1 &=& 0 \ , \\
\left( \frac{P_{20}}{D_{20}} + \frac{P_{21}}{D_{21}} \frac{P_{10}}{P_{11}} \right) y^0 &=&0 \;.
\end{eqnarray}
If $P_{10} \neq 0$ then from (\ref{chappy}) $y^1 \neq 0$ (as not both homogeneous coordinates can vanish simultaneously).  We can divide (\ref{chappy2}) by $y^1$ and multiply up by $P_{10}$. We then obtain (\ref{thischap}) once more (a similar argument can be made in the case where $P_{11} \neq 0$). However, by the definition of our procedure for forming the section of the second normal bundle factor, that is the rational function $p_2$, we know that it can have no poles when $p_1=0$. We have just demonstrated that (\ref{thischap}) {\it is} $p_2$ evaluated on the solution to $p_1$ and thus it can have no poles. The only way in which this can happen is if the numerators and denominators of the full expression have cancelling factors leaving a simply polynomial expression. This can easily be verified explicitly in any given example.

Thus (\ref{typeIIeg}) describes the same manifold as (\ref{thetwofour}) for any value of $i$ and the infinite class can be viewed as generalized ineffective splittings of this standard CICY.

\subsubsection*{Type III}

The configuration matrices of Type III also all describe regular CICYs. As before, we exemplify the structure with a single case, in this instance the second example in Table \ref{t:(1,1)14}, 
\begin{equation} \label{type3eg1}
\left[\begin{array}{c||c|c}
\mathbb{P}^{1} & 2 + i & -i\\
\mathbb{P}^{4} & 1 & 4
\end{array}\right] \ , \qquad i \in \mathbb{Z}_{>0} \ .
\end{equation}

A series of operations on this infinite class of configuration matrices can demonstrate that (\ref{type3eg1}) describes the same manifold for all $i$. We begin by performing an ineffective split as follows
\begin{eqnarray} \label{spliteg11}
\left[\begin{array}{c||c|c}
\mathbb{P}^{1} & 2 + i & -i\\
\mathbb{P}^{4} & 1 & 4
\end{array}\right] \to 
\left[ \begin{array}{c||cc|c}  \mathbb{P}^1 & 0 & 2+i & -i  \\
 \mathbb{P}^1 & 1&1& 0 \\
  \mathbb{P}^4 & 1 & 0 & 4 \end{array} \right] \ .
\end{eqnarray}
In peforming such a split in the gCICY case, all of the subtleties mentioned in Section \ref{redun} must be considered. Once we have the configuration matrix in the form (\ref{spliteg11}) we can use the following postulated identity \cite{Candelas:1987kf}
\begin{eqnarray} \label{postident}
\left[ \begin{array}{c||cc} \mathbb{P}^1 &  1 & a \\ \mathbb{P}^n &1& n b \\ Y& 0 &M \end{array} \right] \to \left[ \begin{array}{c||c} \mathbb{P}^1 & a+b \\\mathbb{P}^{n-1} &n b \\ Y&M \end{array} \right] \ .
\end{eqnarray}
This identity was used in the original investigations of CICY three-folds \cite{Candelas:1987kf}, although it has never, to our knowledge, been proven. Here we follow those authors in making use of the identity, leaving a rigorous proof of its validity to future work. Applying (\ref{postident}) in this case we find the following
\begin{eqnarray} \label{spliteg12}
\left[\begin{array}{c||c|c}
\mathbb{P}^{1} & 2 + i & -i\\
\mathbb{P}^{4} & 1 & 4
\end{array}\right] \to 
\left[ \begin{array}{c||cc|c}  \mathbb{P}^1 & 0 & 2+i & -i  \\
 \mathbb{P}^1 & 1&1& 0 \\
  \mathbb{P}^4 & 1 & 0 & 4 \end{array} \right] \to \left[ \begin{array}{c||c|c} \mathbb{P}^1 & 1 & 1 \\ \mathbb{P}^3 & 0 &4 \\ \mathbb{P}^1 & 2+i & -i\end{array}  \right] \ .
\end{eqnarray}

Finally, we may apply an ineffective contraction in the first $\mathbb{P}^1$ direction of (\ref{spliteg12}) to arrive at the following chain
\begin{eqnarray} \label{spliteg13}
\left[\begin{array}{c||c|c}
\mathbb{P}^{1} & 2 + i & -i\\
\mathbb{P}^{4} & 1 & 4
\end{array}\right] \to 
\left[ \begin{array}{c||cc|c}  \mathbb{P}^1 & 0 & 2+i & -i  \\
 \mathbb{P}^1 & 1&1& 0 \\
  \mathbb{P}^4 & 1 & 0 & 4 \end{array} \right] \to \left[ \begin{array}{c||c|c} \mathbb{P}^1 & 1 & 1 \\ \mathbb{P}^3 & 0 &4 \\ \mathbb{P}^1 & 2+i & -i\end{array}  \right] \to \left[ \begin{array}{c||c} \mathbb{P}^3 & 4\\\mathbb{P}^1 & 2 \end{array} \right] \ .
\end{eqnarray}
Thus, we finally see that the entire infinite class of configuration matrices (\ref{type3eg1}) all simply describe a single ordinary CICY.

A very similar analysis applies to the other cases of Type III. The fourth example in Table \ref{t:(1,1)113} and second example in Table \ref{t:(1,1)122} turn out to be equivalent to the configuration matrix,
\begin{eqnarray}
\left[ \begin{array}{c||c} \mathbb{P}^1 & 2 \\\mathbb{P}^1 & 2 \\\mathbb{P}^2 &3\end{array} \right]\;.
\end{eqnarray}
The fifth example in Table \ref{t:(1,1)1112} turns out to be equivalent to the tetraquadric:
\begin{eqnarray}
\left[ \begin{array}{c||c} \mathbb{P}^1 & 2 \\ \mathbb{P}^1 & 2 \\ \mathbb{P}^1 & 2 \\ \mathbb{P}^1 & 2 \end{array} \right]\;.
\end{eqnarray}

With the demonstration that this last type of infinite set of configuration matrices actually corresponds to a finite number of Calabi-Yau three-folds we have proven that the gCICYs of codimension $(1,1)$ constitute a finite dataset of manifolds.

\subsubsection{New Calabi-Yau three-folds}\label{new_geom}
It is interesting to determine which of the manifolds in Tables \ref{t:(1,1)14}-\ref{t:(1,1)11111}  are genuinely new Calabi-Yau three-folds, never before seen in other datasets. Steps can be taken in towards this goal by computing topological invariants. We have studied the Hodge data and basis independent quantities computed from the Chern classes and intersection numbers (see Ref.~\cite{Gray:2014fla} for an analogue of this for the four-fold cases). We find that eight of the codimension (1,1) gCICYs have never appeared before, at least as far as our search of the literature could determine. They are certainly of a different topological type to anything seen in the CICY or the Kreuzer-Skarke datasets. The eight new manifolds are those with
\beq
(h_{1,1},h_{2,1})=(2,46),(3,31),(3,39),(3,27),(3,42),(3,33),(4,20),(5,29) 
\eeq
 from Tables~\ref{t:(1,1)14}-\ref{t:(1,1)11111}. Note that many of these Hodge pairs have certainly appeared in the literature before. The manifolds are distinguished by the more subtle topological properties mentioned above and illustrated in Subsection \ref{cherny}.

\subsection{Codimension (2,1) generalized CICYs}\label{21egs}
The codimension two examples in Section \ref{subsec:codim11} can be straightforwardly generalized to codimension three. We once again focus on negative codimension one cases, i.e, codimension $(2,1)$ examples with the following configuration matrix:
\begin{equation}\label{21general}
X_{(2,1)}=
\left[\begin{array}{c||cc|c}
\mathbb{P}^{n_1} & a^{1}_1 & a^{1}_2  & b^{1}\\
\mathbb{P}^{n_2} & a^{2}_1 & a^ {2}_2  & b^{2}\\
\vdots & \vdots &\vdots & \vdots\\
\mathbb{P}^{n_N} & a^{m}_1 & a^{m}_2  & b^{m}\\
\end{array}\right] \ ,
\end{equation}
where the first column $a^{i}_1, a^{i}_2 \geq 0$, but $b^{i}$ are allowed to assume negative integer values. As described previously, we require the third column to define an anticanonical hypersurface in ${\cal M}$. That is, $\mathcal{L}\equiv\mathcal{O}_{\mathcal{M}}(b^{1}_3,\ldots,b^{m}_3)={K_{\cal M}}^{-1}$ with
\begin{equation}
\mathcal{M}=
\left[\begin{array}{c||cc}
\mathbb{P}^{n_1} & a^{1}_1 & a^{1}_2  \\
\mathbb{P}^{n_2} & a^{2}_2 & a^ {2}_2 \\
\vdots & \vdots &\vdots \\
\mathbb{P}^{n_N}& a^{m}_1 & a^{m}_2 \\
\end{array}\right] \ .
\end{equation}
 In defining these CY $3$-folds there are $10$ possible ambient spaces for this class of generalized CICY three-folds: 
 \bea
 & \mathbb{P}^5 \times \mathbb{P}^1,~~\mathbb{P}^4 \times \mathbb{P}^2,~~\mathbb{P}^3 \times \mathbb{P}^3,~~\mathbb{P}^4 \times \mathbb{P}^1 \times \mathbb{P}^1 \ , \\
 & \mathbb{P}^3 \times \mathbb{P}^2\times \mathbb{P}^1,~~\mathbb{P}^2 \times \mathbb{P}^2 \times \mathbb{P}^2,~~\mathbb{P}^3 \times \mathbb{P}^1 \times \mathbb{P}^1 \times \mathbb{P}^1 \ , \\
 &\mathbb{P}^2 \times \mathbb{P}^2 \times \mathbb{P}^1 \times \mathbb{P}^1,~~\mathbb{P}^2 \times \mathbb{P}^1 \times \mathbb{P}^1 \times \mathbb{P}^1 \times \mathbb{P}^1,~~\mathbb{P}^1\times \mathbb{P}^1 \times \mathbb{P}^1 \times \mathbb{P}^1 \times \mathbb{P}^1 \times \mathbb{P}^1 \ .
 \eea
In each of these embedding products of projective spaces, there are many more classes of generalized configuration matrices than in codimension $(1,1)$ case.  

Beginning with these ambient spaces, we construct the generalized CICYs in several steps:
\begin{itemize}
 \item To ensure global sections on ${\cal M}$, the negative number $b^{i }$ can only appear in two $\mathbb{P}^1$ factors or one $\mathbb{P}^2$ (see Appendix \ref{line_cohom} for an analysis of the Koszul sequences and the Bott Theorem \eref{bottformula} leading to this requirement). Moreover, since the last column corresponds to an algebraic constraint arising from the vanishing of a global section of a line bundle on ${\cal M}$, we explicitly check in each case that  $h^0(\mathcal{M},\cO_{\cal M}({b^i}))>0$.

 \item We choose $b^{i}<0$ consistent with vanishing first Chern class for the gCICY.
For example, in a configuration matrix, if one of the rows corresponds to an ambient $\mathbb{P}^2$ factor (for example $\left[\begin{array}{c||cccc}
\mathbb{P}^{2} & 0 &3& 0
\end{array}\right]$) we will construct a configuration to be $\left[\begin{array}{c||cc|c}
\mathbb{P}^{2} & 0 &3+n & -n
\end{array}\right]$. For the present, preliminary scan we will restrict the integer range of $n$, allowing it to vary from $1$ to $4$, i.e. (\ref{gmatrix}), 
\begin{equation}
\label{gmatrix}
X_{(2,1)}=\left[\begin{array}{c||cc|c}
\vdots &  \vdots & \vdots & \vdots \\
\mathbb{P}^{2} & 0 & 3 & 0 \\
\vdots &  \vdots & \vdots & \vdots \\
\end{array}\right] 
\quad
\Rightarrow
\quad
X'_{(2,1)}=\left[\begin{array}{c||cc|c}
\vdots &  \vdots & \vdots & \vdots \\
\mathbb{P}^{2} & 0 & 3+n & -n \\
\vdots &  \vdots & \vdots & \vdots \\
\end{array}\right] \ ,
\end{equation}
the negative entries appearing in the generalized configuration is greater or equal to $-4$.
In this scan,  since the negative entries appearing in the generalized configuration are greater or equal to $-4$, for each $n$ we generate $4$ new generalized CICYs in one class of the generalized configuration matrices listed in Tables \ref{t:(2,1)411} - \ref{t:(2,1)111111}.   In this way, $34,192$ classes of generalized configuration matrices were generated for the initial codimension $(2,1)$ scan.
 \item To remove some singular/non-reduced, non-CY geometries, we also verify explicitly that the trivial line bundle cohomology on $X$ is $h^*(X,\cO)=\{1,0,0,1\}$. 
 So in this scanning, we have already ruled out the Type I infinite class discussed in Section \ref{infinite_sec} from the starting point.
  \end{itemize}


Besides the above requirements, there is another constraint that arises from the Koszul sequence, \eref{koz_here}. This provides certain bounds on the magnitude of non-zero entries in the rows of ${\cal M}$ and can be used to rule out certain non-CY configuration matrices. This additional constraint is described in Appendix \ref{bonus_constraint}. Under all these constraints, we constructed all the classes of generalized configuration matrices of this type, which could be analyzed in a reasonable computing time\footnote{The primary time constraints arises from the calculation of line bundle cohomology.  To complete the initial scan in finite time, we leave for future scans any line bundle cohomology whose calculation time was greater than five minutes.}.

Unlike in the codimension $(1,1)$ case and the ordinary CICY dataset, in the case of codimension $(2,1)$ configuration matrices we find many  manifolds with positive Euler number $\chi>0$. This preliminary scan yielded $57$ spaces with $\chi>0$ and $2,676$ spaces with $\chi \leq 0$  out of $34,192$ classes of generalized configuration matrices \eref{21general}, which satisfied all the requirements discussed above.  The distribution of these spaces in terms of embedding projective spaces is classified in Table \ref{(2,1)distri}  \footnote{There may still exist some redundancies in counting of these spaces like ineffective splits, some identity and reducedness, generally described in sections \ref{redun} and \ref{classcases}. Although some of these geometries are singular, we assume the method in determining the topological quantity  described in section \ref{subsec:topology} still applied, and we assume all of them are Calabi-Yau when calculating Hodge number. } .  Due to the rapidly increasing number of spaces in this class of configuration matrices, we leave a full classification of such geometries to future work.
\begin{table}[!h]
\begin{center}
\begin{tabular}{|c|c|c|c|}\hline
Embedding & \# of classes of generalized & \# of spaces with & \# of spaces with\\
   projective spaces   & configuration matrices & positive $\chi$ & non-positive $\chi$ \\\hline\hline
     $\mathbb{P}^5\times\mathbb{P}^1$ & 168& 0 & 28 \\\hline   
     $\mathbb{P}^4\times\mathbb{P}^2$ & 210 & 0 & 6 \\\hline   
   $\mathbb{P}^4\times\mathbb{P}^1\times\mathbb{P}^1$ & 1,197 & 3 & 226 \\\hline   
   $\mathbb{P}^3\times\mathbb{P}^2\times\mathbb{P}^1$ &1,800 & 2 & 261 \\\hline
   $\mathbb{P}^2\times\mathbb{P}^2\times\mathbb{P}^2$ & 550 & 0 & 12 \\\hline
   $\mathbb{P}^3\times\mathbb{P}^1\times\mathbb{P}^1\times\mathbb{P}^1$ & 4,410 & 17 & 528 \\\hline
   $\mathbb{P}^2\times\mathbb{P}^2\times\mathbb{P}^1\times\mathbb{P}^1$ & 5,235 & 9 & 511 \\\hline
   $\mathbb{P}^2\times\mathbb{P}^1\times\mathbb{P}^1\times\mathbb{P}^1\times\mathbb{P}^1$ & 12,180 & 16 & 754 \\\hline
   $\mathbb{P}^1\times\mathbb{P}^1\times\mathbb{P}^1\times\mathbb{P}^1\times\mathbb{P}^1\times\mathbb{P}^1$ & 8,442 & 10 & 350 \\\hline\hline
   Total & 34,192 & 57 & 2,676 \\\hline
   
\end{tabular}
\caption{The distribution of codimension $(2,1)$ spaces embedded in products of projective spaces.}  \label{(2,1)distri}
\end{center}
\end{table}

\subsubsection*{Codimension $(2,1)$ spaces with positive Euler number}

From Table \ref{(2,1)distri} we see that there is no single example with positive Euler number in $\mathbb{P}^5 \times \mathbb{P}^1$,
$\mathbb{P}^4 \times \mathbb{P}^2$ and
$\mathbb{P}^2 \times \mathbb{P}^2 \times \mathbb{P}^2$ which passes all the criteria (out of $168$, $210$ and $550$ classes of generalized configuration matrices scanned respectively). The distribution of the $57$ spaces is explicitly shown in  Table \ref{t:(2,1)411} - \ref{t:(2,1)111111}.  Once again in these tables,  $\CIRCLE$ indicates a smooth manifold, $\Circle$ a singular manifold and finally, $\LEFTcircle$ indicates that the smoothness check has timed out in the current run. For all the generically singular geometries we found, the singularities are not of a minimal order in that the ``normal form'', eq.~(\ref{nfXA}), and its exterior derivative both vanish. 

In terms of topological data, none of the positive Euler number examples appears in the regular CICY list \cite{Candelas:1987kf}, though many of the Hodge number pairs appear in the Kreuzer-Skarke list \cite{Kreuzer:2000xy} (however, as seen already in co-dimension $(1,1)$ case, two spaces with the same Hodge number does not guarantee they have the same Chern classes or triple intersection numbers). In some cases, we are not yet able to determine the Hodge pairs (again due to the slow computation of line bundle cohomology. The integer uncertainty in these Hodge pairs is denoted $``x"$ in the Tables below).

\begin{table}[!h]
\begin{center}
\begin{tabular}{|c|c|c|c|}\hline
$X$&$\chi$&\footnotesize{($h^{1,1}(X), h^{1,2}(X)$)}& Smoothness \\\hline\hline &&&\\[-4.2mm]
\tiny${\left[
\begin{array}{c||cc|c}
\IP^4 & 1 & 1 & 3 \\
\IP^1 & 0 & 2 & 0\\
\IP^1 & 1 & 3 & -2
\end{array}
\right]}$&$12$& $(6+x,x)$ & \Circle  \\\hline&&&\\[-4.2mm]

\tiny${\left[
\begin{array}{c||cc|c}
\IP^4 & 1 & 1 & 3 \\
\IP^1 & 0 & 2 & 0\\
\IP^1 & 1 & 4 & -3
\end{array}
\right]}$&$60$& $(30+x,x)$ & \Circle \\\hline\hline&&&\\[-4.2mm]
\tiny${\left[
\begin{array}{c||cc|c}
\IP^4 &1 & 1 & 3 \\
\IP^1 &0 & 2 & 0\\
\IP^1 &2 & 3 & -3
\end{array}
\right]}$&$24$& $(12+x,x)$ & \LEFTcircle \\\hline

\end{tabular}
\caption{$3$ results out of  $1,197$ classes of  generalized configuration matrices scanned in
  $\mathbb{P}^4\times\mathbb{P}^1\times\mathbb{P}^1$. $\Circle$, $\LEFTcircle$ indicate singular and undetermined manifolds respectively.  In the Hodge numbers, $x$ denotes an undetermined non-negative integer.}  \label{t:(2,1)411}
\end{center}
\end{table}

\begin{table}[!h]
\begin{center}
\begin{tabular}{|c|c|c|c|}\hline
$X$&$\chi$&\footnotesize{($h^{1,1}(X), h^{1,2}(X)$)}& Smoothness\\\hline\hline&&&\\[-4.2mm]

\tiny${\left[
\begin{array}{c||cc|c}
\IP^3 &1 & 0 & 3 \\
\IP^2 &1 & 3 & -1\\
\IP^1 & 0 & 1 & 1
\end{array}
\right]}$&$12$& $(6+x,x)$ & \Circle  \\\hline\hline&&&\\[-4.2mm]
\tiny${\left[
\begin{array}{c||cc|c}
\IP^3 & 1 & 0 & 3 \\
\IP^2 & 0 & 2 & 1\\
\IP^1 & 2 & 3 & -3
\end{array}
\right]}$&$18$& $(9+x,x)$ & \LEFTcircle \\\hline

\end{tabular}
\caption{$2$ results out of $1,800$ classes of  generalized configuration matrices scanned in
  $\mathbb{P}^3\times\mathbb{P}^2\times\mathbb{P}^1$.  $\Circle$, $\LEFTcircle$ indicate singular and undetermined manifolds respectively.  In the Hodge numbers, $x$ denotes an undetermined non-negative integer.} \label{t:(2,1)321}
\end{center}
\end{table}

\begin{table}[!h]
\begin{center}
\begin{tabular}{|c|c|c|c|}\hline
$X$&$\chi$&\footnotesize{($h^{1,1}(X), h^{1,2}(X)$)}& Smoothness \\\hline\hline&&&\\[-4.2mm]

\tiny${\left[
\begin{array}{c||cc|c}
\IP^3 & 2 & 0 & 2 \\
\IP^1 &0 & 1 & 1\\
\IP^1 &0 & 2 & 0\\
\IP^1 &1 & 2 & -1
\end{array}
\right]}$&$24$& $(12+x,x)$ & \Circle  \\\hline&&&\\[-4.2mm]
\tiny${\left[
\begin{array}{c||cc|c}
\IP^3 &1 & 1 & 2\\
\IP^1 &0 & 0 & 2 \\
\IP^1 &0 & 2 & 0\\
\IP^1 &1 & 3 & -2
\end{array}
\right]}$&$32$& $(16+x,x)$ & \Circle \\\hline&&&\\[-4.2mm]
\tiny${\left[
\begin{array}{c||cc|c}
\IP^3 & 1 & 0 & 3\\
\IP^1 & 0 & 1 & 1 \\
\IP^1 & 0 & 2 & 0\\
\IP^1 & 1 & 3 & -2
\end{array}
\right]}$&$72$& $(36+x,x)$ & \Circle \\\hline&&&\\[-4.2mm]
\tiny${\left[
\begin{array}{c||cc|c}
\IP^3 &1 & 0 & 3\\
 \IP^1 &0 & 1 & 1 \\
\IP^1 &0 & 2 & 0\\
\IP^1 &1 & 4 & -3
\end{array}
\right]}$&$144$& $(72+x,x)$ & \Circle \\\hline\hline&&&\\[-4.2mm]
\tiny${\left[
\begin{array}{c||cc|c}
\IP^3 & 2 & 0 & 2\\
\IP^1 & 0 & 1 & 1 \\
\IP^1 & 0 & 4 & -2\\
\IP^1 &1 & 1 & 0
\end{array}
\right]}$&$24$& $(27,15)$ & \LEFTcircle \\\hline&&&\\[-4.2mm]
\tiny${\left[
\begin{array}{c||cc|c}
\IP^3 & 2 & 0 & 2\\
\IP^1 & 0 & 0 & 2 \\
\IP^1 & 1 & 1 & 0\\
\IP^1 & 1 & 5 & -4
\end{array}
\right]}$&$16$& $(69,61)$ & \LEFTcircle \\\hline&&&\\[-4.2mm]
\tiny${\left[
\begin{array}{c||cc|c}
\IP^3 & 1 & 0 & 3 \\
\IP^1 & 0 & 1 & 1\\
\IP^1 & 1 & 2 & -1\\
\IP^1 & 0 & 3 & -1
\end{array}
\right]}$&$72$& $(36+x,x)$ & \LEFTcircle \\\hline&&&\\[-4.2mm]
\tiny${\left[
\begin{array}{c||cc|c}
\IP^3 &1 & 0 & 3\\
\IP^1 & 0 & 1 & 1 \\
\IP^1 & 0 & 5 & -3\\
\IP^1 & 1 & 1 & 0
\end{array}
\right]}$&$36$& $(32,14)$ & \LEFTcircle \\\hline&&&\\[-4.2mm]
\tiny${\left[
\begin{array}{c||cc|c}
\IP^3 & 1 & 0 & 3\\
\IP^1 & 0 & 1 & 1 \\
\IP^1 & 0 & 1 & 1\\
\IP^1 & 1 & 5 & -4
\end{array}
\right]}$&$36$& $(32,14)$ & \LEFTcircle \\\hline&&&\\[-4.2mm]
\tiny${\left[
\begin{array}{c||cc|c}
\IP^3 & 1 & 0 & 3\\
\IP^1 & 0 & 1 & 1 \\
\IP^1 & 0 & 2 & 0\\
\IP^1 & 2 & 3 & -3
\end{array}
\right]}$&$72$& $(36+x,x)$ & \LEFTcircle \\\hline&&&\\[-4.2mm]
\tiny${\left[
\begin{array}{c||cc|c}
\IP^3 & 2 & 0 & 2 \\
 \IP^1 &0 & 0 & 2 \\
\IP^1 & 0 & 6 & -4\\
\IP^1 &1 & 1 & 0
\end{array}
\right]}$&$16$& $(75,67)$ & \LEFTcircle \\\hline&&&\\[-4.2mm]
\tiny${\left[
\begin{array}{c||cc|c}
\IP^3 &1 & 1 & 2\\
\IP^1 & 0 & 0 & 2 \\
\IP^1 &2 & 0 & 0\\
\IP^1 &2 & 4 & -4
\end{array}
\right]}$&$32$& $(16+x,x)$ & \LEFTcircle \\\hline&&&\\[-4.2mm]
\tiny${\left[
\begin{array}{c||cc|c}
\IP^3 & 1 & 0 & 3\\
\IP^1 & 0 & 1 & 1 \\
\IP^1 & 0 & 5 & -3\\
\IP^1 & 2 & 1 & -1
\end{array}
\right]}$&$36$& $(31,13)$ & \LEFTcircle \\\hline&&&\\[-4.2mm]
\tiny${\left[
\begin{array}{c||cc|c}
\IP^3 & 1 & 0 & 3\\
\IP^1 & 0 & 1 & 1 \\
\IP^1 & 0 & 6 & -4\\
\IP^1 & 1 & 1 & 0
\end{array}
\right]}$&$72$& $(48,12)$ & \LEFTcircle \\\hline&&&\\[-4.2mm]
\end{tabular}
\end{center}
\end{table}

\begin{table}[H]
\begin{center}
\begin{tabular}{|c|c|c|c|}\hline
$X$&$\chi$&\footnotesize{($h^{1,1}(X), h^{1,2}(X)$)}& Smoothness \\\hline\hline&&&\\[-4.2mm]

\tiny${\left[
\begin{array}{c||cc|c}
\IP^3 & 2 & 0 & 2\\
\IP^1 & 0 & 1 & 1 \\
\IP^1 & 0 & 1 & 1\\
\IP^1 & 1 & 4 & -3
\end{array}
\right]}$&$24$& $(27,15)$ & \LEFTcircle \\\hline&&&\\[-4.2mm]
\tiny${\left[
\begin{array}{c||cc|c}
\IP^3 &1 & 0 & 3\\
\IP^1 & 0 & 4 & -2 \\
\IP^1 & 1 & 0 & 1\\
\IP^1 & 2 & 1 & -1
\end{array}
\right]}$&$48$& $(76, 52)$ & \LEFTcircle \\\hline&&&\\[-4.2mm]
\tiny${\left[
\begin{array}{c||cc|c}
\IP^3 &1 & 0 & 3\\
\IP^1 & 0 & 5 & -3 \\
\IP^1 & 1 & 0 & 1\\
\IP^1 & 2 & 1 & -1
\end{array}
\right]}$&$96$& $(116, 68)$ & \LEFTcircle \\\hline

\end{tabular}
\caption{17 results out of $4,410$ classes of  generalized configuration matrices scanned in
  $\mathbb{P}^3\times\mathbb{P}^1\times\mathbb{P}^1\times\mathbb{P}^1$.   $\Circle$, $\LEFTcircle$ indicate singular and undetermined manifolds respectively. In the Hodge numbers, $x$ denotes an undetermined non-negative integer.} \label{t:(2,1)3111}
\end{center}
\end{table}

\begin{table}[!h]
\begin{center}
\begin{tabular}{|c|c|c|c|}\hline
$X$&$\chi$&\footnotesize{($h^{1,1}(X), h^{1,2}(X)$)}& Smoothness \\\hline\hline&&&\\[-4.2mm]

\tiny${\left[
\begin{array}{c||cc|c}
\IP^2 & 0 & 0 & 3 \\
\IP^2 &1 & 1 & 1\\
\IP^1 & 2 & 0 & 0\\
\IP^1 & 2 & 1 & -1
\end{array}
\right]}$&$36$& $(18+x,x)$ & \Circle  \\\hline&&&\\[-4.2mm]
\tiny${\left[
\begin{array}{c||cc|c}
\IP^2 &0 & 2 & 1\\
\IP^2 &1 & 0 & 2\\
\IP^1 &0 & 0 & 2 \\
\IP^1 &1 & 3 & -2
\end{array}
\right]}$&$16$& $(8+x,x)$ & \Circle \\\hline&&&\\[-4.2mm]
\tiny${\left[
\begin{array}{c||cc|c}
\IP^2 & 0 & 1 & 2 \\
\IP^2 &1 & 0 & 2\\
\IP^1 & 0 & 2 & 0\\
\IP^1 & 1 & 3 & -2
\end{array}
\right]}$&$16$& $(8+x,x)$ & \Circle  \\\hline\hline&&&\\[-4.2mm]
\tiny${\left[
\begin{array}{c||cc|c}
\IP^2 & 3 & 1 & -1 \\
\IP^2 & 0 & 0 & 3\\
\IP^1 & 1 & 0 & 1\\
\IP^1 & 0 & 1 & 1
\end{array}
\right]}$&$18$& $(9+x,x)$ & \LEFTcircle  \\\hline&&&\\[-4.2mm]
\tiny${\left[
\begin{array}{c||cc|c}
\IP^2 & 0 & 1 & 2 \\
\IP^2 & 1 & 0 & 2\\
\IP^1 & 0 & 6 & -4\\
\IP^1 & 1 & 1 & 0
\end{array} 
\right]}$&$16$& $(32,24)$ & \LEFTcircle \\\hline&&&\\[-4.2mm]
\tiny${\left[
\begin{array}{c||cc|c}
\IP^2 & 0 & 0 & 3 \\
\IP^2 & 1 & 1 & 1\\
\IP^1 &1 & 3 & -2\\
\IP^1 & 2 & 0 & 0
\end{array} 
\right]}$&$36$& $(18+x,x)$ & \LEFTcircle \\\hline&&&\\[-4.2mm]
\tiny${\left[
\begin{array}{c||cc|c}
\IP^2 & 0 & 0 & 3 \\
\IP^2 &1 & 1 & 1\\
\IP^1 &0 & 5 & -3\\
\IP^1 &2 & 0 & 0
\end{array} 
\right]}$ & $36$ & $(18+x,x)$ & \LEFTcircle \\\hline&&&\\[-4.2mm]
\tiny${\left[
\begin{array}{c||cc|c}
\IP^2 & 1 & 0 & 2 \\
\IP^2 &1 & 0 & 2\\
\IP^1 & 2 & 1 & -1\\
\IP^1 &0 & 4 & -2
\end{array} 
\right]}$ & $12$ & $(75, 69)$ & \LEFTcircle \\\hline&&&\\[-4.2mm]
\tiny${\left[
\begin{array}{c||cc|c}
\IP^2 & 1 & 0 & 2 \\
\IP^2 &1 & 0 & 2\\
\IP^1 &2 & 1 & -1\\
\IP^1 &0 & 5 & -3
\end{array} 
\right]}$ & $48$ & $(111,87)$ & \LEFTcircle \\\hline
\end{tabular}
\caption{9 results out of $5,235$ classes of  generalized configuration matrices scanned in
  $\mathbb{P}^2\times\mathbb{P}^2\times\mathbb{P}^1\times\mathbb{P}^1$.  $\Circle$, $\LEFTcircle$ indicate singular and undetermined manifolds respectively. In the Hodge numbers, $x$ denotes an undetermined non-negative integer.} \label{t:(2,1)2211}
\end{center}
\end{table}

\begin{table}[H]
\begin{center}
\begin{tabular}{|c|c|c|c|}\hline
$X$&$\chi$&\footnotesize{($h^{1,1}(X), h^{1,2}(X)$)}& Smoothness \\\hline\hline&&&\\[-4.2mm]
\tiny${\left[
\begin{array}{c||cc|c}
\IP^2& 1 & 0 & 2 \\
\IP^1 &1 & 0 & 1\\
\IP^1 &0 & 1 & 1\\
\IP^1 &0 & 2 & 0\\
\IP^1 &1 & 2 & -1
\end{array}
\right]}$&$24$& $(12+x,x)$ & \Circle \\\hline&&&\\[-4.2mm]

\end{tabular}
\end{center}
\end{table}

\begin{table}[H]
\begin{center}
\begin{tabular}{|c|c|c|c|}\hline
$X$&$\chi$&\footnotesize{($h^{1,1}(X), h^{1,2}(X)$)}& Smoothness \\\hline\hline&&&\\[-4.2mm]

\tiny${\left[
\begin{array}{c||cc|c}
\IP^2 & 1 & 1 & 1 \\
\IP^1 & 2 & 1 & -1\\
\IP^1 & 2 & 0 & 0\\
\IP^1 &0 & 0 & 2\\
\IP^1 &0 & 0 & 2
\end{array}
\right]}$&$32$& $(16+x,x)$ & \Circle \\\hline&&&\\[-4.2mm]
\tiny${\left[
\begin{array}{c||cc|c}
\IP^2 & 1 & 0 & 2\\
\IP^1 & 0 & 0 & 2 \\
\IP^1 & 0 & 1 & 1\\
\IP^1 & 0 & 2 & 0\\
\IP^1 & 1 & 3 & -2
\end{array}
\right]}$&$64$& $(32+x,x)$ & \Circle \\\hline\hline&&&\\[-4.2mm]
\tiny${\left[
\begin{array}{c||cc|c}
\IP^2 &1 & 0 & 2\\
\IP^1 & 0 & 0 & 2 \\
\IP^1 & 0 & 1 & 1\\
\IP^1 &0 & 5 & -3\\
\IP^1 &1 & 1 & 0
\end{array}
\right]}$&$32$& $(29,13)$ & \LEFTcircle \\\hline
\tiny${\left[
\begin{array}{c||cc|c}
\IP^2 &1 & 1 & 1\\
\IP^1 & 0 & 0 & 2 \\
\IP^1 &0 & 0 & 2\\
\IP^1 &0 & 5 & -3\\
\IP^1 &2 & 0 & 0
\end{array}
\right]}$&$32$& $(16+x,x)$ & \LEFTcircle \\\hline&&&\\[-4.2mm]
\tiny${\left[
\begin{array}{c||cc|c}
\IP^2 &1 & 1 & 1\\
 \IP^1 &0 & 0 & 2 \\
 \IP^1 &0 & 0 & 2\\
 \IP^1 &1 & 3 & -2\\
 \IP^1 &2 & 0 & 0
\end{array}
\right]}$&$ 32 $& $(16+x,x)$ & \LEFTcircle \\\hline&&&\\[-4.2mm]
\tiny${\left[
\begin{array}{c||cc|c}
\IP^2 &1 & 0 & 2\\
  \IP^1 &0 & 1 & 1 \\
 \IP^1 &0 & 4 & -2\\
 \IP^1 &1 & 0 & 1\\
 \IP^1 &1 & 1 & 0
\end{array}
\right]}$&$ 24 $& $(24,12)$ & \LEFTcircle \\\hline&&&\\[-4.2mm]
\tiny${\left[
\begin{array}{c||cc|c}
\IP^2 &1 & 0 & 2\\
 \IP^1 & 0 & 0 & 2 \\
 \IP^1 &1 & 0 & 1\\
 \IP^1 &1 & 1 & 0\\
 \IP^1 &1 & 5 & -4
\end{array}
\right]}$&$ 16 $& $ (40,32)$ & \LEFTcircle \\\hline&&&\\[-4.2mm]
\tiny${\left[
\begin{array}{c||cc|c}
 \IP^2 &0 & 0 & 3 \\
 \IP^1 &1 & 0 & 1\\
 \IP^1 &1 & 0 & 1\\
 \IP^1 &1 & 1 & 0\\
 \IP^1 &1 & 4 & -3
\end{array}
\right]}$&$36$& $(33,15)$ & \LEFTcircle \\\hline&&&\\[-4.2mm]
\tiny${\left[
\begin{array}{c||cc|c}
 \IP^2 &0 & 0 & 3 \\
 \IP^1 &1 & 0 & 1\\
 \IP^1 &1 & 0 & 1\\
 \IP^1 &1 & 1 & 0\\
 \IP^1 &1 & 5 & -4
\end{array}
\right]}$&$72$& $(55,19)$ & \LEFTcircle \\\hline&&&\\[-4.2mm]
\tiny${\left[
\begin{array}{c||cc|c}
 \IP^2 &3 & 1 & -1 \\
 \IP^1 &1 & 0 & 1\\
 \IP^1 &0 & 1 & 1\\
 \IP^1 &0 & 0 & 2\\
 \IP^1 &0 & 0 & 2
\end{array}
\right]}$&$16$& $(8+x,x)$ & \LEFTcircle \\\hline&&&\\[-4.2mm]
\tiny${\left[
\begin{array}{c||cc|c}
\IP^2 &1 & 0 & 2\\
 \IP^1 & 0 & 0 & 2 \\
 \IP^1 &0 & 1 & 1\\
 \IP^1 &0 & 1 & 1\\
 \IP^1 &1 & 5 & -4
\end{array}
\right]}$&$32$& $(29,13)$ & \LEFTcircle \\\hline&&&\\[-4.2mm]
\tiny${\left[
\begin{array}{c||cc|c}
\IP^2 &1 & 0 & 2\\
 \IP^1 & 0 & 0 & 2 \\
 \IP^1 &0 & 6 & -4\\
 \IP^1 &1 & 0 & 1\\
 \IP^1 &1 & 1 & 0
\end{array}
\right]}$&$16$& $(58,50)$ & \LEFTcircle \\\hline&&&\\[-4.2mm]
\tiny${\left[
\begin{array}{c||cc|c}
\IP^2 & 0 & 0 & 3\\
 \IP^1 &0 & 5 & -3\\
 \IP^1 &1 & 0 & 1\\
 \IP^1 &1 & 0 & 1\\
 \IP^1 &1 & 1 & 0
\end{array}
\right]}$&$36$& $(53,35)$ & \LEFTcircle \\\hline&&&\\[-4.2mm]
\tiny${\left[
\begin{array}{c||cc|c}
 \IP^2 &0 & 0 & 3 \\
 \IP^1 &0 & 6 & -4\\
 \IP^1 & 1 & 0 & 1\\
 \IP^1 &1 & 0 & 1\\
 \IP^1 &1 & 1 & 0
\end{array}
\right]}$&$72$& $(75,39)$ & \LEFTcircle \\\hline&&&\\[-4.2mm]
\tiny${\left[
\begin{array}{c||cc|c}
 \IP^2 &1 & 0 & 2\\
 \IP^1 & 0 & 1 & 1 \\
 \IP^1 &0 & 1 & 1\\
 \IP^1 &1 & 0 & 1\\
 \IP^1 &1 & 4 & -3
\end{array}
\right]}$&$24$& $(24,12)$ & \LEFTcircle \\\hline

\end{tabular}
\caption{16 results out of $12,180$ classes of  generalized configuration matrices scanned in
  $\mathbb{P}^2\times\mathbb{P}^1\times\mathbb{P}^1\times\mathbb{P}^1\times\mathbb{P}^1$.  $\Circle$, $\LEFTcircle$ indicate singular and undetermined manifolds respectively.  In the Hodge numbers, $x$ denotes an undetermined non-negative integer.} \label{t:(2,1)21111}

\end{center}
\end{table}

\begin{table}[H]
\begin{center}
\begin{tabular}{|c|c|c|c|}\hline
$X$&$\chi$&\footnotesize{($h^{1,1}(X), h^{1,2}(X)$)}& Smoothness \\\hline\hline&&&\\[-4.2mm]
\tiny${\left[
\begin{array}{c||cc|c}
 \IP^1 & 2 & 1 & -1 \\
 \IP^1 &2 & 0 & 0\\
 \IP^1 &1 & 0 & 1\\
 \IP^1 &0 & 1 & 1\\
 \IP^1 &0 & 1 & 1\\
 \IP^1 &0 & 1 & 1
\end{array}
\right]}$&$16$& $(8+x,x)$ & \Circle \\\hline&&&\\[-4.2mm]
\tiny${\left[
\begin{array}{c||cc|c}
 \IP^1 & 2 & 1 & -1 \\
 \IP^1 &2 & 0 & 0\\
 \IP^1 &1 & 0 & 1\\
 \IP^1 &0 & 1 & 1\\
 \IP^1 &0 & 1 & 1\\
 \IP^1 &0 & 0 & 2
\end{array}
\right]}$&$32$& $(16+x,x)$ & \Circle \\\hline\hline
\tiny${\left[
\begin{array}{c||cc|c}
 \IP^1 &0 & 0 & 2 \\
 \IP^1 &0 & 0 & 2\\
 \IP^1 &0 & 5 & -3\\
 \IP^1 &1 & 0 & 1\\
 \IP^1 &1 & 0 & 1\\
 \IP^1 &1 & 1 & 0
\end{array}
\right]}$&$32$& $(50,34)$ & \LEFTcircle \\ \hline
\tiny${\left[
\begin{array}{c||cc|c}
 \IP^1 & 0 & 1 & 1 \\
 \IP^1 &0 & 1 & 1\\
 \IP^1 &1 & 0 & 1\\
 \IP^1 &1 & 0 & 1\\
 \IP^1 &1 & 0 & 1\\
 \IP^1 &1 & 4 & -3
\end{array}
\right]}$&$16$& $(25,17)$ & \LEFTcircle \\\hline&&&\\[-4.2mm]
\tiny${\left[
\begin{array}{c||cc|c}
  \IP^1 &0 & 0 & 2 \\
 \IP^1 &0 & 1 & 1\\
 \IP^1 &0 & 1 & 1\\
 \IP^1 &1 & 0 & 1\\
 \IP^1 &1 & 0 & 1\\
 \IP^1 &1 & 4 & -3
\end{array}
\right]}$&$32$& $(26,10)$ & \LEFTcircle \\\hline&&&\\[-4.2mm]
\tiny${\left[
\begin{array}{c||cc|c}
  \IP^1 &0 & 1 & 1 \\
 \IP^1 &0 & 4 & -2\\
 \IP^1 &1 & 0 & 1\\
 \IP^1 &1 & 0 & 1\\
 \IP^1 &1 & 0 & 1\\
 \IP^1 &1 & 1 & 0
\end{array}
\right]}$&$16$& $(25,17)$ & \LEFTcircle \\\hline&&&\\[-4.2mm]
\tiny${\left[
\begin{array}{c||cc|c}
 \IP^1 &0 & 0 & 2\\
 \IP^1 &0 & 1 & 1\\
 \IP^1 &0 & 4 & -2\\
 \IP^1 &1 & 0 & 1\\
 \IP^1 &1 & 0 & 1\\
 \IP^1 &1 & 1 & 0
\end{array}
\right]}$&$32$& $(26,10)$ & \LEFTcircle \\\hline&&&\\[-4.2mm]
\tiny${\left[
\begin{array}{c||cc|c}
 \IP^1 &0 & 0 & 2 \\
 \IP^1 &0 & 0 & 2\\
 \IP^1 &1 & 0 & 1\\
 \IP^1 &1 & 0 & 1\\
 \IP^1 &1 & 1 & 0\\
 \IP^1 &1 & 4 & -3
\end{array}
\right]}$&$32$& $(32,16)$ & \LEFTcircle \\\hline&&&\\[-4.2mm]
\tiny${\left[
\begin{array}{c||cc|c}
 \IP^1 &0 & 0 & 2 \\
 \IP^1 &0 & 0 & 2\\
 \IP^1 &1 & 0 & 1\\
 \IP^1 &1 & 0 & 1\\
 \IP^1 &1 & 1 & 0\\
 \IP^1 &1 & 5 & -4
\end{array}
\right]}$&$64$& $(52,20)$ & \LEFTcircle \\\hline&&&\\[-4.2mm]
\tiny${\left[
\begin{array}{c||cc|c}
 \IP^1 &0 & 0 & 2 \\
 \IP^1 &0 & 0 & 2\\
 \IP^1 &0 & 6 & -4\\
 \IP^1 &1 & 0 & 1\\
 \IP^1 &1 & 0 & 1\\
 \IP^1 &1 & 1 & 0
\end{array}
\right]}$&$64$& $(70, 38)$ & \LEFTcircle \\\hline

\end{tabular}
\caption{$10$ results out of $8,441$ classes of  generalized configuration matrices scanned in
  $\mathbb{P}^1\times\mathbb{P}^1\times\mathbb{P}^1\times\mathbb{P}^1\times\mathbb{P}^1\times\mathbb{P}^1$.  $\Circle$, $\LEFTcircle$ indicate singular and undetermined manifolds respectively. In the Hodge numbers, $x$ denotes an undetermined non-negative integer.} \label{t:(2,1)111111}

\end{center}
\end{table}

\subsubsection*{Codimension $(2,1)$ spaces with non-positive Euler number}

The initial scan produced $2,676$ co-dimension $(2,1)$ examples with non-positive Euler number out of the $34,192$ classes of generalized configuration matrices considered. Of these, the exact Hodge number pairs could be determined for $2,469$. For these geometeries then, we can compare the Hodge number pairs with those found in the literature to date: we find $319$ pairs with $162$ different Euler number which are not in the regular CICY list \cite{Candelas:1987kf} and  among them,  $129$ pairs with $24$ different Euler number not present in the Kreuzer-Skarke list \cite{Kreuzer:2000xy}. Further more, there are $16$ geometries with new Hodge numbers not appearing elsewhere in the literature \cite{cyexplorer}. These $16$ Hodge pairs are distributed in $8$ different Euler numbers. These  examples with new Hodge number are listed in Table \ref{(2,1)newhodge}.

\begin{table}[H]
\begin{center}
\begin{tabular}{|c|c|c|c|}\hline
\footnotesize$(h^{1,1}(X), h^{1,2}(X))$ &  $X$  \\\hline\hline&\\[-4.2mm]
$(1, 91)$ & 
\tiny${\left[
\begin{array}{c||cc|c}
\IP^2 &1 & 1 & 1\\
\IP^2 &0 & 3 & 0\\
\IP^1 &0 & 0 & 2 \\
\IP^1 &1 & 2 & -1\\
\end{array}
\right]}$ 
\\\hline&\\[-4.2mm]
$(1, 109)$ & 
\tiny${\left[
\begin{array}{c||cc|c}
\IP^2&1 & 0 & 2\\
\IP^2 &0 & 3 & 0\\
\IP^1 &0 & 1 & 1 \\
\IP^1 &1 & 3 & -2\\
\end{array}
\right]}$  \\\hline&\\[-4.2mm]
$(2,  98)$ & 
\tiny${\left[
\begin{array}{c||cc|c}
\IP^2 &1 & 0 & 2\\
\IP^1 & 0 & 2 & 0 \\
\IP^1 &0 & 1 & 1\\
\IP^1 &0 & 2 & 0\\
\IP^1 & 1 & 3 & -2\\
\end{array}
\right]}$ 
 \\\hline&\\[-4.2mm]
$(6, 18)$ & 
\tiny${\left[
\begin{array}{c||cc|c}
\IP^2 &0 & 1 & 2\\
\IP^1 &0 & 1 & 1 \\
\IP^1 &1 & 0 & 1\\
\IP^1 &1 & 0 & 1\\
\IP^1 &1 & 3 & -2\\
\end{array}
\right]}$ ,
\tiny${\left[
\begin{array}{c||cc|c}
\IP^2 &0 & 1 & 2\\
\IP^1 &0 & 3 & -1 \\
\IP^1 &1 & 1 & 0\\
\IP^1 &1 & 0 & 1\\
\IP^1 &1 & 0 & 1\\
\end{array}
\right]}$ ,
\tiny${\left[
\begin{array}{c||cc|c}
\IP^2 &0 & 1 & 2\\
\IP^1 &0 & 0 & 2 \\
\IP^1 &1 & 1 & 0\\
\IP^1 &1 & 3 & -2\\
\IP^1 &1 & 0 & 1\\
\end{array}
\right]}$ 
 \\\hline&\\[-4.2mm]
$(10, 19)$ & 
\tiny${\left[
\begin{array}{c||cc|c}
\IP^2 &0 & 0 & 3 \\
\IP^2 &1 & 1 & 1\\
\IP^1 &0 & 1 & 1\\
\IP^1 &1 & 3 & -2\\
\end{array}
\right]}$,
\tiny${\left[
\begin{array}{c||cc|c}
\IP^2 &0 & 0 & 3 \\
\IP^2 &1 & 1 & 1\\
\IP^1 &1 & 0 & 1\\
\IP^1 &1 & 5 & -4\\
\end{array}
\right]}$  ,
\tiny${\left[
\begin{array}{c||cc|c}
\IP^2 &0 & 0 & 3 \\
\IP^2 &1 & 1 & 1\\
\IP^1 &1 & 0 & 1\\
\IP^1 &2 & 3 & -3\\
\end{array}
\right]}$  ,
\tiny${\left[
\begin{array}{c||cc|c}
\IP^2 &0 & 0 & 3 \\
\IP^2 &2 & 2 & -1\\
\IP^1 &0 & 1 & 1\\
\IP^1 &1 & 0 & 1\\
\end{array}
\right]}$  
\\\hline&\\[-4.2mm]
$(9, 13)$  & 
\tiny${\left[
\begin{array}{c||cc|c}
\IP^3 &2 & 0 & 2\\
\IP^1 &0 & 1 & 1 \\
\IP^1 &0 & 1 & 1\\
\IP^1 &1 & 3 & -2\\
\end{array}
\right]}$ ,
\tiny${\left[
\begin{array}{c||cc|c}
\IP^3 &2 & 0 & 2\\
\IP^1 &0 & 3 & -1 \\
\IP^1 &0 & 1 & 1\\
\IP^1 &1 & 1 & 0\\
\end{array}
\right]}$ 
 \\\hline&\\[-4.2mm]
$(9, 15)$  & 
\tiny${\left[
\begin{array}{c||cc|c}
\IP^3 &1 & 0 & 3\\
\IP^1 &0 & 1 & 1 \\
\IP^1 &1 & 1 & 0\\
\IP^1 &1 & 3 & -2\\
\end{array}
\right]}$ 
  \\\hline&\\[-4.2mm]
  $(10, 14)$ & 
\tiny${\left[
\begin{array}{c||cc|c}
\IP^2 &1 & 0 & 2\\
\IP^1 &0 & 1 & 1 \\
\IP^1 &0 & 1 & 1\\
\IP^1 &1 & 3 & -2\\
\IP^1 &1 & 0 & 1\\
\end{array}
\right]}$,   
\tiny${\left[
\begin{array}{c||cc|c}
\IP^2 &1 & 0 & 2\\
\IP^1 &0 & 3 & -1 \\
\IP^1 &0 & 1 & 1\\
\IP^1 &1 & 1 & 0\\
\IP^1 &1 & 0 & 1\\
\end{array}
\right]}$ ,
\tiny${\left[
\begin{array}{c||cc|c}
\IP^2 &1 & 0 & 2\\
\IP^1 &0 & 0 & 2 \\
\IP^1 &0 & 1 & 1\\
\IP^1 &1 & 1 & 0\\
\IP^1 &1 & 3 & -2\\
\end{array}
\right]}$ 
\\\hline

\end{tabular}
\caption{The Hodge pairs and configuration matrices of novel codimension (2,1) examples. These new Hodge pairs do not appear in the regular CICY list \cite{Candelas:1987kf},  Kreuzer-Skarke list \cite{Kreuzer:2000xy} or elsewhere in the known literature \cite{cyexplorer}. }  \label{(2,1)newhodge}
\end{center}
\end{table}

\section{Physics Applications and Outlook}\label{phys_outlook}
In this work we have introduced a new construction of Calabi-Yau spaces that has the potential to yield very large datasets of manifolds of a variety of different dimensions. We have seen already that this construction has yielded new manifolds as well as previously unknown Hodge number pairs.  If we take one for each infinite family in the codimension (1,1) dataset, we constructed  $28$ spaces. For codimension (2,1) case, there are $57$ spaces with positive Euler number and $2,676$ with non-positive Euler number.
In total we constructed  $2,761$ spaces. 

In the discussion below we turn briefly to some of the physics applications of this new dataset of manifolds. Many of the topics here deserve to be the subject of future work and we highlight the ways that this dataset will be particularly applicable to many physical problems, as well as areas where interesting open questions remain.

\subsection*{Fibration structure} \label{fibstruct}
The obvious fibration structures exhibited by generalized CICYs follow a very similar form to those seen in standard semi-positive configuration matrices. Consider a configuration matrix which can be put in the following form by row and column permutations:
\begin{equation}\eqlabel{mrfib}
{ X}= \left[\begin{array}{c||cc}  {\cal A}_1 & 0 & {\cal F} \\ 
{\cal A}_2 & {\cal B} & {\cal T} \end{array}\right] .
\end{equation}
where  ${\cal A}_1$ and ${\cal A}_2$ are two products of  projective spaces, while ${\cal F}, {\cal B}$ and ${\cal T}$ are block sub-matrices. Such a configuration matrix, in the case of semi-positive configuration matrices, describes a fibration of the manifold given by $F=\left[ {\cal A}_1 || {\cal F}\right]$ over the base $B=\left[ {\cal A}_2 || {\cal B}\right]$, where the variation of the fibre over the base is determined by the matrix ${\cal T}$. It has been observed, for example, that almost all known CICY three- and four-folds are fibered with Calabi-Yau manifolds of every lower dimension multiple times  \cite{US_FUTURE2}.

As we have mentioned, a similar structure is exhibited by generalized CICYs. Consider, for example, the configuration \eref{eg2} for a gCICY three-fold discussed in the introduction. In this example, we can rearrange the matrix and make the division as follows,
\begin{equation}
X =
\left[\begin{array}{c||cc|cc}
\mathbb{P}^{5} & 3 & 1 & 1 & 1\\
\mathbb{P}^{1} & 1 & 1 & 1 & -1\\\hdashline
\mathbb{P}^{1} & 1 & 1 & -1 & 1\\
\end{array}\right] \;.
\end{equation}
Here the base is ${B}=  \mathbb{P}^1$ and the fibre, $F$, is given as
\begin{eqnarray}
{F}= \left[\begin{array}{c||ccc|c}
\mathbb{P}^{5} & 3 & 1 & 1 & 1\\
\mathbb{P}^{1} & 1 & 1 & 1 & -1\\
\end{array}\right],
\end{eqnarray}
which is a K3 surface. 
A difference with the normal CICY situation is that we must check that any mixed-sign line bundles involved in defining the fiber and the base, ${\cal O}_{\cM'}(1,-1)$ in this case, with
$\tiny{
{\cal M'}=\left[\begin{array}{c||cccc}
\mathbb{P}^{5} & 3 & 1 & 1 \\
\mathbb{P}^{1} & 1 & 1 & 1 \\
\end{array}\right]}
$,
have global sections such that the associated manifolds do indeed exist. Computation of the line bundle cohomology indeed leads to $h^0({\cal M'}, {\cal O}_{\cM'}(1,-1))= 2>0$. As a further check, the Hodge number computation reveals that $h^{0,0}({F})=1$ and $h^{1,1}({F})=20$ as they should. The gCICY three-fold at hand is also elliptically fibered, which can similarly be seen from the division, 
\begin{equation}
X =
\left[\begin{array}{c||cc|cc}
\mathbb{P}^{5} & 3 & 1 & 1 & 1\\\hdashline
\mathbb{P}^{1} & 1 & 1 & 1 & -1\\
\mathbb{P}^{1} & 1 & 1 & -1 & 1\\
\end{array}\right],
\end{equation}
that makes manifest the fibration of the elliptic curve,
\begin{equation}
{F}=\left[\begin{array}{c||cccc}
\mathbb{P}^{5} & 3 & 1 & 1 &1\\
\end{array}\right]\ , 
\end{equation} over the base ${B}=\mathbb{P}^1 \times \mathbb{P}^1$. Note that the elliptic and K3 fibration structures here are nested.

As another example, the following gCICY four-fold,
\begin{equation}\label{cy3fib}
X =
\left[\begin{array}{c||c|c}
\mathbb{P}^{3} & 2 & 2\\
\mathbb{P}^{1} & 0 & 2\\
\mathbb{P}^{1} & 0 & 2\\
\mathbb{P}^{1} & 3 & -1\\
\end{array}\right],
\end{equation}
similarly has a nested fibration structure of elliptic, K3, and Calabi-Yau three-fold fibers, over the bases $\IP^1 \times \IP^1 \times \IP^1$, $\IP^1 \times \IP^1$, and $\IP^1$, respectively.
The following divisions of the configuration~(\ref{cy3fib}) in turn make such fibrations manifest:
\begin{equation}
\left[\begin{array}{c||c|c}
\mathbb{P}^{3} & 2 & 2\\ \hdashline
\mathbb{P}^{1} & 0 & 2\\
\mathbb{P}^{1} & 0 & 2\\
\mathbb{P}^{1} & 3 & -1\\
\end{array}\right] ; \quad
\left[\begin{array}{c||c|c}
\mathbb{P}^{3} & 2 & 2\\
\mathbb{P}^{1} & 0 & 2\\ \hdashline
\mathbb{P}^{1} & 0 & 2\\
\mathbb{P}^{1} & 3 & -1\\
\end{array}\right] ; \quad
\left[\begin{array}{c||c|c}
\mathbb{P}^{3} & 2 & 2\\
\mathbb{P}^{1} & 0 & 2\\
\mathbb{P}^{1} & 0 & 2\\ \hdashline
\mathbb{P}^{1} & 3 & -1\\
\end{array}\right] . 
\end{equation}

It seems likely that the generalized CICYs exhibit the same rich fibration structure that has recently been investigated in other constructions \cite{Gray:2013mja,Gray:2014kda,Gray:2014fla,Taylor:2015isa,Johnson:2014xpa,Martini:2014iza,Morrison:2012np,Morrison:2012js,Anderson:2014gla}. Naturally, a rigorous confirmation of such a claim must await a full classification of the dataset.

\subsubsection*{Discrete Symmetries, Torsion and Wilson Lines}
One application of this simple, algebraic set of manifolds lies in the ease with which its discrete automorphisms can be classified. Discrete symmetries of CY manifolds have a wide number of applications in string compactifications ranging from Type II orientifold actions \cite{Cvetic:2010ky} to heterotic Wilson lines \cite{Braun:2004xv,Ovrut:2012wg,Anderson:2009mh,Apruzzi:2014dza} to discrete symmetries in F-theory (see for example \cite{Garcia-Etxebarria:2014qua,Karozas:2014aha,Mayrhofer:2014haa,Mayrhofer:2014laa,Anderson:2014yva,Cvetic:2015moa,Grimm:2015ona} for recent work). The role of cohomological torsion in Mirror symmetry \cite{Braun:2007tp,Braun:2007xh,Braun:2007vy} is also an ongoing subject of investigation.

For the ordinary CICYs, the set of freely acting discrete symmetries was recently classified in \cite{Braun:2010vc}. The structure of this classification took a two step approach: 
\begin{enumerate}
\item Classify discrete symmetries descending from the ambient projective space factors (i.e. whose action was manifest on the homogenous coordinates of a product of projective spaces)
\item Consider CICYs that are invariant under those actions (i.e. equivariant normal bundles) and for which the induced symmetry action is fixed point free.
\end{enumerate}

For the manifolds presented here, the same procedure directly applies. Once again, the ambient space symmetry actions are known and their action on the defining relations -- even the ``rational" sections described here -- can be readily studied. Since the dataset includes many manifolds with small Hodge numbers already, it is intriguing future area of study to consider the quotient manifolds $X/\Gamma$ under the already classified finite discrete groups $\Gamma$.

\subsubsection*{Computability and simple algebraic construction}
The two most commonly used datasets of CY manifolds -- the CICY list \cite{Candelas:1987kf} and the Kreuzer-Skarke list \cite{Kreuzer:2000xy} -- have played an important role in the development of string theory largely due to the simple algebraic nature of the constructions. The arenas of ordinary projective or toric geometry provide a rich toolkit which can be used to calculate the necessary topological and cohomological structure of string backgrounds (i.e. bundle valued cohomology, Chern classes, intersection numbers, etc.)~\cite{cicy, Blumenhagen:2010pv,Blumenhagen:2010ed,Blumenhagen:2011xn, Kreuzer:2002uu, Altman:2014bfa, Gao:2013pra}.

For illustration, one such recent application in heterotic string theory includes systematic, algorithmic searches for heterotic Standard Model vacua on smooth CY $3$-folds \cite{Anderson:2007nc,Anderson:2008uw,He:2009wi, Anderson:2011ns,He:2011rs,Anderson:2012yf,Anderson:2013xka,Anderson:2014hia,He:2013ofa,Lin:2014lya}. These scans over literally hundreds of billions of heterotic vacua have only been possible due to the simple description of the CICY geometry and vector bundles over it and the suitability of these algebraic constructions for analysis using computational algebraic geometry. We emphasize here that this dataset is on an equal footing in terms of computational ease. Indeed, one could also follow a similar program of Calabi-Yau manifold construction to the one discussed here for the case of general Toric ambient spaces.

\subsubsection*{M-theory on CY 4-folds and instantons}
The simple nature of ordinary CICYs in products of projective spaces have also allowed for the easy extraction of certain general properties of instanton physics in heterotic string theory, M-theory and F-theory (see for example \cite{Beasley:2003fx,Beasley:2005iu}). We explore one aspect of this here involving the physics of instantons in $3$-dimensional compactifications of M-theory.

In \cite{Witten:1996bn}, Witten pointed out that for M-theory on a CY $4$-fold there is a crucial condition that must be satisfied for any instanton to contribute non-trivially to the N=1 superpotential in $3$-dimensions. In particular, it is necessary that 
\beq
\chi(D, {\cal O}(D))=1 \ .
\eeq
In \cite{Witten:1996bn} it is argued that for ordinary CICYs in products of projective space, there are no $D$'s with the required properties and thus, the superpotential is identically zero. This argument is based on fact that the set of effective divisors on the CICY consist of those that descend from the ambient product of projective spaces. As we have demonstrated in Section \ref{kahlersec}, in the case of gCICYs the effective cone of $X$ is generically much larger than that of the ambient space ${\cal A}= \mathbb{P}^{n_1} \times \ldots \mathbb{P}^{n_m}$. As a result, there are many more divisors available to satisfy the condition given above. A full exploration of such effects and the geometry of gCICY fourfolds is an exciting prospect for future study.


\subsubsection*{Relationship to GLSMs?}
Positive CICYs have played an important role as the vacuum solutions of Gauged Linear Sigma Models (GLSMs) \cite{Witten:1993yc}. It is an intriguing question whether or not gCICYs could also be realized in this way? At first pass, it seems that the answer should in fact be no, since the generalization under consideration is precisely the fact that these manifolds cannot be viewed as the complete intersection of a set of strictly polynomial equations in an ambient product of projective spaces. Phrased differently, they are not realizable as complete intersection manifolds on the ambient product of projective spaces. Instead they only exhibit such a structure on ${\cal M}$ (where all negative entry line bundles have global sections) which does not carry manifest toric/$U(1)$ actions on its coordinates.

Despite this obstacle however, there remain open questions. While the manifolds cannot be realized as a complete intersection of {\it polynomial} equations on ${\cal A}= \mathbb{P}^{n_1} \times \ldots \mathbb{P}^{n_m}$, they can be described by a complete intersection of (suitably non-singular) {\it rational} functions in these coordinates as described in detail in Section \ref{secsmooth}. The question arises then, could such a system of rational conditions arise from the F-terms associated to a holomorphic superpotential of a GLSM? If so, it would perhaps provide a novel generalization of the solutions studied to date in the literature. We leave this as an intriguing topic of future investigation.

\section*{Acknowledgements}
 The authors would like to thank A. Lukas for helpful conversations and the Universidad Aut\'onoma de Madrid for hospitality during the final stage of this work. The work of LA (and XG in part) is supported by NSF grant PHY-1417337 and that of JG (and SJL in part) is supported by NSF grant PHY-1417316. The work of FA was in part supported by the German Research Foundation (DFG) and the RTG 1463 ``Analysis, Geometry and String Theory". FA would like to thank Virginia Tech for hospitality during various stages of this work.

\appendix

\section{Line Bundles and Cohomology}\label{line_cohom}
An important tool in the computation of vector bundle-valued cohomology on CICYs or gCICYs is the calculation of line bundle cohomology. In this section we focus on the cohomology of ${\cal O}_{X}(D)$ where $D$ is a divisor obtained by restriction from the ambient space (i.e. the divisor is ``favorable" in the sense used in \cite{Anderson:2007nc,Anderson:2008uw,Anderson:2011ns,Anderson:2012yf,Anderson:2013xka}).

The Bott-Borel-Weil Theorem (see \cite{Eastwood} for the form used here) is a powerful tool that can be used to calculate the cohomology of $V$, a holomorphic homogeneous vector bundle over some Flag manifold, $M$. Here we will apply this technology in the simple case of projective space to obtain the so-called Bott-formula \cite{hartshorne} for cohomology of line bundles on a single projective space:

\beq\label{bottformula}
h^{q}(\mathbb{P}^{n}, \mathcal{O}_{\IP^n}(k))=\left\{
\begin{array}
[c]{ll}%
\binom{k+n}{n} & q=0\quad k>-1\\
1 & q=n\quad k=-n-1\\
\binom{-k-1}{-k-n-1} & q=n\quad k<-n-1\\
0 & \mbox{otherwise}
\end{array}
\right. \ .
\eeq
The computation of line bundle cohomology described by the Bott-Borel-Weil theorem is easily generalized to products of projective space using the K\"unneth formula \cite{Hubsch:1992nu} which gives the cohomology of bundles over a direct product of
spaces. For products of projective spaces it states that:
\beq\label{kunneth}
H^n(\IP^{n_1} \times \ldots \times \IP^{n_m}, \cO(q_1, \ldots, q_m)) =
\bigoplus_{k_1+\ldots+k_m = n} H^{k_1}(\IP^{n_1},\cO(q_i)) \times
\ldots \times H^{k_m}(\IP^{n_m},\cO(q_m)) \ ,
\eeq
With this in hand, we can compute the cohomology of line bundles over the ambient space. 

Next, given the Koszul sequence \eref{koszul}, it is clear that the cohomology $H^\ast(X, L)$ can be determined in terms of $H^\ast({\cal M}, L)$ for any line bundle $L$ obtained by restriction from ${\cal M}$. Here we review briefly the techniques for calculating line bundle cohomology on a general (non-CY) complete intersection manifold ${\cal M}$ of the form given in \eref{conf}.

\beq\label{koszulA}
0 \to {\cal V} \otimes \wedge^K N_X^* \to {\cal V} \otimes \wedge^{K-1} N_X^*
\to \ldots \to {\cal V} \otimes N_X^* \to {\cal V} \to {\cal V}|_X \to 0 \ .
\eeq

We can break the sequence \eref{koszulA} into a series of short exact sequences as 
\bea
0 \to {\cal V} \otimes \wedge^K N_X^* \to {\cal V} \otimes \wedge^{K-1} N_X^*
\to \mathcal{K}_1 \to 0 \ , \\
0 \to \mathcal{K}_1 \to {\cal V} \otimes \wedge^{K-2} N_X^* \to \mathcal{K}_2 \to 0  \ , \\
\ldots \\
0 \to \mathcal{K}_{K-1} \to {\cal V} \to {\cal V}|_X \to 0 \ ,
\eea
and each of these short exact sequences will give rise to a long exact sequence in cohomology:
\bea \label{long_exact}
0&\to& H^0({\cal A}, {\cal V} \otimes \wedge^K N_X^*) \to H^0({\cal A}, {\cal V} \otimes \wedge^{K-1} N_X^*)
\to H^0({\cal A}, \mathcal{K}_1) \ , \\
0 &\to& H^0({\cal A},\mathcal{K}_1) \to H^0({\cal A}, {\cal V} \otimes \wedge^{K-2} N_X^*) \to H^0({\cal A}, \mathcal{K}_2) \to \ldots  \ , \\
\ldots \\
0 &\to& H^0({\cal A}, \mathcal{K}_{K-1}) \to H^0({\cal A}, {\cal V}) \to H^0(X, {\cal V}|_X) \to \ldots \ .
\eea
To find $H^*(X, {\cal V}|_{X})$ we must determine the various cohomologies in \eref{long_exact}. It is easy to see that for higher co-dimensional spaces or tensor powers of bundles, this decomposition of sequences is a laborious process. Fortunately, the analysis of these arrays of exact sequences is dramatically simplified by the use of spectral sequences. Spectral sequences are completely equivalent to the collection of exact sequences described above, but designed for explicit cohomology computation. Since there are many good reviews of spectral sequence available in the literature \cite{Hubsch:1992nu, Distler:1987ee, Anderson:2008ex}, we will only discuss the essential features in the following paragraphs. 

To obtain the necessary cohomology of $V|_X$ from \eref{koszulA}, we define a tableaux
\beq\label{leray}
E^{j,k}_{1}(V) := H^j(A,  V \otimes \wedge^{k} N_X^*), \qquad
k = 0, \ldots, K; \  j=0, \ldots, \dim(A) = \sum_{i=1}^m n_i \ .
\eeq
which forms the first term of a {\it spectral sequence} \cite{hartshorne, GH}. Here the spectral sequence of line bundle cohomology is a complex defined by differential maps $d_i : E^{j,k}_i \to
E^{j-i+1,k-i}_i$ for $j = 1,2,\ldots$ {\it ad infinitum} where $d_i \circ d_i =0$. The higher terms in the spectral sequence are defined by
\beq \label{spec iterate}
E^{j,k}_{i+1}(V)= \frac{ker(d_{i}: E^{j,k}_{i}(V) \rightarrow E^{j-i+1,k-i}_{i}(V))}{Im(d_{i}: E^{j+i-1,k+i}_{i}(V) \rightarrow E^{j,k}_{i}(V))} \ .
\eeq
Since the number of terms in the Koszul sequence  \eref{koszulA} is finite, there exists a limit to the spectral sequence. That is, the sequence of tableaux converge after a finite number of steps to $E^{j,k}_{\infty}(V)$. The actual cohomology of the bundle $V$ is constructed from this limit tableaux:
\beq\label{hodge_converge}
h^q(X, V|_X) = \sum^{K}_{m=0} \text{rank} E^{q+m,m}_{\infty}(V) \ .
\eeq
where $h^q(X, V|_X)= \text{dim} (H^q(X, V|_X))$. In practice, this sequence converges rapidly. For other discussions of the calculation of vector/line bundle cohomology in similar contexts, see \cite{Anderson:2008ex, Blumenhagen:2010pv,Blumenhagen:2010ed,Blumenhagen:2011xn}.


\section{Trivial Bundle Cohomology and Filtering out Non-CY Configuration Matrices}\label{bonus_constraint}
As illustrated in Section \ref{secsmooth}, some configuration matrices of the form \eref{conf} do not give rise to Calabi-Yau manifolds -- due to the fact that the zero-locus of the defining equations produces a non-reduced scheme whose classical variety\footnote{i.e. the variety associated to the radical of the ideal \cite{hartshorne}.} does not satisfy the necessary CY conditions. In this section, we mention a simple criteria for configuration matrices that {\it always} give rise to such non-CY geometries. All such configuration matrices have been omitted from the scans carried out in this work.

A simple filter for ``non-CY" configurations can be obtained by considering the cohomology of the trivial bundle. In the case of a CY manifold the cohomology is $h^{\*}({\cal O}_{X})=(1,0,0,1)$. Here we present a class of manifolds which will generically produce a trivial bundle with different cohomology. Any $(p,1)$ configuration matrix of the form
\bea\label{conf_again}
\left[ \,\mathbf n\, ||\, \{\mathbf a_\alpha\}\,|\, \{\mathbf{b}_\mu\}\,\right] = \def\arraystretch{1.2}\left[\ba{c||ccc|ccc} 
{n_1}&  a^{1}_{1} & a^{1}_{r} &\cdots & b^1_1  \\
{n_2} &  a^{2}_{1} & 0  &\cdots & b^2_1 \\ 
\vdots &  \vdots &\ddots&\vdots & \vdots & \ddots  \\
{n_m} & a^{m}_{1} &a^{m}_{r} &\cdots  & b^m_1  \\ 
\ea\right]  \ , 
\eea
which has a column containing a zero and {\it another entry $a^{s}_{r}$ in the same column satisfying}
\beq\label{prob_conf}
a^{s}_{r} \geq n_{r}+1
\eeq  
for its respective ambient $\mathbb{P}^{n_r}$ dimension.

To see why such configuration matrices are problematic, as in Section \ref{kosz_sec}, we can consider the Koszul sequence:
 \beq\label{koz_here}
 0 \to N_{X}^{\vee} \to {\cal O}_{M} \to O_X \to 0 \ ,
 \eeq
where the cohomology of the line bundles on $M$ is given in terms of the ambient product of projective spaces via
\bea\label{trivM}
0 \to  \wedge^K N_M^* \to \wedge^{K-1} N_M^*
\to \mathcal{K}_1 \to 0 \ , \\
0 \to \mathcal{K}_1 \to \wedge^{K-2} N_M^* \to \mathcal{K}_2 \to 0 \ , \\
\ldots \\
0 \to \mathcal{K}_{K-2} \to {N_{M}}^* \to  \mathcal{K}_{K-1}  \ , \\
0 \to \mathcal{K}_{K-1} \to {\cal O}_{\cal A} \to {\cal O}|_M \to 0 \ .
\eea
The key observation here is that if the configuration matrix of ${\cal M}$ contains a column as described above then generically $h^i({\cal M}, {\cal O}_{\cal M}) \neq 0$ for some $i>0$. When this result is combined with the long exact sequence associated to \eref{koz_here}, it leads to trivial bundle cohomology on $X$ that is incompatible with the CY condition. 

This is most simply illustrated via an example. Consider the following configuration matrix:
\begin{equation}
X=\left[\begin{array}{c|cc|c}
\pp^{4} & 1 & 0 & 4\\
\pp^{1} & 1 & 2 & -1 \\
\pp^{1} & 0 & 3 & -1
\end{array}\right] \ ;
\end{equation}
Here the sequence \eref{trivM} takes the form
\begin{align}
& 0 \to {\cal O}(-1,-3,-3) \to {\cal O}(-1,-1,0)\oplus {\cal O}(0,-2,-3) \to \mathcal{K}_1 \to 0 \ , \\
& 0 \to \mathcal{K}_1 \to {\cal O}_{\cal A} \to {\cal O}_{M} \to 0 \ .
\end{align}
Unlike in the case of good configuration matrices, the line bundle ${\cal O}(0,-2,-3)$ does not have entirely vanishing or top cohomology. Rather, $H^2({\cal A}, {\cal O}(0,-2,-3))=2$. As a result, the long exact sequence in cohomology associated to this sequence yields
\beq
h^*(M, {\cal O}_M)=(1,2,0,0,0) \ .
\eeq
Combining this with the Koszul sequence \eref{koz_here} for $X \subset M$
\beq
0 \to {\cal O}(-4,1,1) \to {\cal O}_M \to {\cal O}_{X} \to 0 \ ,
\eeq
and considering the long exact sequence in cohomology, we arrive at the final conclusion:
\beq
h^*(X,{\cal O}_{X})=(1,2,0,3) \ .
\eeq
This fails to agree with the $h^*(X,{\cal O}_{X})=(1,0,0,1)$ of a CY manifold. It is straightforward to verify that similar problematic cohomology will arise for $h^*(X, {\cal O}_X)$ for any configuration of the form given in \eref{conf_again} and \eref{prob_conf}.

\section{Numerical Method for Section Construction}\label{numsec}
In this Appendix, we discuss a numerical method to construct the global sections of a line bundle for defining a gCICY three-fold. For a configuration matrix, eq.~(\ref{conf}), let us explain how the sections,
\beq
q_1 \in H^0(\cM, \cO_\cM (\mathbf b_1)) \ , 
\eeq
can be systematically constructed on a computer. For simplicity, only the construction of $q_1$ is discussed here, but the method straight-forwardly generalizes to sections coming from any other columns of the configuration matrix. 

As described in Section~\ref{secsmooth}, let us consider the rational form,~(\ref{q}),
\beq\label{q1}
q_1 = \frac{N(\mathbf x_1, \cdots, \mathbf x_m)}{D(\mathbf x_1, \cdots, \mathbf x_m)} \ , 
\eeq
for a section $q_1 \in H^0(\cM, \cO_\cM(\mathbf b_1))$, 
where $N$ and $D$ are a polynomial in $\mathbf x_1, \cdots, \mathbf x_m$, of multi-degree $\left[ \mathbf b_1\right]_+$ and $\left[ \mathbf b_1\right]_-$, respectively. We should then require that the numerator $N$ vanishes on the divisor $D=0$ of $\cM$. 

At the practical level on a computer, we first choose a generic denominator polynomial $D$ and write $N$ as a linear combination,
\beq
N(\mathbf x_1, \cdots, \mathbf x_m) = \sum\limits_{{\rm deg}\, \mathbf m = \left[\mathbf b_1 \right]_+} c_{\mathbf m} \,\cdot \mathbf m(\mathbf x_1, \cdots, \mathbf x_m) \ , 
\eeq
of all the monomials $\mathbf m$ of the right multi-degree, with their coefficients, $c_{\mathbf m}$, being a free parameter. We then intersect the divisor $D=0$ of $\cM$, whose complex dimension is, 
\beq
{\rm dim}_\IC \, \cM - 1 = L+2 \ , 
\eeq
with an appropriate number of generic multi-linear hypersurfaces, 
\beq\label{hypers}
h_i(\mathbf x_1, \cdots, \mathbf x_m)=0\ , \quad{\text{for~~}} i=1, \cdots, L+2 \ , 
\eeq
so that the resulting solution set upon the slicing may only consist of a finite number of points,
\beq
\mathcal I_{\mathbf h}=\{ \, x \in \cA ~|~ D(x) =0 \ ,~p_\alpha(x) = 0 ~\text{for}~  1 \leq \alpha \leq K \ , ~ h_i(x) = 0 ~\text{for}~ 1\leq i \leq L+2 \} \ ,
\eeq
which can be obtained numerically on a computer. Here, the subscript $\mathbf h$ collectively denotes the choice of generic hypersurfaces,~(\ref{hypers}). 
Then we demand that the numerator polynomial, $N$, should vanish when evaluated at each point $x=(\mathbf x_1, \cdots, \mathbf x_m) \in \mathcal I_{\mathbf h}$, thereby obtaining a system of linear equations on the coefficient parameters, $c_{\mathbf m}$.  
The procedure should be repeated with different generic slicing choices, $\mathbf h$, and the additional constraints should augment the system until it becomes big enough so that yet another ${\mathbf h}$ choice does not give rise to any more independent constraints. Once the system saturates in the aforementioned sense, we end up with a subspace of the available linear combinations for $N$, each of which will lead to a globally holomorphic section, $q_1 = N/D$. 
Note that such a linear system on $c_{\mathbf m}$ leads to a necessary condition for $q_1$ to be a global section over $\cM$. It is also sufficient, however, given the genericity of the slicing choices. 

Let us illustrate our numerical method with the gCICY in eq.~(\ref{e.g.singular}), 
\begin{equation}
X=\left[\begin{array}{c|cc|c}
\pp^{3} & 2 & 0 &2\\
\pp^{1} & 0 & 1&1\\
\pp^{1} & 0 & 2 & 0 \\
\pp^{1} & 1 & 2 & -1 
\end{array}\right] \ ; \quad \cM=
\left[\begin{array}{c|cc}
\pp^{3} & 2 & 0  \\
\pp^{1} & 0 & 1 \\
\pp^1 & 0 & 2 \\
\pp^{1} & 1 & 2 
\end{array}\right] \ ,
\end{equation} 
by constructing sections, 
\beq
q \in H^0(\cM, \cO_\cM(2,1,0,-1)) \ ,  
\eeq 
for the last column of the configuration matrix.
To begin with, we fix the complex structure of $\cM$ by choosing two sections, $p_1$ and $p_2$, as generic polynomials with the right multi-degree. Moving on to $q$ for the remaining complex structure of $X$, let us start with the rational form,~(\ref{q1}), where the multi-degrees of $N$ and $D$ are $(2,1,0,0)$ and $(0,0,0,1)$, respectively. Then we choose a generic denominator, say,
\beq
D=3 \, x_4^0 -7 x_4^1 \ , 
\eeq
that is linear in $\mathbf x_4$ and write the numerator as a linear combination,
\beq
N = \sum\limits_{A=1}^{20} c_{A} \,\cdot \mathbf m_A(\mathbf x_1, \mathbf x_2) \ , 
\eeq
of the $20$ monomials $\mathbf m_A$ with bi-degree $(2,1)$ in $\mathbf x_1$ and $\mathbf x_2$.
The last choice we should make is $L+2=3$ generic multi-linear polynomials, $h_1$, $h_2$ and $h_3$, each of which is a generic linear combination of 32 multi-linear monomials. 
For each generic choice of $\mathbf h=(h_1, h_2, h_3)$, an $18$-point set $\mathcal I_{\mathbf h}$ is obtained numerically. Consequently, we are given the corresponding $18$ linear constraints on the $20$ coefficient parameters, $c_A$. However, these $18$ are not linearly independent in that when the system is solved there remains $8$ free parameters. Thus, we repeat the procedure with another generic choice of $\mathbf h$ and add the resulting $18$ linear constraints to the constraint system. Now the system has $36$ equations on $20$ variables and ends up leading to a $2$-parameter family of solutions. By repeating it with yet another $\mathbf h$ choice, the system now consists of $54$ linear constraints. However, the solution set is not constrained any further and is still parametrized by the same $2$ free parameters. This means that we end up with the $2$-dimensional subspace of linear combinations for $N$, which in turn results in the corresponding $2$-dimensional space of sections for $q=N/D$. 
An independent computation of line bundle cohomology leads to $h^0(\cM, \cO_\cM(2,1,0,-1))=2$, which indicates the completeness of our numerical section construction method for this example.



\begin{thebibliography}{99}


\bibitem{Hubsch:1986ny} 
  T.~Hubsch,
  ``Calabi-yau Manifolds: Motivations and Constructions,''
  Commun.\ Math.\ Phys.\  {\bf 108}, 291 (1987).

\bibitem{Candelas:1987kf} 
  P.~Candelas, A.~M.~Dale, C.~A.~Lutken and R.~Schimmrigk,
  ``Complete Intersection Calabi-Yau Manifolds,''
  Nucl.\ Phys.\ B {\bf 298}, 493 (1988).

\bibitem{Green:1986ck} 
  P.~Green and T.~Hubsch,
  ``Calabi-yau Manifolds as Complete Intersections in Products of Complex Projective Spaces,''
  Commun.\ Math.\ Phys.\  {\bf 109}, 99 (1987).

\bibitem{Candelas:1987du} 
  P.~Candelas, C.~A.~Lutken and R.~Schimmrigk,
  ``Complete Intersection Calabi-yau Manifolds. 2. Three Generation Manifolds,''
  Nucl.\ Phys.\ B {\bf 306}, 113 (1988).

  \bibitem{Anderson:2011ns} 
  L.~B.~Anderson, J.~Gray, A.~Lukas and E.~Palti,
  ``Two Hundred Heterotic Standard Models on Smooth Calabi-Yau Threefolds,''
  Phys.\ Rev.\ D {\bf 84}, 106005 (2011)
  [arXiv:1106.4804 [hep-th]].

  \bibitem{Anderson:2011ty} 
  L.~B.~Anderson, J.~Gray, A.~Lukas and B.~Ovrut,
  ``The Atiyah Class and Complex Structure Stabilization in Heterotic Calabi-Yau Compactifications,''
  JHEP {\bf 1110}, 032 (2011)
  [arXiv:1107.5076 [hep-th]].
  
 \bibitem{Anderson:2012yf} 
  L.~B.~Anderson, J.~Gray, A.~Lukas and E.~Palti,
  ``Heterotic Line Bundle Standard Models,''
  JHEP {\bf 1206}, 113 (2012)
  [arXiv:1202.1757 [hep-th]].

\bibitem{Anderson:2013qca} 
  L.~B.~Anderson, J.~Gray, A.~Lukas and B.~Ovrut,
  ``Vacuum Varieties, Holomorphic Bundles and Complex Structure Stabilization in Heterotic Theories,''
  JHEP {\bf 1307}, 017 (2013)
  [arXiv:1304.2704 [hep-th]].

\bibitem{Anderson:2013xka} 
  L.~B.~Anderson, A.~Constantin, J.~Gray, A.~Lukas and E.~Palti,
  ``A Comprehensive Scan for Heterotic SU(5) GUT models,''
  JHEP {\bf 1401}, 047 (2014)
  [arXiv:1307.4787 [hep-th]].

  \bibitem{Anderson:2014hia} 
  L.~B.~Anderson, A.~Constantin, S.~J.~Lee and A.~Lukas,
  ``Hypercharge Flux in Heterotic Compactifications,''
  Phys.\ Rev.\ D {\bf 91}, no. 4, 046008 (2015)
  [arXiv:1411.0034 [hep-th]].
 
  \bibitem{Buchbinder:2014qca} 
  E.~I.~Buchbinder, A.~Constantin and A.~Lukas,
  ``Heterotic QCD axion,''
  Phys.\ Rev.\ D {\bf 91}, no. 4, 046010 (2015)
  [arXiv:1412.8696 [hep-th]].

\bibitem{Brunner:1996bu} 
  I.~Brunner, M.~Lynker and R.~Schimmrigk,
  ``Unification of M theory and F theory Calabi-Yau fourfold vacua,''
  Nucl.\ Phys.\ B {\bf 498}, 156 (1997)
  [arXiv: hep-th/9610195].

\bibitem{Gray:2013mja} 
  J.~Gray, A.~S.~Haupt and A.~Lukas,
  ``All Complete Intersection Calabi-Yau Four-Folds,''
  JHEP {\bf 1307}, 070 (2013)
  [arXiv:1303.1832 [hep-th]].

\bibitem{Gray:2014kda} 
  J.~Gray, A.~Haupt and A.~Lukas,
  ``Calabi-Yau Fourfolds in Products of Projective Space,''
  Proc.\ Symp.\ Pure Math.\  {\bf 88}, 281 (2014).
  
\bibitem{Gray:2014fla} 
  J.~Gray, A.~S.~Haupt and A.~Lukas,
  ``Topological Invariants and Fibration Structure of Complete Intersection Calabi-Yau Four-Folds,''
  JHEP {\bf 1409}, 093 (2014)
  [arXiv:1405.2073 [hep-th]].

\bibitem{Candelas:1993dm} 
  P.~Candelas, X.~De La Ossa, A.~Font, S.~H.~Katz and D.~R.~Morrison,
  ``Mirror symmetry for two parameter models. 1.,''
  Nucl.\ Phys.\ B {\bf 416}, 481 (1994)
  [arXiv: hep-th/9308083].
  
\bibitem{Green:1987rw} 
  P.~Green and T.~Hubsch,
  ``Polynomial Deformations and Cohomology of Calabi-yau Manifolds,''
  Commun.\ Math.\ Phys.\  {\bf 113}, 505 (1987).
  
\bibitem{Candelas:1994hw} 
  P.~Candelas, A.~Font, S.~H.~Katz and D.~R.~Morrison,
  ``Mirror symmetry for two parameter models. 2.,''
  Nucl.\ Phys.\ B {\bf 429}, 626 (1994)
  [hep-th/9403187].

  
  \bibitem{Mavlyutov}
  A.~Mavlyutov,
  ``Embedding of Calabi-Yau deformations into toric varieties", [math/0309240].
  
  
\bibitem{2004InMat.157..621M} 
A. ~E. Mavlyutov, 
 ``Deformations of Calabi-Yau hypersurfaces arising from deformations of toric varieties ,''
Inventiones Mathematicae, 157, 621 .
[arXiv:math/0309239]


\bibitem{Mori88}
S. Mori,
 ``Flip Theorem and the Existence of Minimal Models for 3-Folds ,''
Journal of the American Mathematical Society
Vol. 1, No. 1 (Jan., 1988), pp. 117-253.

\bibitem{clemens89}
H. Clemens, J. Kollar, S. Mori,
 ``Higher-dimensional complex geometry ,'' Asterique (166) 1989
 
 
 \bibitem{hacon}
  C.~Birkar, P.~ Cascini, C.~D.~Hacon, J.~McKernan,
 ``Existence of minimal models for varieties of log general type", J. Amer. Math. Soc. {\bf 23} (2010), 405-468.

 
 \bibitem{Fujino09}
O. Fujino, 
 ``New developments in the theory of minimal models, " 
 Sugaku (Mathematical Society of Japan) 61 2009 (2) pp:162�186
 
\bibitem{Morrison:2012np} 
  D.~R.~Morrison and W.~Taylor,
  ``Classifying bases for 6D F-theory models,''
  Central Eur.\ J.\ Phys.\  {\bf 10}, 1072 (2012)
  [arXiv:1201.1943 [hep-th]].

\bibitem{Grassi:2014zxa} 
  A.~Grassi, J.~Halverson, J.~Shaneson and W.~Taylor,
  ``Non-Higgsable QCD and the Standard Model Spectrum in F-theory,''
  JHEP {\bf 1501}, 086 (2015)
  [arXiv:1409.8295 [hep-th]].

\bibitem{Morrison:2014lca} 
  D.~R.~Morrison and W.~Taylor,
  ``Non-Higgsable clusters for 4D F-theory models,''
  JHEP {\bf 1505}, 080 (2015)
  [arXiv:1412.6112 [hep-th]].
 
  \bibitem{US_FUTURE}
  L.~B.~Anderson, F.~Apruzzi, X.~Gao, J.~Gray, and S.~J.~Lee, To appear.

  
\bibitem{Kreuzer:2000xy} 
  M.~Kreuzer and H.~Skarke,
  ``Complete classification of reflexive polyhedra in four-dimensions,''
  Adv.\ Theor.\ Math.\ Phys.\  {\bf 4}, 1209 (2002)
  [arXiv: hep-th/0002240].


 \bibitem{batyrev_grassman}
  V.~Batyrev, I.~ Ciocan-Fontanine, B.~Kim, D.~ van Straten,
 ``Conifold Transitions and Mirror Symmetry for Calabi-Yau Complete Intersections in Grassmannians",
 Nucl. Phys. B514 (1998) 640 [alg-geom/9710022].

\bibitem{batyrev_flag}
V.~Batyrev, I.~Ciocan-Fontanine, B.~Kim, D.~van Straten, 
``Mirror Symmetry and Toric Degenerations of Partial Flag Manifolds", Acta Math. 184 (2000), no. 1, 1-39
[arXiv:math/9803108].

\bibitem{Batyrev:2008rp} 
  V.~Batyrev and M.~Kreuzer,
  ``Constructing new Calabi-Yau 3-folds and their mirrors via conifold transitions,''
  Adv.\ Theor.\ Math.\ Phys.\  {\bf 14}, 879 (2010)
  [arXiv:0802.3376 [math.AG]].

  
  \bibitem{Klemm:1992bx}
  A.~Klemm and R.~Schimmrigk,
  ``Landau-Ginzburg string vacua,''
  Nucl.\ Phys.\ B {\bf 411}, 559 (1994)
  [arXiv: hep-th/9204060].
 

\bibitem{2008arXiv0802.3669K} 
G. Kapustka and M. Kapustka,
``A cascade of determinantal Calabi--Yau threefolds ,''
[arXiv:0802.3669] 


\bibitem{Hori:2006dk} 
  K.~Hori and D.~Tong,
  ``Aspects of Non-Abelian Gauge Dynamics in Two-Dimensional N=(2,2) Theories,''
  JHEP {\bf 0705}, 079 (2007)
  [hep-th/0609032].
  
\bibitem{Donagi:2007hi} 
  R.~Donagi and E.~Sharpe,
  ``GLSM's for partial flag manifolds,''
  J.\ Geom.\ Phys.\  {\bf 58}, 1662 (2008)
  [arXiv:0704.1761 [hep-th]].

\bibitem{Hori:2011pd} 
  K.~Hori,
  ``Duality In Two-Dimensional (2,2) Supersymmetric Non-Abelian Gauge Theories,''
  JHEP {\bf 1310}, 121 (2013)
  [arXiv:1104.2853 [hep-th]].

   
\bibitem{Jockers:2012zr} 
  H.~Jockers, V.~Kumar, J.~M.~Lapan, D.~R.~Morrison and M.~Romo,
  ``Nonabelian 2D Gauge Theories for Determinantal Calabi-Yau Varieties,''
  JHEP {\bf 1211}, 166 (2012)
  [arXiv:1205.3192 [hep-th]].
  
\bibitem{Jockers:2012dk}
  H.~Jockers, V.~Kumar, J.~M.~Lapan, D.~R.~Morrison and M.~Romo,
  ``Two-Sphere Partition Functions and Gromov-Witten Invariants,''
  Commun.\ Math.\ Phys.\  {\bf 325} (2014) 1139
  [arXiv:1208.6244 [hep-th]].


  
  \bibitem{rodland2}
E.~A.~Rodland, 
``The Pfaffian Calabi?Yau, its mirror, and their link to the Grassmannian G(2, 7)", 
Compositio Math. 122 (2000) 135?149, arXiv:math.AG/9801092.

\bibitem{Lynker:1998pb} 
  M.~Lynker, R.~Schimmrigk and A.~Wisskirchen,
  ``Landau-Ginzburg vacua of string, M theory and F theory at c = 12,''
  Nucl.\ Phys.\ B {\bf 550}, 123 (1999)
  [hep-th/9812195].

\bibitem{Hubsch:1992nu} 
  T.~Hubsch,
  ``Calabi-Yau manifolds: A Bestiary for physicists,''
  \bibitem{huybrechts}
  D.~Huybrechts,
  ``Complex Geometry, An Introduction", Springer, 2004.


  
  \bibitem{hartshorne}
 R.~Hartshorne, 
 ``Algebraic Geometry, Springer," GTM 52, Springer-Verlag, 1977. 
 
  
 \bibitem{GH}
 P.~Grifiths, J.~Harris,
``Principles of algebraic geometry," 1978.
 
 \bibitem{Yau:1986gu}
  S.~T.~Yau,
  ``Compact Three-dimensional Kahler Manifolds With Zero Ricci Curvature,''
  In *Argonne/chicago 1985, Proceedings, Anomalies, Geometry, Topology*, 395-406.
 
 
 \bibitem{yau}
 S.~T.~Yau,
 ``On Ricci curvature of a compact K\"ahler manifold and complex Monge-Amp\'ere equation I." Comm. Pure and App. Math. 31 (1979), 339-411.
 
\bibitem{Anderson:2007nc} 
  L.~B.~Anderson, Y.~H.~He and A.~Lukas,
  ``Heterotic Compactification, An Algorithmic Approach,''
  JHEP {\bf 0707}, 049 (2007)
  [arXiv: hep-th/0702210 ].
  
\bibitem{Anderson:2008uw} 
  L.~B.~Anderson, Y.~H.~He and A.~Lukas,
  ``Monad Bundles in Heterotic String Compactifications,''
  JHEP {\bf 0807}, 104 (2008)
  [arXiv:0805.2875 [hep-th]].

   \bibitem{Bott}
  R.~Bott,
  ``On a theorem of Lefschetz", Mich. Math J. {\bf 6}, (1959), 79.

 \bibitem{stringvacua}
  J.~Gray, Y.~H.~He, A.~Ilderton, and A.~Lukas, 
  ``STRINGVACUA. A Mathematica package for studying vacuum configurations in string phenomenology," Computer Physics Communications, vol. 180, no. 1, pp. 107-119, 2009.
  
\bibitem{Anderson:2008ex} 
  L.~B.~Anderson,
  ``Heterotic and M-theory Compactifications for String Phenomenology,''
  [arXiv:0808.3621 [hep-th]]. Oxford University DPhil. Thesis (2008).

\bibitem{Donagi:2004qk} 
  R.~Donagi, Y.~H.~He, B.~A.~Ovrut and R.~Reinbacher,
 ``Moduli dependent spectra of heterotic compactifications,''
  Phys.\ Lett.\ B {\bf 598}, 279 (2004)
  [hep-th/0403291].
  
\bibitem{Donagi:2004ub} 
  R.~Donagi, Y.~H.~He, B.~A.~Ovrut and R.~Reinbacher,
  ``The Spectra of heterotic standard model vacua,''
  JHEP {\bf 0506}, 070 (2005)
  [hep-th/0411156].
  
\bibitem{Braun:2005xp} 
  V.~Braun, Y.~H.~He, B.~A.~Ovrut and T.~Pantev,
  ``Moduli dependent mu-terms in a heterotic standard model,''
  JHEP {\bf 0603}, 006 (2006)
  [hep-th/0510142].
  
\bibitem{Anderson:2009ge} 
  L.~B.~Anderson, J.~Gray, D.~Grayson, Y.~H.~He and A.~Lukas,
  ``Yukawa Couplings in Heterotic Compactification,''
  Commun.\ Math.\ Phys.\  {\bf 297}, 95 (2010)
  [arXiv:0904.2186 [hep-th]].
  
   
\bibitem{Anderson:2009mh} 
  L.~B.~Anderson, J.~Gray, Y.~H.~He and A.~Lukas,
  ``Exploring Positive Monad Bundles And A New Heterotic Standard Model,''
  JHEP {\bf 1002}, 054 (2010)
  [arXiv:0911.1569 [hep-th]].


 \bibitem{cyexplorer}
  B.~Jurke,
  CY Explorer Website,  http://cyexplorer.benjaminjurke.net.
  
  
     \bibitem{US_FUTURE2}
  L.~B.~Anderson, X.~Gao, J.~Gray, and S.~J.~Lee, To appear.

\bibitem{Anderson:2014gla} 
  L.~B.~Anderson and W.~Taylor,
  ``Geometric constraints in dual F-theory and heterotic string compactifications,''
  JHEP {\bf 1408}, 025 (2014)
  [arXiv:1405.2074 [hep-th]].
  
\bibitem{Taylor:2015isa} 
  W.~Taylor and Y.~N.~Wang,
  ``Non-toric Bases for Elliptic Calabi-Yau Threefolds and 6D F-Theory Vacua,''
 [ arXiv:1504.07689 [hep-th]].
  
\bibitem{Johnson:2014xpa} 
  S.~B.~Johnson and W.~Taylor,
  ``Calabi-Yau threefolds with large $h^{2,1}$,''
  JHEP {\bf 1410}, 23 (2014)
  [arXiv:1406.0514 [hep-th]].
  
\bibitem{Martini:2014iza} 
  G.~Martini and W.~Taylor,
  ``6D F-theory models and elliptically fibered Calabi-Yau threefolds over semi-toric base surfaces,''
  JHEP {\bf 1506}, 061 (2015)
  [arXiv:1404.6300 [hep-th]].

\bibitem{Morrison:2012js} 
  D.~R.~Morrison and W.~Taylor,
  ``Toric bases for 6D F-theory models,''
  Fortsch.\ Phys.\  {\bf 60}, 1187 (2012)
  [arXiv:1204.0283 [hep-th]].

\bibitem{Cvetic:2010ky} 
  M.~Cvetic, I.~Garcia-Etxebarria and J.~Halverson,
  ``On the computation of non-perturbative effective potentials in the string theory landscape: IIB/F-theory perspective,''
  Fortsch.\ Phys.\  {\bf 59}, 243 (2011)
  [arXiv:1009.5386 [hep-th]].
  
\bibitem{Apruzzi:2014dza} 
  F.~Apruzzi, F.~F.~Gautason, S.~Parameswaran and M.~Zagermann,
  ``Wilson lines and Chern-Simons flux in explicit heterotic Calabi-Yau compactifications,''
  JHEP {\bf 1502}, 183 (2015)
  [arXiv:1410.2603 [hep-th]].

\bibitem{Braun:2004xv} 
  V.~Braun, B.~A.~Ovrut, T.~Pantev and R.~Reinbacher,
  ``Elliptic Calabi-Yau threefolds with Z(3) x Z(3) Wilson lines,''
  JHEP {\bf 0412}, 062 (2004)
  [hep-th/0410055].

\bibitem{Ovrut:2012wg} 
  B.~A.~Ovrut, A.~Purves and S.~Spinner,
  ``Wilson Lines and a Canonical Basis of SU(4) Heterotic Standard Models,''
  JHEP {\bf 1211}, 026 (2012)
  [arXiv:1203.1325 [hep-th]].
  
\bibitem{Anderson:2014yva} 
  L.~B.~Anderson, I.~Garcia-Etxebarria, T.~W.~Grimm and J.~Keitel,
  ``Physics of F-theory compactifications without section,''
  JHEP {\bf 1412}, 156 (2014)
  [arXiv:1406.5180 [hep-th]].
  
\bibitem{Cvetic:2015moa} 
  M.~Cvetic, R.~Donagi, D.~Klevers, H.~Piragua and M.~Poretschkin,
  ``F-Theory Vacua with $Z_3$ Gauge Symmetry,''
  arXiv:1502.06953 [hep-th].

\bibitem{Grimm:2015ona} 
  T.~W.~Grimm, T.~G.~Pugh and D.~Regalado,
  ``Non-Abelian discrete gauge symmetries in F-theory,''
  arXiv:1504.06272 [hep-th].

  
\bibitem{Garcia-Etxebarria:2014qua} 
  I.~Garcia-Etxebarria, T.~W.~Grimm and J.~Keitel,
  ``Yukawas and discrete symmetries in F-theory compactifications without section,''
  JHEP {\bf 1411}, 125 (2014)
  [arXiv:1408.6448 [hep-th]].

\bibitem{Karozas:2014aha} 
  A.~Karozas, S.~F.~King, G.~K.~Leontaris and A.~Meadowcroft,
  ``Discrete Family Symmetry from F-Theory GUTs,''
  JHEP {\bf 1409}, 107 (2014)
  [arXiv:1406.6290 [hep-ph]].

\bibitem{Mayrhofer:2014haa} 
  C.~Mayrhofer, E.~Palti, O.~Till and T.~Weigand,
  ``Discrete Gauge Symmetries by Higgsing in four-dimensional F-Theory Compactifications,''
  JHEP {\bf 1412}, 068 (2014)
  [arXiv:1408.6831 [hep-th]].

\bibitem{Mayrhofer:2014laa} 
  C.~Mayrhofer, E.~Palti, O.~Till and T.~Weigand,
  ``On Discrete Symmetries and Torsion Homology in F-Theory,''
  JHEP {\bf 1506}, 029 (2015)
  [arXiv:1410.7814 [hep-th]].

\bibitem{Braun:2007tp} 
  V.~Braun, M.~Kreuzer, B.~A.~Ovrut and E.~Scheidegger,
  ``Worldsheet instantons, torsion curves, and non-perturbative superpotentials,''
  Phys.\ Lett.\ B {\bf 649}, 334 (2007)
  [arXiv: hep-th/0703134].
  
\bibitem{Braun:2007xh} 
  V.~Braun, M.~Kreuzer, B.~A.~Ovrut and E.~Scheidegger,
  ``Worldsheet instantons and torsion curves, part A: Direct computation,''
  JHEP {\bf 0710}, 022 (2007)
  [arXiv: hep-th/0703182 ].
  
\bibitem{Braun:2007vy} 
  V.~Braun, M.~Kreuzer, B.~A.~Ovrut and E.~Scheidegger,
  ``Worldsheet Instantons and Torsion Curves, Part B: Mirror Symmetry,''
  JHEP {\bf 0710}, 023 (2007)
  [arXiv:0704.0449 [hep-th]].
  


\bibitem{Braun:2010vc} 
  V.~Braun,
  ``On Free Quotients of Complete Intersection Calabi-Yau Manifolds,''
  JHEP {\bf 1104}, 005 (2011)
  [arXiv:1003.3235 [hep-th]].


\bibitem{Blumenhagen:2010pv} 
  R.~Blumenhagen, B.~Jurke, T.~Rahn and H.~Roschy,
  ``Cohomology of Line Bundles: A Computational Algorithm,''
  J.\ Math.\ Phys.\  {\bf 51}, 103525 (2010)
  [arXiv:1003.5217 [hep-th]].
  
\bibitem{Blumenhagen:2010ed} 
  R.~Blumenhagen, B.~Jurke, T.~Rahn and H.~Roschy,
  ``Cohomology of Line Bundles: Applications,''
  J.\ Math.\ Phys.\  {\bf 53}, 012302 (2012)
  [arXiv:1010.3717 [hep-th]].
  
\bibitem{Blumenhagen:2011xn} 
  R.~Blumenhagen, B.~Jurke and T.~Rahn,
  ``Computational Tools for Cohomology of Toric Varieties,''
  Adv.\ High Energy Phys.\  {\bf 2011}, 152749 (2011)
  [arXiv:1104.1187 [hep-th]].

   
\bibitem{Kreuzer:2002uu} 
  M.~Kreuzer and H.~Skarke,
  ``PALP: A Package for analyzing lattice polytopes with applications to toric geometry,''
  Comput.\ Phys.\ Commun.\  {\bf 157}, 87 (2004)
  [arXiv: math/0204356].
  
\bibitem{cicy}
  L.~B.~Anderson, J.~Gray, Y.-H.~He, S.~J.~Lee, and A.~Lukas, ``CICY package," based on methods described in [arXiv:0911.1569], [arXiv:0911.0865], [arXiv:0805.2875], [arXiv: hep-th/0703249], [arXiv: hep-th/0702210].
  
  
\bibitem{Gao:2013pra} 
  X.~Gao and P.~Shukla,
  ``On Classifying the Divisor Involutions in Calabi-Yau Threefolds,''
  JHEP {\bf 1311}, 170 (2013)
  [arXiv:1307.1139 [hep-th]].

  
\bibitem{Altman:2014bfa} 
  R.~Altman, J.~Gray, Y.~H.~He, V.~Jejjala and B.~D.~Nelson,
  ``A Calabi-Yau Database: Threefolds Constructed from the Kreuzer-Skarke List,''
  JHEP {\bf 1502}, 158 (2015)
  [arXiv:1411.1418 [hep-th]].
  
  
\bibitem{He:2009wi} 
  Y.~H.~He, S.~J.~Lee and A.~Lukas,
  ``Heterotic Models from Vector Bundles on Toric Calabi-Yau Manifolds,''
  JHEP {\bf 1005}, 071 (2010)
  [arXiv:0911.0865 [hep-th]].

  
\bibitem{He:2011rs} 
  Y.~H.~He, M.~Kreuzer, S.~J.~Lee and A.~Lukas,
  ``Heterotic Bundles on Calabi-Yau Manifolds with Small Picard Number,''
  JHEP {\bf 1112}, 039 (2011)
  [arXiv:1108.1031 [hep-th]].

\bibitem{He:2013ofa} 
  Y.~H.~He, S.~J.~Lee, A.~Lukas and C.~Sun,
  ``Heterotic Model Building: 16 Special Manifolds,''
  JHEP {\bf 1406}, 077 (2014)
  [arXiv:1309.0223 [hep-th]].
  
\bibitem{Lin:2014lya} 
  H.~Lin, B.~Wu and S.~T.~Yau,
  ``Heterotic String Compactification and New Vector Bundles,''
  arXiv:1412.8000 [hep-th].


\bibitem{Beasley:2003fx} 
  C.~Beasley and E.~Witten,
  ``Residues and world sheet instantons,''
  JHEP {\bf 0310}, 065 (2003)
  [arXiv: hep-th/0304115].
 
   
\bibitem{Beasley:2005iu} 
  C.~Beasley and E.~Witten,
  ``New instanton effects in string theory,''
  JHEP {\bf 0602}, 060 (2006)
  [arXiv: hep-th/0512039].
   


\bibitem{Witten:1996bn} 
  E.~Witten,
  ``Nonperturbative superpotentials in string theory,''
  Nucl.\ Phys.\ B {\bf 474}, 343 (1996)
  [arXiv: hep-th/9604030].
  
\bibitem{Witten:1993yc} 
  E.~Witten,
  ``Phases of N=2 theories in two-dimensions,''
  Nucl.\ Phys.\ B {\bf 403}, 159 (1993)
  [arXiv: hep-th/9301042].
  
 \bibitem{Eastwood}
  M.~G.~ Eastwood,
  ``The generalized Penrose-Ward Transform", Math. Proc. Camb. Phil. Soc. {\bf 97} (1985), 165.


\bibitem{Distler:1987ee} 
  J.~Distler and B.~R.~Greene,
  ``Aspects of (2,0) String Compactifications,''
  Nucl.\ Phys.\ B {\bf 304}, 1 (1988).
  

  
    
  
   
 
    
\end{thebibliography}
\end{document}